\documentclass[journal]{IEEEtran}

\usepackage[table,xcdraw]{xcolor}
\usepackage{graphicx}
\usepackage{tikz}
\usepackage[edges]{forest}
\usetikzlibrary{trees,positioning,shapes,shadows,arrows.meta}
\usetikzlibrary{mindmap, shadows}
\usetikzlibrary{matrix, fit, backgrounds}
\usetikzlibrary{calc, positioning}
\usepackage{caption}
\usepackage{amsmath}   
\usepackage{textcomp}  
\usepackage{enumitem}
\usepackage{booktabs}
\usepackage{amssymb}
\usepackage{tabularx}
\usepackage{multirow} 
\usepackage[blocks]{authblk}
\usepackage{xltabular}
\usepackage{makecell}
\usepackage[acronym,toc]{glossaries}
\usepackage{subcaption}
\usepackage{tabularray}

\usepackage[most]{tcolorbox}

\usepackage{enumitem}
\usepackage{hyperref}

\usepackage{cite}

\definecolor{quantumink}{RGB}{0, 45, 70} 

\newtcolorbox{ieeeinsight}[1]{
    enhanced,
    sharp corners,                          
    boxrule=0pt,                            
    frame hidden,                           
    borderline west={2.5pt}{0pt}{gray!80!black}, 
    coltitle=quantumink,                         
    fonttitle=\sffamily\scshape\small,            
    title={#1},
    colback=blue!5!white,
    boxed title style={         
        colback=white, 
        frame hidden,
        boxrule=0pt
    },
    attach title to upper,                  
    after title={\vspace{1mm}\hfill\break}, 
    left=3mm,                               
    right=2mm,
    top=5mm,
    bottom=2mm,
    fontupper=\linespread{1}\selectfont, 
}

\definecolor{headerblue}{HTML}{E3F2FD} 
\definecolor{subheadergray}{HTML}{F5F5F5}
\definecolor{pqcblue}{RGB}{52, 122, 235}
\definecolor{qkdgreen}{RGB}{46, 139, 87}
\definecolor{hyborange}{RGB}{242, 158, 41}

\newtcolorbox{mybox}[1][]{
    colback=gray!15,
    frame hidden,
    enhanced,
    #1
}

\tikzset{
    basic/.style  = {draw, text width=3cm, align=center, font=\sffamily, rectangle},
    root/.style   = {basic, rounded corners=2pt, thin, align=center, fill=green!30},
    onode/.style = {basic, thin, rounded corners=2pt, align=center, fill=green!60,text width=3cm,},
    tnode/.style = {basic, thin, align=left, fill=pink!60, text width=15em, align=center},
    xnode/.style = {basic, thin, rounded corners=2pt, align=center, fill=blue!20,text width=5cm,},
    wnode/.style = {basic, thin, align=left, fill=pink!10!blue!80!red!10, text width=6.5em},
    edge from parent/.style={draw=black, edge from parent fork right}

}

\forestset{
    direction switch/.style={
        forked edges,
        for tree={
            calign=last,
            edge+=thick, 
            font=\sffamily,
        },
        where level>=0{folder, grow'=0}{},
    },
}

\forestset{
    direction switch2/.style={
        forked edges,
        for tree={
            calign=last,
            edge+=thick, 
            font=\sffamily,
        },
        where level>=1{folder, grow'=0}{},
    },
}

\definecolor{MaturityConceptual}{HTML}{0057B8} 
\definecolor{MaturityPoC}{HTML}{D97706}        
\definecolor{MaturityReal}{HTML}{008000}       

\author[*,$\diamond$]{Ricardo Parizotto}
\author[*,$\S$]{Preeti Yadav}
\author[$\dagger$,$\ddagger$]{Marcus Freire}
\author[$\dagger$,$\ddagger$]{Anderson Tomkelski}
\author[$\dagger$,$\ddagger$]{Maycon Peixoto}
\author[$\S$,$\P$]{Emmanuel Zambrini Cruzeiro}
\author[$\|$]{Israat Haque}

\affil[$\diamond$]{Universidade Federal da Fronteira Sul (UFFS), Brazil}
\affil[$\dagger$]{Universidade Federal da Bahia (UFBA), Brazil}
\affil[$\ddagger$]{SENAI CIMATEC, Brazil}
\affil[$\S$]{Instituto de Telecomunicações (IT), Portugal}
\affil[$\P$]{Instituto Superior Técnico (IST), Universidade de Lisboa, Portugal}
\affil[$\|$]{Dalhousie University, Canada}

\begin{document}

\noindent

 \title{ A Survey on Quantum-Safe Cryptographic Mechanisms: Building Blocks and Applications }

\maketitle

\renewcommand{\thefootnote}{\fnsymbol{footnote}}
\footnotetext[1]{Ricardo and Preeti contributed equally to this work.}

\begin{abstract}

The possible emergence of fault-tolerant quantum computers may enable widely used cryptographic algorithms to be broken. These algorithms for public key encryption and digital signatures could be exposed if an efficient algorithm is employed. This poses a threat to existing infrastructure, motivating the development of mechanisms that can withstand quantum-computer cybersecurity risks. However, the transition to new security mechanisms is still underway and faces many challenges, including identifying which applications are under threat and implementing agile, scalable migration processes. In this work, we survey the state of the art in quantum-safe cryptographic mechanisms, existing applications targeting migration, and open challenges. We examine the building blocks commonly used to integrate quantum-safe mechanisms into existing applications and categorize them by the security mechanisms they employ. Next, we systematically review existing migrations into quantum-safe schemes and categorize them by application domain, including Telecommunications, the Internet of Things, and Blockchains. Finally, we summarize the challenges that must be addressed to enable a successful migration to quantum-resistant mechanisms, as well as the motivations for pursuing this migration.

\end{abstract}

\section{Introduction}
\label{sec:introduction}

The improvements in the development of quantum computers hold great promise. From computers with only a few qubits developed in the last decade to the state-of-the-art computers having over 1,000 qubits, quantum computation is a growing field. Estimates are that the number of qubits could surpass 1 million by 2035. Although there are still many open challenges, including the lack of efficient error-correction mechanisms, many applications have already been discussed, such as decrypting messages or simulating chemical compounds. In particular, the potential emergence of large, fault-tolerant quantum computers, also known as cryptographically relevant quantum computers (CRQCs), could enable the breaking of widely used cryptographic algorithms. Among these algorithms are Rivest-Shamir-Adleman (RSA) and elliptic curve cryptography (ECC), which form the basis of existing cryptographic applications, such as financial transactions or message communication. The security of these mechanisms relies on the complexity of factoring large integers or computing discrete logarithms.  These problems, however, could be solved by Shor's algorithm \cite{shor1994algorithms} in a CRQC, posing a threat to existing applications. This context underscores the critical need to migrate to quantum-safe cryptographic mechanisms. 

The terms \textit{quantum-safe cryptography} and \textit{quantum-resistant cryptography} are often used interchangeably in the literature. They generally refer to cryptographic schemes that remain secure even in the presence of reliable quantum computers \cite{etsiquantum2016}. These terms serve as general concepts that include several approaches to securing communications against quantum-enabled adversaries.\footnote{In this work, we use the terms quantum-safe and quantum-resistant interchangeably.}
More broadly, quantum-safe mechanisms can be designed either by relying on computational hardness assumptions or by exploiting the physical principles of quantum mechanics \cite{aquina2025critical}. For instance, post-quantum cryptography (PQC) relies on mathematical problems that are believed to remain computationally hard even for quantum computers. A key advantage of PQC is that it can be deployed incrementally within existing communication infrastructures. In practice, given that the full capabilities of future quantum computers remain uncertain, PQC is often used to refer to cryptographic primitives designed to be resilient against known quantum algorithms, such as Shor’s algorithm \cite{bernstein2025post,mashatan2021complex}. Another approach is quantum cryptography, which encodes information in physical systems governed by quantum physics. Quantum cryptography leverages phenomena such as the no-cloning theorem and quantum entanglement to provide information-theoretic security.

Recently, many institutions started standardizing quantum-safe mechanisms. The National Institute of Standards and Technology (NIST) has advocated for PQC and standardized cryptographic systems to enable an efficient, agile transition. The standards for key encapsulation are specified in FIPS 203 (ML-KEM), while the digital signature standards are specified in FIPS 204 (ML-DSA) and FIPS 205 (SH-DSA). NIST has also been driving the migration to new standards, contributing to crypto-agility techniques that enable rapid, safe application transitions. The European Telecommunications Standards Institute (ETSI), on the other hand, works to provide standards for the industry adoption of a quantum cryptography mechanism called quantum key distribution (QKD) \cite{saez2024current}, such as ETSI GS QKD 014 \cite{distribution2019protocol} and ETSI GS QKD 004 \cite{etsi2020quantum}. The focus of ETSI is to create standards for computer systems to interact with QKD devices and ensure interoperability. The International Telecommunication Union (ITU) also works on QKD standards, focusing on the architecture of QKD networks \cite{ITUtx}. In addition, the Internet Engineering Task Force (IETF) standardized TLS 1.3, providing the foundation for migrating the protocol to PQC. 

However, there are numerous challenges and limitations related to the adoption/migration of both PQC and quantum cryptography. Due to the complexity of algorithms, the latency for key generation, encryption, and verification can negatively impact applications. Additionally, memory and energy consumption can become limiting factors, particularly for resource-constrained devices, as is often the case in IoT and embedded industrial systems. Lastly, it is important to emphasize that there are no known barriers to creating quantum algorithms capable of breaking PQC algorithms. This remains an active research area and is a critical factor in making adoption trustworthy for applications.

While numerous studies have investigated PQC and quantum-based quantum-safe mechanisms, most focus on PQC, QKD, or specific application domains (e.g., blockchains and digital identity systems). Although these studies provide valuable insights, they largely examine quantum-safe solutions in isolation or within narrowly defined contexts. Consequently, the literature still lacks a comprehensive and systematic view of how quantum-safe mechanisms are being deployed across diverse applications and system environments. This fragmented perspective leaves several key questions insufficiently addressed. For example, \textit{What techniques and building blocks are under practical investigation for deployment}?  \textit{Which applications are currently considered primary candidates for migration to quantum-safe solutions}? \textit{Which mechanisms are most suitable for specific application requirements and what functionalities do they best support}? Such questions are increasingly emerging in the literature as the transition toward quantum-safe infrastructures gains momentum. Addressing these issues requires a holistic perspective that consolidates current developments and systematically analyzes the deployment landscape of quantum-safe mechanisms across different application domains.

This work addresses the identified gaps by presenting a comprehensive survey of quantum-safe mechanisms, including PQC and quantum cryptography, and their integration into diverse applications and systems. We combine a systematic review with an exploratory analysis to achieve this goal. Specifically, we conduct a systematic review of major computer science digital libraries (e.g., ACM Digital Library and IEEE Xplore), focusing on research on PQC and quantum cryptography, as well as on their integration into real-world applications and system architectures. Based on this analysis, we examine how quantum-safe techniques are being adopted across different sectors and technological domains. The contributions of this survey are summarized below.

\begin{itemize}

    \item \textbf{Categorization of migration approaches and building blocks:} We analyze how quantum-safe migration is occurring across different sectors and highlight the most influential techniques being integrated into cryptographic protocols and system functionalities.

    \item \textbf{Systematic analysis of application domains:} We identify and categorize applications adopting quantum-safe techniques and organize them into a structured taxonomy of application domains through a systematic review of major computer science digital libraries.
    
    \item \textbf{Classification of adoption challenges:} We investigate the challenges associated with the adoption of quantum-safe mechanisms and classify them according to aspects like efficiency, deployment, and security challenges.

\end{itemize}

\textbf{Organization.} The organization of this survey is structured as follows. Section~\ref{sec:related} presents related surveys in this area. Next, Section~\ref{sec:methodology} presents our search methodology, focusing on how the systematic search was conducted and how the literature investigation was conducted. Section~\ref{sec:concepts} reviews fundamental concepts related to quantum safe cryptography, including post-quantum cryptography and quantum cryptography. Section~\ref{sec:blocks} categorizes specific building blocks considered in existing reported integrations and presents insights obtained. Section~\ref{sec:domains} presents our literature survey, focusing on a systematic analysis of the applied domains. Furthermore, Section~\ref{sec:challenges} presents a list of open research challenges we identified during the elaboration of this survey. Finally, we summarize our lessons learned and conclude.

\section{Related Work}
\label{sec:related}

\begin{table*}[t]
\centering
\caption{Comparative analysis of related literature on quantum-safe cryptographic techniques and migration approaches.}
\label{tab:related_work}
\footnotesize
\begin{tblr}{
  colspec = {l c X[1.2,l] X[l] X[l] c c c}, 
  hlines, 
  vlines, 
  row{1} = {font=\bfseries, bg=gray!10}, 
  row{Z} = {font=\bfseries}, 
  rowsep = 1pt,
}
\textbf{Reference} & \textbf{Year} & \textbf{Context} & \textbf{Migration Focus} & \textbf{Application Scope} & \textbf{PQC} & \textbf{QKD} & \textbf{Challenges} \\
\hline 

Alnahawi et al. \cite{alnahawi2024comprehensive} & 2024 & TLS Protocols & Protocol integration & TLS systems & \checkmark & - & - \\

Baseri et al. \cite{baseri2024navigating} & 2024 & Network Architectures & Implementation & Networking environments & \checkmark & - & - \\

Sowa et al. \cite{sowa2024post} & 2024 & Networking & Protocol inventory & Communication protocols & \checkmark & - & - \\

Näther \cite{nather2024migrating} & 2024 & Software Migration & Migration process & General systems & \checkmark & - & \checkmark \\

Tan et al. \cite{tan2022challenges} & 2022 & Digital Signatures & Adoption constraints & Signature-based applications & \checkmark & - & \checkmark \\

Giron et al. \cite{giron2023post} & 2023 & Key Exchange & Hybrid schemes & Cryptographic protocols & \checkmark & - & - \\

Chhetri et al. \cite{chhetri2025post} & 2025 & Quantum-safe security & Algorithm taxonomy & IoT systems & \checkmark & \checkmark & \checkmark \\

Aquina et al. \cite{aquina2025critical} & 2025 & Hybrid security & Critical analysis & Conceptual security models & - & \checkmark & \checkmark \\

Xu et al. \cite{xu2023overview} & 2023 & Quantum-safe cryptography & Algorithm overview & Cryptographic primitives & \checkmark & \checkmark & - \\

Sharath et al. \cite{sharath2025quantum} & 2025 & Quantum-safe techniques & Algorithm overview & Cryptographic primitives & \checkmark & \checkmark & - \\
\hline

\textbf{This Survey} & 2026 & Multi-domain systems & Migration landscape & Cross-domain applications & \checkmark & \checkmark & \checkmark \\

\end{tblr}
\end{table*}

This section presents existing literature reviews on PQC, QKD, and their adoption in various applications. We also provide a comparative analysis of these reviews with ours in Table~\ref{tab:related_work}.

\textbf{Application and Domain Specific Studies.} Several studies investigate the integration of quantum-safe techniques within specific protocols or technological contexts. For example, some works focus on incorporating post-quantum mechanisms into networking protocols such as Transport Layer Security (TLS) \cite{alnahawi2024comprehensive}. Other studies examine broader yet still domain-specific settings, including networking environments and communication infrastructures \cite{baseri2024navigating,sowa2024post}. However, the range of applications migrating toward post-quantum security extends far beyond individual protocols or networking contexts. Such transitions often require migration efforts that involve multiple system components and architectural building blocks rather than protocol-level modifications alone. A broader perspective on migration is presented in \cite{nather2024migrating}, where the authors provide an overview of migration techniques, identify key migration steps, and discuss major challenges associated with transition. Their work also emphasizes the importance of standardization efforts and the development of crypto inventories as part of the migration process.

\textbf{PQC Adoption.} Several studies examine challenges associated with the adoption of PQC. For instance, the challenges related to the deployment of PQC-based digital signatures are explored in \cite{tan2022challenges}. The authors categorize real-world applications that rely on signature schemes, outline key adoption constraints, and propose a framework to guide the transition based on these constraints. However, their analysis focuses exclusively on digital signatures and does not consider key encapsulation mechanisms (KEMs), which play a central role in many cryptographic protocols. The broader research challenges related to the adoption of PQC are discussed in \cite{ott2019identifying}; however, this work does not address QKD.
A more recent survey is presented in \cite{chhetri2025post}, which provides an overview of quantum-safe security with a particular focus on PQC. The authors propose a taxonomy of post-quantum cryptographic algorithms, including both standardized schemes and those that have not yet been incorporated into standardization efforts. In addition, they discuss the integration of PQC into existing protocols, the challenges associated with deploying PQC on constrained devices, and the results of the performance evaluation. The survey also briefly discusses the distribution of quantum keys as a complementary approach to quantum-safe security.

Several studies provide broader overviews of quantum-safe security mechanisms, including both PQC approaches and QKD techniques \cite{sharath2025quantum, xu2023overview}. These works typically summarize the main categories of quantum-safe cryptographic algorithms and discuss representative solutions proposed in the literature. Although such surveys offer valuable background on the design and characteristics of quantum-safe techniques, they primarily focus on algorithmic aspects rather than examining how these mechanisms are integrated into real-world applications and system architectures.

\textbf{QKD and Hybrid Approaches.} The hybrid cryptographic schemes that combine multiple quantum-safe techniques are increasingly being considered as part of migration strategies toward quantum-resistant security. In \cite{giron2023post}, the authors examine the adoption of hybrid schemes for post-quantum key exchange. Their study identifies and categorizes different interpretations of the term hybrid and provides performance insights into these approaches. A broader critical analysis of both QKD and PQC is presented in \cite{aquina2025critical}. This work examines the assumptions underlying quantum-resistant security strategies and explores how QKD and PQC can be combined to strengthen security guaranties. The authors also critically assess the applicability of QKD, arguing that in many cases it may be either insufficient or unnecessary, particularly in scenarios that require only authenticity. Nevertheless, they highlight the potential relevance of QKD for transport-layer security. However, the study does not provide an in-depth discussion of PQC applications.

\section{Scope and Search Methodology}
\label{sec:methodology}

We conducted a systematic search to identify studies that integrate quantum-safe techniques into cybersecurity systems and applications. The search process is guided by three research questions (RQs):

\begin{enumerate}[label=\textbf{RQ\arabic*:}, leftmargin=1.2cm]

\item What quantum-safe cryptographic techniques and architectural building blocks are used when migrating communication protocols and cybersecurity systems into PQC and quantum cryptography?

\item Which application domains and system types are adopting or being redesigned to incorporate PQC or quantum cryptography, how do these applications incorporate specific cryptographic functionalities, and what is their level of maturity?

\item What technical limitations and deployment challenges arise when integrating PQC or quantum cryptography into real-world systems and applications?
\end{enumerate}

These research questions guide the search methodology to the literature that examines the integration of quantum-safe techniques into practical systems and applications. We define an inclusion scoring system to refine the results obtained from the search process based on the RQs. The scoring system assigns higher scores to papers that provide evidence of practical integration, clearly identify application domains, and discuss technical challenges associated with deployment. In addition to the systematic search, we include complementary references, such as relevant preprints and foundational papers identified through exploratory search and snowballing. These additional sources are included when they are considered influential in the field and achieve high scores according to the defined inclusion criteria.

\subsection{Selection Criteria}

We define a search string composed of two groups of terms combined using the Boolean operator \textbf{AND}, aiming to retrieve studies that address the integration of quantum-safe technologies into cybersecurity systems and applications. The first group includes terms related to the target technologies and their common variants, such as Quantum Key Distribution (QKD) and Post-Quantum Cryptography (PQC). The second group focuses on terms associated with system integration and interoperability aspects, including migration, compatibility, and protocol integration.

Table~\ref{tab:terms} summarizes the terms included in each group. The use of multiple synonyms and related expressions aims to increase coverage and reduce the risk of missing relevant studies.

\begin{table}[t!]
\centering
\caption{Terms used in our query.}
\label{tab:terms}
\resizebox{.45\textwidth}{!}{%
\begin{tabular}{|c|c|}
\hline
\textbf{Set 1}                 & \textbf{Set 2}         \\ \hline \hline
Quantum Key Distribution       & Integration            \\ 
QKD                            & Interoperability       \\ 
Quantum Cryptography           & Migration              \\ 
Quantum Key Exchange           & Framework              \\ 
Post-Quantum Cryptography      & Compatibility          \\ 
PQC                            & Communication protocol \\ 
Quantum-Resistant Cryptography & Oblivious Transfer          \\ 
Quantum-Safe Cryptography      &                        \\ \hline
\end{tabular}}
\end{table}

We queried three primary digital libraries—IEEE Xplore, Scopus, and the ACM Digital Library—covering publications from 2020 to 2026. These sources were selected because they provide broad coverage of peer-reviewed research in computer science and cybersecurity. The search was restricted to articles whose \textit{abstract} contained at least one term from \textbf{Set 1} and one term from \textbf{Set 2}. This constraint allows us to identify studies that simultaneously address quantum-safe technologies (e.g., QKD or PQC) and aspects related to their integration into systems or protocols. When supported by the database, additional filters were applied to restrict the results to the computer science domain (e.g., Scopus filter \texttt{SUBJAREA=COMP}) and to the publication years of interest. These constraints help ensure that the retrieved studies are both recent and relevant to the technical integration of quantum-safe mechanisms in cybersecurity systems.

\begin{figure}[t!]
  \centering
  \includegraphics[width=.85\linewidth]{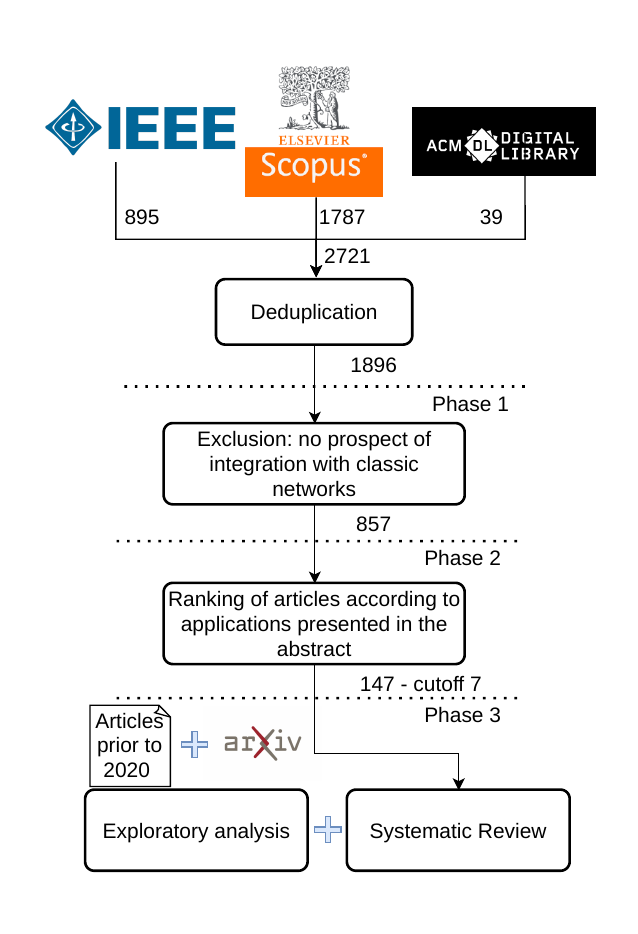}
  \caption{Retrieval and screening pipeline with counts per phase (2020–2025).}
  \label{fig:flow}
\end{figure}

\subsection{Search Screening and Pre-processing}

We refined the set of retrieved studies by applying a structured screening process instead of evaluating all papers returned by the search string. This process relied on a combination of exclusion and inclusion criteria to identify the most relevant works for our survey. In this section, we describe the screening procedure and summarize key observations from the resulting dataset.

The overall search results and screening pipeline, including the number of papers at each stage and the exclusion phases applied, are illustrated in Figure~\ref{fig:flow}. The initial search returned 2,721 records across the three selected databases. We first performed a deduplication step based on normalized titles, DOI identifiers (when available), and fuzzy title similarity, resulting in 1,896 unique entries.
The screening process then proceeded in two phases (see Figure~\ref{fig:flow}). In the first phase, we excluded papers that had a purely theoretical focus without a clear path toward integration into applications, studies outside the scope of networking and security communications, non-scholarly document types, and results outside the focus of this review. This filtering stage reduced the dataset to 857 articles. In the second phase, we conducted an abstract-level screening and applied an inclusion scoring scheme aligned with the scope of this survey and the defined research questions. The scoring system considered the following criteria:

\begin{enumerate}[label=(\roman*)]
\item \textbf{Evidence of application involving quantum cryptography and/or PQC}—including prototypes, experimental demonstrations, simulations, or architectural frameworks \textit{[2 points if present; 0 otherwise]};
\item \textbf{Discussion of technical challenges}—quantitative analysis \textit{[1 point]} and/or qualitative discussion \textit{[1 point]};
\item \textbf{Clearly identified application domains}—\textit{[0.5 points each]}, assigned when the abstract provided sufficient information to classify the work within an application context.
\end{enumerate}

This scoring step yielded 147 articles above the predefined cutoff. From this subset, papers whose full text was accessible through our institution and demonstrated clear evidence of practical integration formed the core dataset, totaling 54 studies. We complemented the core of papers that result from the query with 74 additional papers retrieved non-systematically by backward/forward snowballing, venue and author tracking, and preprints or proceedings from sources such as arXiv, IACR ePrint, and USENIX.

All records, screening decisions, and inclusion scores were stored in standardized spreadsheets containing unique identifiers, bibliographic metadata, source database, query version, publication year, keywords, eligibility indicators, and deduplication fields. This structured dataset enabled efficient cross-source merging, transparent removal of duplicates, consistent scoring and labeling for subsequent thematic analysis, and full traceability and reproducibility of the methodological process.

Finally, the selected papers were analyzed individually and categorized according to a taxonomy of application domains, disciplines, and employed techniques. Each study was classified within these categories, allowing us to identify the main sectors adopting quantum-safe mechanisms and to highlight specific applications that integrate such techniques within each discipline.

\section{Background}
\label{sec:concepts}


The emergence of reliable large-scale quantum computers poses a fundamental threat to modern public-key cryptography. Most public-key cryptography systems currently in use rely on mathematical problems that are believed to be computationally infeasible to solve with classical computers, e.g., factorization of large integers and the discrete logarithm problem. The assumed hardness of these problems allows widely deployed cryptographic algorithms, including RSA, ECC, and the Diffie–Hellman key exchange \cite{perlner2009quantum}. However, this assumption no longer holds in the presence of sufficiently powerful quantum computers, which could efficiently solve these problems and compromise the security of these widely deployed systems. 

Shor’s algorithm \cite{shor1994algorithms} demonstrates that integer factorization and discrete logarithm problems can be solved exponentially faster on a sufficiently powerful quantum computer, which poses a fundamental threat to the security of current asymmetric cryptographic systems. A detailed overview of quantum attacks against classical cryptographic schemes is provided in \cite{bonnetain2019quantum}. In such scenarios, an adversary could exploit publicly available information, e.g., public keys or digital certificates, to efficiently derive the corresponding private keys and break the confidentiality and authenticity of these schemes.

In contrast, symmetric-key cryptography currently does not face the same level of vulnerability to quantum attacks. Grover’s algorithm \cite{grover1996fast} provides a quadratic speedup for brute-force key search compared to classical approaches. However, this attack model can be mitigated by increasing the key size. For instance, symmetric algorithms such as Advanced Encryption Standard (AES) can maintain quantum-resistant security levels by doubling the key length and compensate for the quadratic speedup offered by quantum search algorithms \cite{bonnetain2019quantum}.

\subsection{Fundamentals of Cryptography}
\label{subsec:fundamentals}

Before discussing quantum-resistant mechanisms, it is useful to briefly review several fundamental cryptographic concepts. We first introduce the core security properties of confidentiality, integrity, authentication, and non-repudiation, commonly used in the current Internet infrastructure, including key exchange, secure data storage, and protected communication. We then briefly discuss advanced techniques that enable computations on data while preserving privacy, commonly referred to as privacy-preserving computation.

\textbf{Confidentiality, Integrity, Authentication, and Non-repudiation.} Confidentiality refers to the property that sensitive information can only be accessed by authorized entities. In practice, confidentiality is typically achieved through encryption (e.g., symmetric or asymmetric) and using access-control mechanisms. Although confidentiality is a central goal of information security, some applications do not require data to remain secret, e.g., public communication over broadcast channels. Nevertheless, it remains essential to protect transmitted or stored data from unauthorized modification, a requirement ensured by data integrity. Integrity ensures that data has not been altered in an unauthorized manner while in transit or at rest. This property can be achieved using cryptographic hash functions (e.g., SHA-256). Cryptographic hash functions provide efficient mechanisms for detecting accidental or malicious modifications, although they cannot authenticate the origin of the data.

Authentication allows a receiver to verify the identity of the entity sending the data or initiating a communication. When combined with integrity protection, authentication ensures that the message originates from a legitimate source and has not been modified in transit. Message authentication can be achieved using message authentication codes (MACs) based on shared symmetric keys or digital signatures based on asymmetric cryptography. Finally, some applications require non-repudiation, meaning that the sender of a message cannot deny having generated or transmitted it. It should be noted that public-key-based digital signatures provide non-repudiation, whereas this feature is not offered by symmetric-key-based MACs. This is because non-repudiation requires a trusted entity, such as the certificate authorities in public-key infrastructure (PKI).

\textbf{Cryptographic Key Exchange.} Cryptographic key exchange refers to the process of establishing and distributing cryptographic keys between two or more parties so that these keys can be used later by cryptographic algorithms to provide security properties such as confidentiality, integrity, and authenticity. In asymmetric cryptography, key exchange is relatively straightforward because only the public key needs to be distributed, and it can be shared over a public channel without compromising security. In practice, however, secure key exchange primarily concerns the distribution of symmetric keys. Symmetric encryption algorithms are significantly more efficient than asymmetric ones and are widely used to protect data transmitted over the Internet. Consequently, the problem of securely establishing shared symmetric keys between distant parties lies at the core of modern information security. One option is to securely exchange symmetric keys via asymmetric-key-based communication; e.g., applications like WhatsApp use this mechanism. 

\textbf{Secure Data Storage vs. Communication.}
Protecting sensitive data stored on devices or servers (a.k.a. data at rest) is a critical aspect of information security. In certain scenarios, cryptographic keys themselves must also be stored using a key management system (KMS), which can manage these keys and corresponding metadata. If appropriate protection mechanisms are not implemented, the compromise of stored keys can impact data encryption and authentication mechanisms or other cryptographic protocols relying on these keys. Thus, stored data can be protected using symmetric or asymmetric encryptions, complemented by additional safeguards such as password-based access control and authorization mechanisms following successful identity authentication.

Secure communication over the Internet typically involves authenticated encryption to ensure that only authorized parties can access transmitted data and that messages cannot be modified by untrusted entities. In addition, receivers must be able to verify that the communication originates from a legitimate sender. As discussed earlier, MACs and digital signatures can provide data integrity and authentication, while digital signatures additionally enable non-repudiation. Confidentiality is commonly achieved using symmetric encryption algorithms such as AES, due to their computational efficiency. 

\textbf{Privacy-Preserving Computation.}
Data can be an extremely valuable resource for extracting insights and supporting data-driven decision-making. However, certain types of data, e.g., personally identifiable information (PII), must remain protected during and after the collaborative computational analysis tasks. Such collaborative computations include privacy-preserving DNA comparisons between patients' datasets, secure computation of advertising revenues, and determining market-clearing prices in auctions such as sugar beet markets \cite{bogetoft2009secure}, among others. Moreover, there are other applications that desire anonymity by choice, e.g., anonymous electronic payments and anonymous queries to a server. In this regard, privacy-preserving computation (PPC) refers to a set of cryptographic techniques that enable the handling of sensitive data while maintaining its privacy during and after processing. 
In PPC, the private data never becomes available in its original form and remains encrypted for all involved parties, irrespective of them receiving individual outputs or a joint common output of the computation.
The need for PPC is further motivated by the strict regulatory frameworks, including Europe's general data protection regulations (GDPR) and others worldwide. 

PPC provides techniques and tools to facilitate computations and data analysis through secure multi-party computation (MPC), homomorphic encryption (HE) and oblivious transfer (OT). In MPC, multiple parties jointly compute a function over their private inputs such that nothing is revealed beyond the final output. That is, each party’s input remains confidential throughout the computation. HE is another PET that allows computations to be performed directly on encrypted data, producing an encrypted result that can later be decrypted to obtain the final output \cite{fontaine2007survey}. In this approach, sensitive data remains encrypted throughout the computation, preventing exposure of plaintext. OT, a fundamental cryptographic primitive, enables two parties to exchange information such that the receiver learns only the selected parts of information without gaining knowledge of the remaining data, while the sender remains unaware of which parts the receiver learned. Note that PPC is considered  part of a broader class of technologies, commonly called privacy-enhancing technologies (PETs). PETs consist of federated learning (FL), differential privacy (DP), trusted-execution environments (TEE), zero-knowledge proofs (ZKP), threshold cryptography, anonymization etc \cite{aunon2024evaluation}. While PETs based on classical cryptographic primitives have been well studied from both research and real-world implementation points of view \cite{PETstandard}, quantum-resistant solutions are still being developed and are therefore in an early stage of maturity.

\begin{figure*}[htb]
    \centering
    \begin{subfigure}[t]{.31\textwidth}
        \centering
        \includegraphics[width=\textwidth]{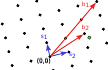}
        \caption{Lattice-based (CVP)}
        \label{fig:latticesproblem}
    \end{subfigure}
    \hfill
    \begin{subfigure}[t]{.31\textwidth}
        \centering
        \includegraphics[width=\textwidth]{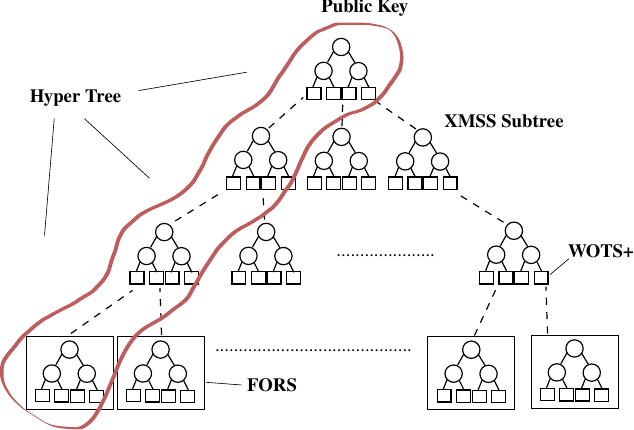}
        \caption{Hash-based (SPHINCS+)}
        \label{fig:sphincs}
    \end{subfigure}
    \hfill
    \begin{subfigure}[t]{.31\textwidth}
        \centering
        \includegraphics[width=0.75\textwidth]{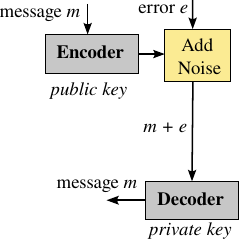}
        \caption{Code-based Cryptography}
        \label{fig:codebase}
    \end{subfigure}
    \caption{Primary cryptographic families selected by NIST for post-quantum standardization. These include lattice-based constructions (e.g., ML-KEM), stateless hash-based signatures (SPHINCS+), and code-based primitives.}
    \label{fig:nist-pq-selections}
\end{figure*}

\subsection{Post-Quantum Cryptography}
\label{subsec:pqc}

A scheme is considered quantum-safe if an attacker, even with access to a sufficiently powerful quantum computer, cannot feasibly forge a valid signature or recover the shared secret established between two legitimate parties. Post-quantum cryptography (PQC) refers to cryptographic schemes designed to remain secure against adversaries equipped with quantum computers capable of executing algorithms such as Shor’s and Grover’s. PQC focuses primarily on public-key cryptographic mechanisms, including digital signatures and key encapsulation mechanisms, which are widely used for authentication and key establishment.  Specifically, PQC schemes rely on computational problems that are believed to be intractable for both classical and quantum computers. Existing PQC proposals are based on several families of hard computational problems, including lattice-based, code-based, hash-based, multivariate, and isogeny-based constructions. In the following, we provide an overview of the main algorithm categories that have been considered in recent standardization efforts.

\textbf{Lattice-based PQC.} 
Lattice-based cryptographic algorithms are widely used for both key encapsulation mechanisms (KEMs) and digital signatures. A lattice is a mathematical structure that represents a discrete set of points in space generated by integer linear combinations of two or more basis vectors. Figure~\ref{fig:latticesproblem} illustrates an example of a lattice generated by two different pairs of basis vectors. Lattices exhibit computational problems that are believed to be hard even for quantum computers, most notably the shortest vector problem (SVP) and the closest vector problem (CVP). Many lattice-based cryptographic constructions rely on problems related to the learning with errors (LWE) problem \cite{regev2009lattices}. Several variants of LWE have been proposed, including module learning with errors (MLWE). The presumed hardness of these problems has motivated the development of a wide range of lattice-based cryptographic schemes.

In lattice-based encryption schemes, the secret key typically corresponds to short basis vectors or secret parameters, while the public key represents an instance of a lattice problem derived from these secrets \cite{regev2009lattices}. An adversary observing only the public key aims to recover the secret parameters that define the lattice structure. The security of these schemes is strengthened by introducing small random errors during the encryption process, which significantly complicates attempts to recover the underlying lattice structure. As a result, an attacker must not only determine the hidden basis vectors that generate the lattice, but also account for the noise introduced in the system. Importantly, many lattice-based constructions rely on reductions showing that solving the underlying average-case problems (such as LWE) is as hard as solving certain worst-case lattice problems. This worst-case to average-case hardness property provides strong theoretical security guaranty and makes lattice-based schemes particularly attractive for post-quantum cryptography.

\textbf{Hash-based.} Hash based cryptography is based on completely different fundamentals to build digital signatures  \cite{srivastava2023overview}. The most fundamental cryptographic systems based on hashes are the Lamport One-Time Signature (L-OTS) \cite{lamport1979constructing} and the Winternits One-Time Signature (WOTS). Hash-based signature algorithms often use a hash function to sign a message and create a signature from a private key. The public key is a set of digests computed by hashing parts of the private key. A client can thus verify if a particular message matches the public key by combining the signature with the received message. Because the private key is revealed in the verification process, such kind of mechanism can be only used one time. The security of Hash based signatures relies in the one-way property of hash functions, making it infeasible to determine the original input given an output hash value.

\textbf{Code-based.} Code-based cryptography is a technique that use code theory techniques for error correction. These techniques often use parity check mechanisms, enabling to identify whether message bits have changed during transmission. In the communication domain, an encoder algorithm processes a message, and the receiver would use a decoder algorithm to check for the presence of errors using a decoder. These mechanisms were adapted for cryptography, by making usage of intentionally added errors to hide messages \cite{redhat_pqc}. Figure \ref{fig:codebase} presents an overview of how we can employ coding algorithms for cryptography. In general, the encoder algorithm is the public key, enabling the encoding of a message and subsequently intentionally adding errors to hide the message. Only the decoder, which is kept in secret as the private key, can decode the message, subsequently identifying and fixing the errors introduced in the message. In practice, this can take form by having a generator algorithm to encode a message, but shuffling and having permutations of this generator, producing the public encoder. Later on, the shuffling and permutations can be undone during the decoding process, revealing the original message.

There are other families of PQC algorithms, which are considered in research, but not standardized at PQC institutions, thus not \textit{(yet)} considered in this review. Multivariate public key cryptography (MPKC), that is based on the hardness of solving nonlinear equations over finite fields. Isogeny based schemes are also an option. Isogenies are mappings between different elliptic curves, and cryptography based on isogenies are based on the hardness of finding such isogenies between different elliptic curves. We point the reader to \cite{bernstein2025post} \cite{de2017mathematics} for more concrete background on MPKC and isogenies, respectively.

\subsubsection{Standards for Post-quantum Cryptography}

Following the recent advances in quantum computing, NIST decided to standardize PQC algorithms. To carry out this standardization, NIST organized a competition to select algorithms that could resist attacks by quantum computers. More than $80$ algorithms were submitted for digital signatures, encryption, or secure key exchange schemes. These algorithms were evaluated over multiple rounds, during which some proposals were not considered adequate and discarded. By the end of 2024, only four of the original 80 algorithms were selected as the main representatives for PQC.

\begin{table}[t!]
\setlength{\tabcolsep}{8pt}
\renewcommand{\arraystretch}{1.45}
\caption{Categorization of post-quantum cryptography algorithms, adapted from \cite{384_hoque2024exploring}.}
\label{tab:algorithms}
\begin{tabular}{l|cc}
\rowcolor[HTML]{E3F2FD} 
\textbf{}     & \multicolumn{2}{c}{\cellcolor[HTML]{E3F2FD}\textbf{Algorithms}}                                                                                        \\
\rowcolor[HTML]{E3F2FD} 
Family        & \multicolumn{1}{c|}{\cellcolor[HTML]{E3F2FD}Digital Signatures} & KEM                                                                                 \\ \hline \hline
\textit{Lattice-based} & \multicolumn{1}{c|}{\textbf{Dilithium}, Falcon}                          & \begin{tabular}[c]{@{}c@{}}\textbf{Kyber}, Saber, NTRU,\\  NTRU Prime, FrodoKEM\end{tabular} \\ \hline
\textit{Code-based }   & \multicolumn{1}{c|}{}                                           & McEliece, HQC, BIKE                                                                 \\ \hline
\textit{Multivariate}  & \multicolumn{1}{c|}{Rainbow, GeMSS}                             &                                                                                     \\ \hline
\textit{Hash-based }   & \multicolumn{1}{c|}{\textbf{SPHINCS+}, PICNIC}                           &                                                                                     \\ \hline
\textit{Isogeny-based} & \multicolumn{1}{c|}{}                                           & SIKE                                                                                \\ 
\end{tabular}
\end{table}

Among the algorithms selected for standardization, the underlying computational problems vary. Table \ref{tab:algorithms} (adapted from \cite{384_hoque2024exploring}) presents an overview of post-quantum cryptography algorithms. We focus in the algorithms that were selected by NIST for standardization. Those include Kyber for KEM, and SPHINCS+ and Dilithium for digital signatures. More recently, a backup algorithm, called HQC, based on error-correcting codes was also selected for standardization. All of these algorithms are based on problems believed to be intractable even for classical computers. Next, we present the background to understand recent standards.  

\textbf{Key Encapsulation Mechanisms.} CRYSTALS-Kyber is a KEM standardized at FIPS 203, identified as Module Lattice Based Key Encapsulation Mechanism (ML-KEM). ML-KEM represents lattices in a polynomial ring $R :=\mathbb{Z}[x][x^2+1]$, in which mechanism security is related to the difficulty of the MLWE problem. In ML-KEM \cite{bos2018crystals}, the private key is a set of random polynomials $s \in R$, which are then used to create the public by combining $s$ with a random matrix $A$ and an error $e$. This end-up with an instance of the MLWE, $pk = As + e$.
ML-KEM employ several optimizations, such as compressing keys and using faster matrix multiplication using the number theoretic transform (NTT) algorithm. 

Another KEM in standardization is the HQC, which stands for Hamming Quasi-Cyclic. The standardization is occurring as FIPS 207, identified as HQC-KEM. This scheme has roots in the McEliece scheme \cite{mceliece1978public}. The McEliece cryptosystem uses a generator matrix for an error-correction mechanism called the Goppa code. This generator matrix is multiplied by an invertible matrix and a permutation matrix to produce the public key. To encrypt a message with this public key, it is possible to encode the message with the public matrix and intentionally introduce errors. To decrypt using the private key, this system unshuffles the matrix generated and shuffled earlier, then uses the Goppa code to correct errors and reveal the message. Differently, in the HQC-KEM, the generator is not shuffled, and security is intrinsic.

\textbf{Digital Signatures.} The hash-bash standard of NIST is SPHINCS+, standardized in FIPS 205, and known as the stateless hash-based digital signature algorithm (SH-DSA). SH-DSA is illustrated in Figure \ref{fig:sphincs}. The underlying primitive of SH-DSA is the Winternitz One-Time Signature (WOTS), which improves on LOTS \cite{mattsson2021quantum}. However, one-time signatures are not enough; SH-DSA mixes other cryptographic structures, such as Merkle Trees, to enable a single public key to be used more than once. In addition, SH-DSA has many layers of sub-trees (a Hypertree), which can be created by the signer on demand, once each leaf of the sub-tree has a unique key associated. The advantages of using a Hypertree cryptographic scheme are that it is not necessary to build the entire tree at once, which also reduces the potential for memory consumption. Finally, combining Forest of Random Subsets (FORS) allows each private key to be used a few times, rather than once. The combination of these building blocks (FORST, Hypertrees) eliminates the need to maintain a state of keys already used, thereby creating a \textit{stateless} scheme. 

Another digital signature scheme standardized in FIPS 204 is CRYSTALS-Dilithium, also known as the Module Lattice Digital Signature Algorithm (ML-DSA). ML-DSA is based on the classical Schnorr signature scheme, which is zero-knowledge, but uses polynomial rings to generate keys instead of only modular arithmetic \cite{ducas2018crystals}. A set of polynomials is used to derive a commitment, challenge, and response vectors that allow efficient verification. These create an instance of the MLWE problem. ML-DSA also employs several optimizations, such as compressing keys and using faster matrix multiplication using the NTT algorithm.

Although NIST has finalized the standardization of the main algorithms, there is still a long way to go before PQC algorithms are deployed in the real world. The migration process, although incremental, can be challenging. Different libraries and frameworks already include implementations of the standardized algorithms, ranging from open-source to commercial solutions. Among the open-source options, the Open Quantum Safe (OQS) project and libpqcrypto stand out. OQS, part of the Linux Foundation, provides C implementations of algorithms and includes support for various existing protocols. Libpqcrypto is maintained by the PQCRYPTO project and supports multiple languages. On the commercial side, solutions include initiatives such as those from Isara and PQShield.

\subsubsection{Other protocols}
\label{sec:other_PQC}
The PQC algorithms discussed so far cover the main security guarantees of confidentiality, data integrity, entity authentication and non-repudiation through the cryptographic tools of KEM and digital signatures. However, within the scope of PPC, quantum-resistant solutions based on PQC are worth discussing \cite{malina2021post}.

One of the most popular and powerful tool of PPC is two/multi-party computation (2PC/MPC), which allows computing arbitrary functions over inputs from two/multiple parties, while keeping the individual inputs hidden and only revealing the final output. Even though there has been a long line of research in this area, through European research projects for example \cite{IMPcCT, MPCPRO}, the focus has mainly been towards the practical feasibility.

One of the most fundamental cryptographic primitives, called oblivious transfer (OT), is a two-party cryptographic protocol where a sender can send data to a receiver based on the receiver's choice, such that the receiver's choice remains hidden from the sender, and the receiver obtains only the data they chose without gaining any information on data that was not requested \cite{rabin2005exchange}. It is well known that implementing a secure OT is sufficient to achieve MPC, ZKP, and bit commitment (BC), among others. For this reason, there has been plenty of research towards achieving efficient OT protocols that remain secure in the post-quantum world. Among the PQC solutions for OT, a proposal for randomized-OT bench-marked their high OT generation rate while relying on the ring learning with errors (RLWE) assumption \cite{branco2021roted}. A recent survey on post-quantum OT protocols can be found here \cite{altana2025survey}. From an application point of view, OT extensions \cite{asharov2017more}, which use a small number of base OTs and extend them to a much larger number of OTs using light symmetric cryptography operations only are very useful. This is particularly relevant for large-scale MPC implementation requiring high OT generation rates, typically of the order of millions-billions OT per second. On the other hand, several post-quantum HE schemes based on the RLWE security assumption were proposed \cite{hoffstein1998ntru, fan2012somewhat, buchmann2016creating, cheon2017homomorphic, multikeytowards}, as well as other code-based alternatives \cite{siddhiprada2025post}. Another proposal for MPC based on fully homomorphic encryption (FHE) from NTRU \cite{lopez2012fly}. Towards the standardization efforts on MPC, a NIST call associated to multi-party threshold cryptography (MPTC) \cite{brandao2023nist} was announced in 2023.

\subsection{Quantum Cryptography}
\label{subsec:quantum}

Quantum cryptography comprises cryptographic primitives whose security is derived, wholly or partially, from constraints imposed by quantum mechanics, such as the no-cloning theorem, measurement disturbance, uncertainty relations, entanglement, and Bell nonlocality. In contrast to PQC, whose security is computational and relies on the assumed hardness of mathematical problems against quantum adversaries, several quantum cryptography protocols can provide information-theoretic security under explicitly stated physical and implementation assumptions. PQC secures classical infrastructures against currently known quantum attacks only, while quantum cryptography provides a long-term solution through key distribution, randomness generation, authentication, and other primitives in settings where quantum resources are available. Due to the contrasting nature of their maturity levels and resources required, quantum cryptography and PQC should be viewed as complementary technologies rather than competing approaches. In this section, we briefly review the main quantum cryptography protocols, as summarized in Table \ref{table:QCrypt}, along with their security model and relation to PQC.

\begin{table*}
\centering
\caption{Main quantum cryptography primitives and their relation to PQC.}
\label{table:QCrypt}
\begin{tblr}{
  width = \linewidth,
  colspec = {Q[98]Q[235]Q[348]Q[258]},
  hline{1-2,3-4,5-6,7-8,10-11} = {-}{},
}
\textbf{Primitive}    & \textbf{Main goal}                                & \textbf{Typical security model}                                        & \textbf{Relation to PQC}                    \\
QKD          & Symmetric key establishment              & {Information-theoretic, authenticated\\~classical channel}    & {PQC can authenticate QKD sessions;\\Hybrid key exchange}  \\
QSDC/QKPC    & Direct confidential communication        & {Protocol-dependent;\\often physical layer assumptions}                                           & {Replace/complement key-distribution\\architectures based on PQC or QKD}          \\
QDS/QIA      & {Authentication, integrity,\\non-repudiation}               & {Information-theoretic\\under network/device assumptions}                                         & {Alternative to PQC digital signatures\\in relevant settings}      \\
QRNG         & Entropy generation                       & Trusted-device, semi-DI, or DI                                & Supplies randomness to PQC and QKD \\
QOT          & {Cryptographic primitive\\for mistrustful-parties scenario}                  & {Impossible without extra assumptions\\e.g., quantum noisy/bounded storage, Hash functions}                          & {Alternative to asymmetric-key crypto\\in privacy-preserving protocols}     \\
DI protocols & {Certification with untrusted~\\devices} & {Bell nonlocality plus no-signalling/\\isolation assumptions} & {Strongest trust model,\\less mature than PQC and QKD}
\end{tblr}
\end{table*}

\subsubsection{Quantum Key Distribution}
Quantum key distribution (QKD), first proposed in 1984 \cite{bennett2014quantum}, is a technique for distributing symmetric encryption keys between network participants using quantum channels. QKD relies on two general steps: information transmission, which is performed across a quantum channel, and post-processing, which is performed over an authenticated classical channel. The post-processing includes sifting, parameter estimation, information reconciliation, error correction and verification, and privacy amplification. The quantum channel consists of a physical medium, such as free space or optical fibers, used to transmit quantum states. In discrete-variable QKD, information is encoded in a finite-dimensional degree of freedom, such as polarization, phase, or time bin. This is achieved via single-photon sources or, more commonly, attenuated coherent pulses together with decoy-state techniques. In continuous-variable QKD, information is encoded in field quadratures of coherent or squeezed optical states and detected using homodyne or heterodyne measurements.

\begin{figure}[t!]
    \centering
    \includegraphics[width=0.45\textwidth]{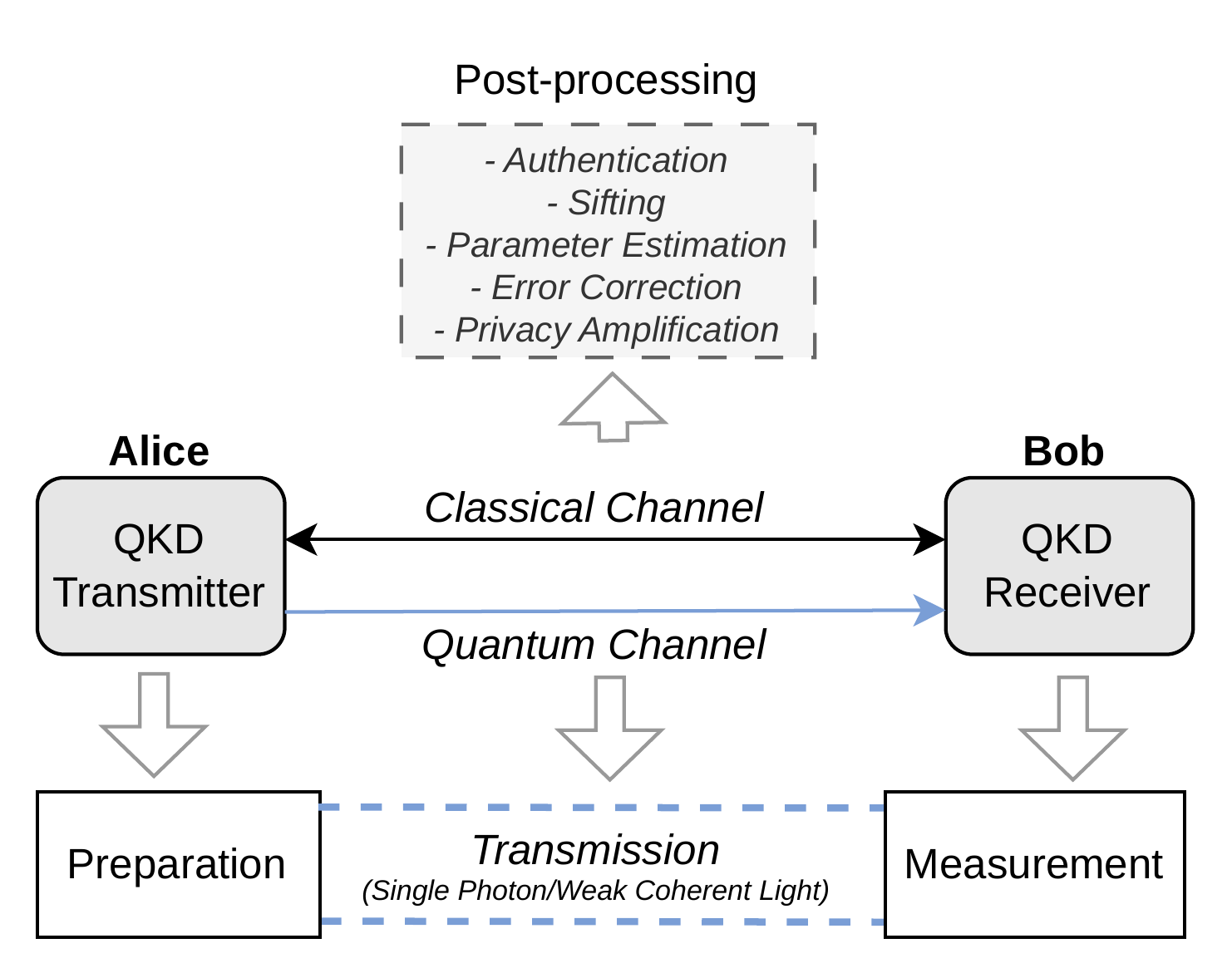}
    \caption{Prepare-and-measure BB84 scheme (inspired by \cite{cao2022evolution}).}
    \label{fig:bb84}
\end{figure}


In the prepare-and-measure version  of the QKD protocols, such as the original BB84 protocol \cite{bennett2014quantum}, a transmitter (traditionally called Alice) prepares quantum states, shown in Figure \ref{fig:bb84}. These states encode the random information to be transmitted using a chosen degree of freedom, such as photon polarization states. For example, $\{ |H\rangle, |V\rangle, |D\rangle, |A\rangle \}$, also called the BB84-states, represent the horizontal, vertical, diagonal, and anti-diagonal linear polarization states, respectively. These encoded states of light are sent across the quantum channel to the receiver (traditionally called Bob), who measures the received photons to decode what is called a raw key. Note that there are other versions of QKD. In entanglement-based QKD, entangled photon pairs are distributed between Alice and Bob, who generate correlated raw keys by measuring their respective subsystems. Another version of QKD called measurement-device-independent (MDI) uses the configuration of a third untrusted party \cite{braunstein2012side}, but here Alice and Bob send independent quantum states to the untrusted receiver in the middle, whose announced measurement outcomes allow them to establish a key while removing detector-side-channel assumptions. 

The parameter estimation step in standard BB84 protocol is critical as it involves determining the quantum bit error rate (QBER) in the channel. Under the standard security assumption, all observed errors are conservatively attributed to the presence of an eavesdropper (traditionally referred to as Eve). Consequently, if the measured QBER exceeds a predefined security threshold, the protocol is aborted, as secure key generation can no longer be guaranteed. Conversely, if the QBER remains below the threshold, Alice and Bob proceed with error correction to their raw keys to arrive at identical keys. The estimated parameters also allow them to upper-bound Eve's information and distill a shorter secret key by privacy amplification to guarantee negligible information leakage to Eve. The classical communication channel, therefore, needs to be authenticated, over which Alice and Bob discuss during the classical postprocessing stage.

There are multiple factors that affect the final secret key rates (SKR) achieved by QKD protocols. The chosen medium determines the signal attenuation for a given degree of freedom used to encode the key bits. Then there are implementation imperfections due to the non-ideal source, modulation, and signal detection, where non-unity detection efficiencies and the dead times of single-photon detectors limit the maximum detection rates and, hence, the achievable secret key rates. Simplified-BB84 scheme with 3-states and 1-decoy maximizes the SKR by optimizing over parameters such as the encoding bases probability, decoy-state intensity levels and probabilities \cite{grunenfelder2018simple}. Over the years, many variations of QKD protocols have been proposed, such as entanglement-based QKD, coherent one-way QKD, differential-phase-shift QKD, and device-independent QKD (DI-QKD). We refer the reader to other surveys in this area for more in-depth details \cite{gisin2002quantum, pirandola2020advances, cao2022evolution, portmann2022security}.\\

\subsubsection{Quantum Direct Communication}

Quantum direct communication refers to the transmission of confidential message directly over a quantum channel, without a pre-distributed cryptographic key. This changes the operational objective relative to QKD where only random key material is exchanged in the quantum transmission, and only leaks random information in case the protocol is aborted before completion. In contrast, because direct communication protocols transmit meaningful messages directly, they must they must be designed so that aborting the protocol does not reveal useful information about the confidential message. This eliminates the need for key generation, distribution and management, particularly relevant in network settings.

Quantum secure direct communication (QSDC) constitutes the largest class of quantum protocols for direct communication where no additional classical information is needed for decoding, as well as the presence of an eavesdropper can be checked in real-time \cite{pan2023free, pan2024evolution}. Most QSDC protocols prevent information leakage by design; through block-based transmission of states, followed by quantum state storage and QBER estimation. The block transmission of photons consists of a two-step (one- or two-way) configuration. That is, photons are sent in batches such that it is first ensured that there was no eavesdropping in the first step by estimating the QBER on randomly sampled quantum states. Only after confirming Eve's absence on the random subset, the secret message is encoded on the remaining photons. The eavesdropping check allows Alice and Bob to estimate whether the channel is compatible with the security requirements of the protocol before message information is decoded in the remaining photons.

Other types of quantum protocols for direct communication protocols were proposed, such as deterministic secure quantum communication (DSQC) and quantum keyless private communication (QKPC). DSQC requires the transmission of classical information after the quantum state transmission to enable message decoding \cite{sun2021deterministic}. QKPC, involving a single step of quantum transmission for free-space line-of-sight connections with restricted eavesdropping, does not provide an active monitoring of eavesdropping. Its security is guaranteed by the physical-layer (PHY) assumption that the legitimate receiver collects a higher-quality signal than an eavesdropper \cite{vazquez2021quantum}. QKPC should be distinguished from QKD and standard QSDC protocols because it offer advantages in terms of transmission rates and experimental setup complexity, while facilitating daylight operations \cite{mendes2025quantum}. \\

\subsubsection{Quantum Authentication}

Remote communications require authentication, where message integrity and entity authentication are mostly necessary, and non-repudiation is desired in certain scenarios of distrustful cryptography. Like many other quantum protocols, QKD itself necessarily requires an authenticated classical communication step. While pre-shared symmetric key-based MACs, such as Wegman-Carter, have traditionally been used in commercial QKD devices; PQC-based digital signature schemes are also being integrated in QKD to achieve a quantum-resistant authentication. As for quantum authentication, quantum digital signatures (QDS) aim to provide all features of classical digital signatures while achieving information-theoretic security.

\begin{figure}
    \centering
    \includegraphics[width=0.45\textwidth]{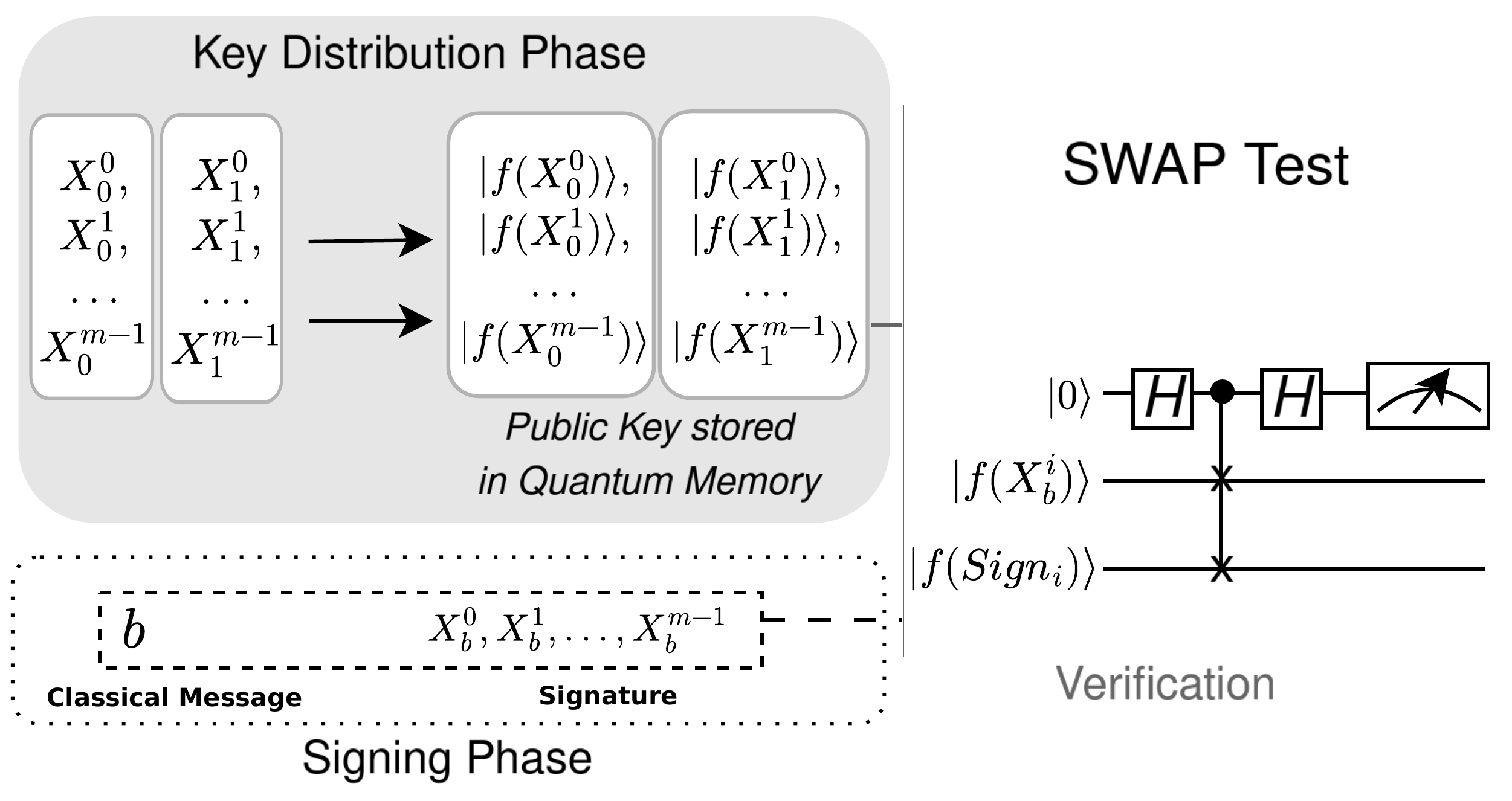}
    \caption{Overview of quantum digital signature scheme.}
    \label{fig:qdsscheme}
\end{figure}

The original QDS scheme \cite{gottesman2001quantum} uses classical information as the private key, while the public key is replaced by quantum states. As shown in Figure \ref{fig:qdsscheme}, the QDS protocol utilizes a pair of classical bit strings $\{X_0^i,X_1^i\}$ as the private key. The Signer applies a public function $f(X)$ to obtain quantum states corresponding to the classical description of their private keys, such as generating BB84 polarization states using a pair of bits. The Signer then distributes these public keys to a number of Verifiers, who store them in quantum memories until the verification phase \cite{heshami2016quantum}. The public keys $\{|f(X_0^i)\rangle, |f(X_1^i)\rangle\}$ can consist of non-orthogonal states, for example. For signing a message bit $b$, the corresponding classical string $X_b^i$ is disclosed during the signing process, i.e., the signed message $(b, X_b^i)$ is sent to a given Verifier. The Verifier employs a SWAP test, using a quantum circuit with a Fredkin gate, a Hadamard gate, and an ancillary qubit, to verify the received message bit $b$. This is done by estimating the fidelity between the quantum state $|f(X_b^i)\rangle$ prepared according to the received classical message and its signature $(b, X_b^i)$, and the corresponding public key stored $\{|f(X_0^i)\rangle, |f(X_1^i)\rangle\}$. Security of the private keys is ensured by the fact that unknown quantum states cannot be copied.

Unlike classical digital signature schemes with easy verification to provide deterministic reject-or-accept outcomes, the verification process in the quantum case is more involved with more than two outcomes, requires communication among parties, as well as needs to consider the noise and imperfections in the quantum communication and storage. The use of quantum memories for storing the public key, which are quantum states, is a substantial practical limitation, considering the current limitations associated to quantum memories in terms of storage times, efficiency and noise \cite{heshami2016quantum}. In addition, performing SWAP tests to verify each communication is also technologically demanding. Lastly, this QDS scheme works under the assumption that the states sent over the quantum channel arrive unaltered, which would require having access to an authenticated quantum channel \cite{gottesman2001quantum}, using quantum MACs \cite{barnum2002authentication}, for example, which brings extra technological cost. It is noteworthy that this QDS scheme requires distributing the public key (quantum states) over quantum channels; therefore, it lacks the ``universal verifiability" feature of classical digital signatures, where the public key can be broadcast.

Over the years, a lot of research has been done to eliminate these impractical requirements \cite{clarke2012experimental, dunjko2014quantum, amiri2016secure, thornton2019continuous}, as well as improve upon security \cite{puthoor2016measurement} and efficiency \cite{li2023one}. This resulted in demonstrations with high communication rates and parties separated by hundreds of kilometers \cite{collins2017experimental, roberts2017experimental}, as well as implementations over metropolitan networks \cite{yin2017experimental, pelet2022unconditionally, du2025chip}. Recently, an experimental quantum e-commerce scheme using QDS based on MDI-QKD was demonstrated \cite{cao2024experimental}, requiring a third party. In contrast, research on quantum identity authentication (QIA) addresses a less demanding scenario \cite{dutta2021short}, limited to verifying that the received information has not been tampered with and originated from the intended entity.\\

\subsubsection{Other Techniques} Beyond the previously discussed quantum cryptography mechanisms, we now discuss additional techniques that make usage of quantum fundamentals. 

\textbf{Quantum Random Number Generators.} Random number generators (RNGs) are essential for many real-world applications, ranging from lotteries to cryptographic protocols for Internet encryption and authentication. RNGs or pseudo-RNGs (PRNGs) are used in most cryptographic algorithms, whether conventional public-key protocols such as RSA and ECC or quantum-resistant protocols such as PQC and QKD. The security of these systems critically depends on the quality and unpredictability of the random numbers generated. While certifying true randomness is difficult, NIST standards SP 800-90B \cite{barker2018sp} and the statistical test suite SP-800-22 \cite{bassham2010sp} are widely used to assess hardware entropy sources and generated random numbers. Quantum random number generators (QRNGs) exploit fundamentally random quantum processes to generate random numbers and provide tools to certify them.

\begin{figure*}[h!]
    \centering
    \includegraphics[width=0.9\linewidth]{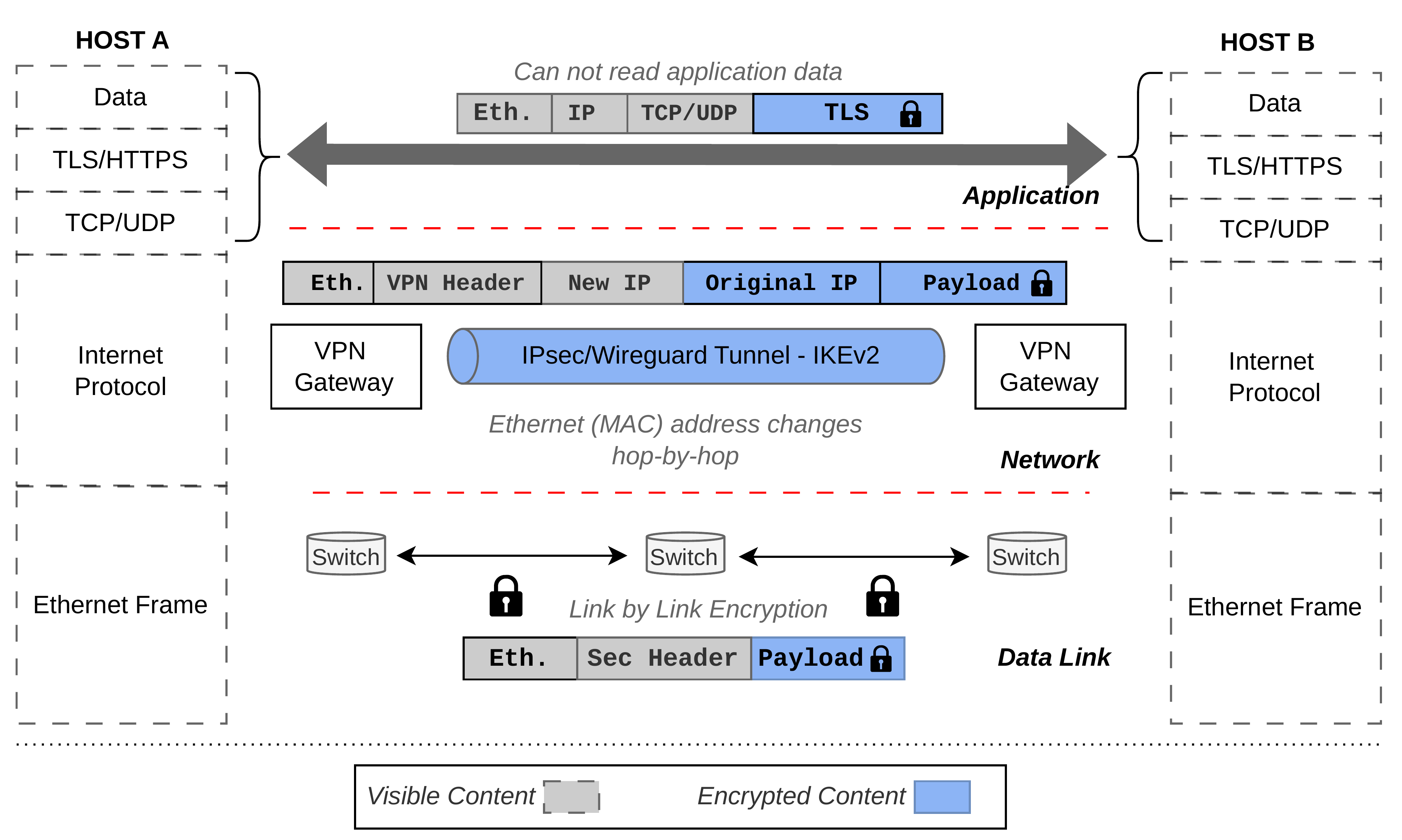}
    \caption{Different security protocols in the TCP/IP model. }
    \label{fig:tcpip}
\end{figure*}

\textbf{Quantum Oblivious Transfer.} Oblivious transfer (OT), as described previously, is a cryptographic primitive where a number of messages are sent, such that the receiver can choose to learn only one of them. The protocol must ensure that the sender remains oblivious to which secret the receiver has extracted, while the receiver cannot learn any information about the remaining secrets. A simple variant of OT is called 1-out-of-2 OT, where Alice sends two messages, and Bob can extract only one of them based on his choice while learning nothing about the other. Quantum oblivious transfer (QOT) \cite{crepeau1994quantum} aims to exploit quantum mechanical principles to achieve information-theoretic security. However, unconditionally secure OT is impossible to achieve \cite{lo1997quantum} in standard quantum mechanics without additional assumptions. This motivated physical assumptions based approaches such as the noisy quantum storage and bounded quantum storage models \cite{damgaard2008cryptography, erven2014experimental}, as well as other computationally-secure quantum protocols based on quantum-secure one-way functions \cite{lemus2025performance}. Using QOT, various protocols for different classes of problems have been proposed over the years, such as private database query, private information retrieval, privacy-preserving decision, among others \cite{chan2014performing, yang2015quantum, gao2019quantum, zhang2020privacy}. A large number of quantum protocols for set computations have been proposed \cite{gomez2025privacy}. A review of QOT protocols can be found in \cite{santos2022quantum}.

\textbf{Device Independence.} Device-independent (DI) quantum cryptography \cite{pironio2009device,zapatero2023advances} is the strongest form of security, requiring minimal assumptions. In DI quantum cryptography, devices are treated as black boxes (with minimum assumptions about their inner workings). The DI framework reduces trust in the internal modeling of source and measurement devices by certifying security from Bell-inequality violations, although it still requires assumptions such as trusted random choices, isolated laboratories, authenticated communication, and control of information leakage. Quantum correlations can then be certified device-independently, enabling, for example, randomness certification based solely on Bell-inequality violations. These ideas can be applied to QRNG \cite{liu2018device}, QKD \cite{nadlinger2022device}, QOT, and other quantum communication protocols. DI quantum cryptography is still in the proof-of-principle stage, although recent advances suggest the technology could mature rapidly \cite{lu2026device}.

There also exist similar trust models which are more relaxed than DI. Notable examples are the semi-device-independent (SDI) and measurement-device-independent (MDI) frameworks.  SDI quantum cryptography \cite{pawlowski2011semi,van2017semi} does not require a complete characterization of the devices, but security is proven under limited assumptions, such as a bound on the dimension of the communicated systems or trusted choices of inputs. SDI protocols offer a more practical route to certifying randomness generation, key distribution, and related tasks while reducing trust on detailed device models. The MDI \cite{xu2014measurement} category of trust models, as the name suggests, removes all assumptions about the measurement devices, preventing detector-side-channel attacks while potentially doubling the transmission distance since the receiver is placed in the middle of the QKD users.

\section{Migration Strategies \& Building Blocks}
\label{sec:blocks}

Aiming to answer our first research question, we now analyze the building blocks crucial to the practical deployment of the quantum-safe techniques investigated. We first analyze the classical communication protocols through the lens of the abstraction level (e.g., TCP/IP layers) at which the investigated cryptographic mechanisms operate. We focus specifically on discussing the migration of classical infrastructure to quantum-safe cryptography mechanisms. Second, we discuss evolutionary and practical approaches for migration into quantum cryptography. Specifically, we focus on hybrid cryptographic techniques that enable incremental deployment of quantum cryptography. Third, we discuss elements and building blocks considered necessary for an even more disruptive change, focusing on aspects specific to quantum communication and the additional infrastructure required, including software-defined networks and key management systems.


\subsection{Deployments for TCP/IP layers}
\label{subsec:netprotocols}

Several secure communication initiatives already incorporate algorithms resilient to quantum attacks. The integration of both PQC and QKD spans multiple layers of the classical TCP/IP model, from the lower layers, such as the data link layer, to the application layer. Figure \ref{fig:tcpip} illustrates the TCP/IP layers and corresponding examples of protocols. We highlight that current deployments either include end-to-end applications (application layer), internet addressing and routing information (network layer), or hop-by-hop messages (data link layer). Next, we analyze such initiatives in a top-down manner (from the application to the data link layer).

\begin{table*}[h!]
\centering
\caption{Classical networking protocols extended with quantum-safe capabilities for secure communication.}
\label{tab:protocols}
\small 
\renewcommand{\arraystretch}{1.3} 
\setlength{\tabcolsep}{10pt}

\resizebox{\textwidth}{!}{%
\begin{tabular}{ll p{4cm} p{3.7cm} p{2.7cm}}
\toprule
\textbf{Layer} & \textbf{Protocol} & \textbf{Application} & \textbf{PQC} & \textbf{QKD} \\
\midrule

\multirow{10}{*}{\textbf{Application}} 
 & TLS & Browsing, Email, Messaging & \cite{sosnowski2023performance, tasopoulos2023energy, 674_burstinghaus2020post, paul2021tpm, sikeridis2020assessing, saribacs2022performance, banerjee2020accelerating} & \cite{garcia2023quantum, 453_buruaga2025hybrid} \\
 \addlinespace[0.5em] 
 
 & QUIC & Browsing, Email, Messaging & \cite{raavi2023post, kempf2024quantum} & -- \\
 \addlinespace[0.5em]
 
 & MQTT & Internet of Things (IoT) & \cite{malina2024quantum} & -- \\
 \addlinespace[0.5em]
 
 & SSH & Remote Connections & \cite{sikeridis2020assessing} & -- \\
 \addlinespace[0.5em]
 
 & DNSSec & Domain Name Translation & \cite{raavi2024securing} & -- \\

\midrule

\textbf{Network} & IPSec & Virtual Private Networks (VPN) & \cite{cano2024integrating, khorkheli2024espq, mirani2022implementation} & \cite{gao2025ipseq, chen2024security, garcia202591} \\
\addlinespace[0.5em]
 & DSRC & Vehicle-to-Vehicle (V2V) & \cite{twardokus2022cryptography, lohmiller2025survey} & -- \\

\midrule

\textbf{Data Link} & MACSec & Point-to-Point Encryption & \cite{cho2021tlv} & \cite{alia2025quantum} \\

\bottomrule
\end{tabular}%
}
\end{table*}

\textbf{Application Layer.} The application layer provides a direct interface with users and the network. Alternative protocols include secure remote access protocols such as SSH and publish-subscribe protocols like MQTT, which are commonly used in IoT systems \cite{malina2024quantum}. Transport Layer Security (TLS), the foundation of modern web browsing, is also going through updates. Recent versions of TLS incorporate PQC for both authentication and encryption. Similarly, QUIC (Quick UDP Internet Connections), which is an alternative to the traditional TCP/TLS stack, is also being targeted for integration with quantum-resistant primitives \cite{kempf2024quantum}. QUIC combines the TLS and TCP handshakes, reducing round-trip time for establishing connections, making it suitable for applications that demand low latency. Efforts to include QKD in TLS are also found in the literature \cite{garcia2023quantum}, but because of its limitations, such as complexity and cost, it has not yet found widespread adoption.

End-to-end message exchange applications, such as email and instant messaging, must ensure protection against eavesdropping. These are application examples targets for PQC migration \cite{847_doberl2023quantum}. While email traditionally relies on protocols like the Simple Mail Transfer Protocol (SMTP), end-to-end encryption can be implemented at the application level, ensuring that only the sender and recipient can decrypt the message. An example is the Signal protocol with Triple Ratchets \cite{dodis2025triple}. It is important to note that end-to-end encryption differs from other application-layer protocol-level security, where intermediaries (e.g., centralized servers) may still access message contents. Another use case of end-to-end encryption is Message Layer Security (MLS) \cite{hashimoto2025exploring}. This can be achieved by using a post-quantum Double Ratchet key exchange to create secure group messaging. This allows for reducing message exchange complexity without revealing the keys to all users if a single one is compromised.

Efforts are also found to apply QKD in email communication \cite{847_doberl2023quantum}, for example, by establishing QKD-based secure links between mail servers while using PQC for client-to-server communication.

\textbf{Network Layer.} The network layer includes network addressing, routing, and packet forwarding. One relevant use case for PQC and QKD is in building Virtual Private Networks (VPNs). VPNs work by tunneling packets to create secure connections on the internet, protecting the network layer content, such as network addresses and internal payloads. Popular implementations such as IPSec \cite{cano2024line, cano2024integrating}, WireGuard \cite{hulsing2021post}, and OpenVPN incorporate PQC on either end-hosts or network devices, providing encryption at the network layer. The use of QKD-generated keys for secure tunneling has also been explored in the literature \cite{gao2025ipseq}, thereby enhancing both authentication and encryption. It is also important to note that common authentication protocols, such as IKEv2, are undergoing migration. This kind of protocol is employed by entities that initiate the establishment of a new VPN using IPsec \cite{pazienza2022analysis}. Efforts to develop PQC for the IKE protocol exist, including from academia and standardization bodies such as the IETF \cite{tjhai2019framework, fluhrer2019postquantum}.


\textbf{Data Link Layer.} To ensure point-to-point encryption at the link layer, protocols like MACsec are being considered for PQC integration. While earlier versions of MACsec relied on pre-shared keys rather than key-encapsulation mechanisms, modern approaches explore the use of PQC for authentication and QKD for key distribution. Specifically, the use of PQC to provide authentication in the MACsec key agreement (MKA) is proposed in \cite{buruaga2025versatile}. Furthermore, the usage of QKD alone is also investigated. Efforts to establish MACsec sessions between switches using keys shared using QKD are documented in experimental papers \cite{alia2025quantum}. 

Emerging communication domains are increasingly investigating PQC at lower layers of the protocol stack. One prominent example is Vehicle-to-Vehicle (V2V) communication. Standards such as IEEE 1609.2-2022, which define secure wireless communication in vehicular networks, are being adapted to support PQC adoption \cite{twardokus2022cryptography, lohmiller2025survey}. Additional applications include secure communications for the Internet of Things, including drone networks and industrial environments \cite{kumar2024future}. For instance, efforts are underway to integrate PQC into the Modbus protocol for SCADA systems. QKD is also under investigation for use in industrial and smart grid systems \cite{alshowkan2022authentication}, with particular emphasis on SCADA-based applications. \\


\subsection{Classical/Quantum Hybrid Mechanisms}
\label{subsec:hybrid}

\begin{figure}[t!]
    \centering
    \includegraphics[width=0.47\textwidth]{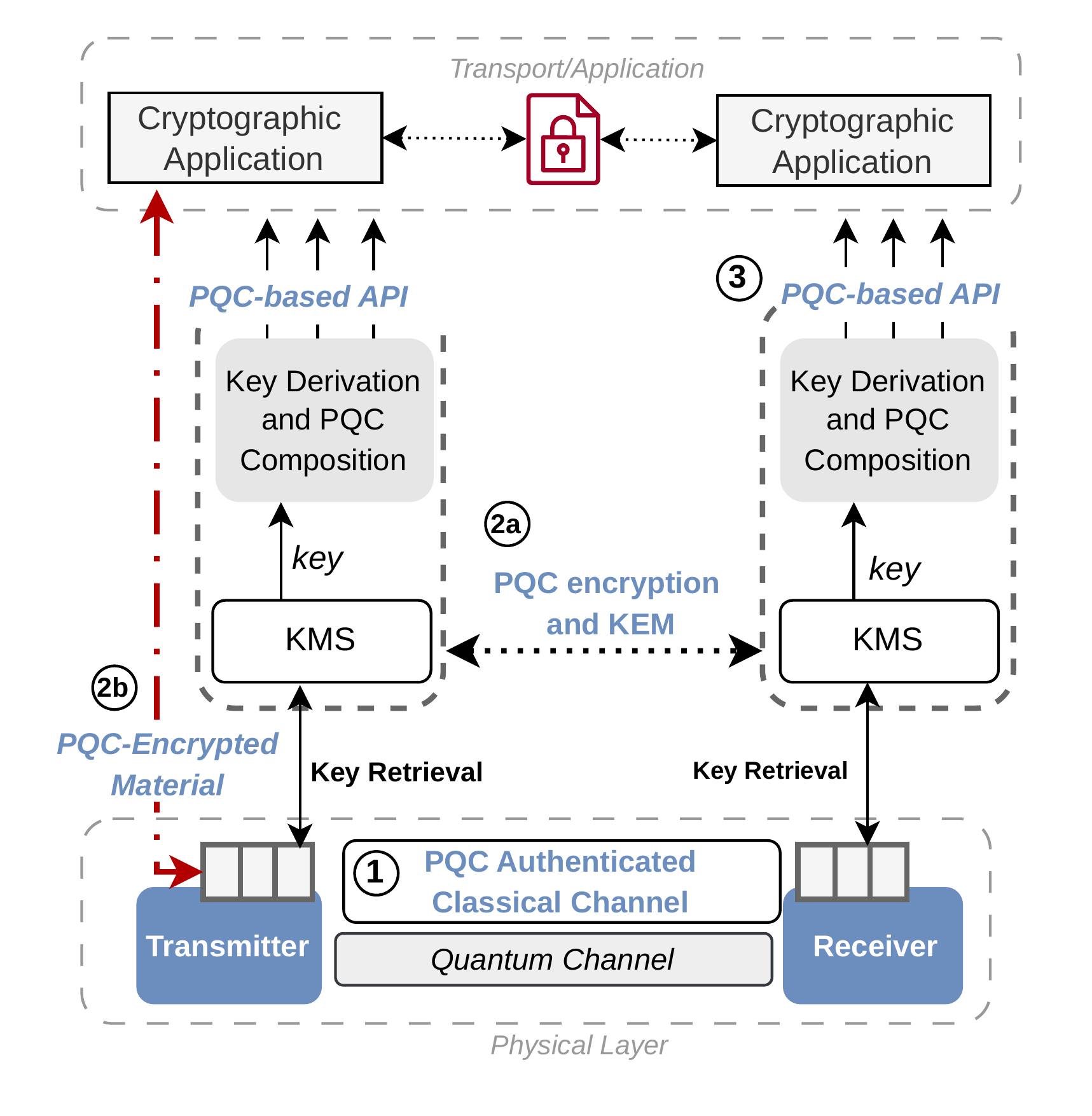}
    \caption{Hybrid mechanisms combining PQC with quantum cryptography: 1) PQC authenticates the classical communication during the quantum protocol; 2a) PQC encryption or KEM strengthens the task at hand in the quantum protocol; 2b) PQC encrypts the material transmitted in the quantum transmitter; 3) PQC secures the API interface post the quantum protocol. }
    \label{fig:hybridpk}
\end{figure}

Beyond having existing research and deployments with PQC independently, hybrid schemes combined with quantum cryptography have been found in the state-of-the-art, most of them focused towards QKD. We identified three existing categories of hybrid cryptographic mechanisms: channel authentication, combining forces and hybrid infrastructures. Figure \ref{fig:hybridpk} illustrates these three hybrid mechanisms across multiple architecture layers for a quantum cryptography protocol integrated in the physical layer of a traditional classical communication infrastructure. We now survey each of these hybrid schemes, presenting a comprehensive overview of how the existing quantum-resistant schemes complement each other in the literature.


\textbf{Authentication.} Quantum protocols provide cryptographic guarantees with information-theoretic security since any eavesdropping attempts on the quantum channel introduce disturbances. The protocols are designed such that these disturbances are guaranteed to be detected by the legitimate parties over a public discussion. Such an interactive public channel is required for most, if not all, quantum protocols, such as QKD and QSDC. This public channel needs to be authenticated so the two parties can be certain of the authenticity the data that they receive from each other to deduce the presence of an eavesdropper. In addition to eavesdropping detection via QBER estimation, other critical procedures are performed over classical authenticated channels, e.g., error correction and privacy amplification in QKD, making them an essential resource. In fact, current commercial quantum cryptography systems arrive with pre-shared keys stored on them, to be used in symmetric key-based authentication, such as information-theoretic secure MACs. However, this is not a very practical solution. Consequently, QKD often faces criticisms in security standards regarding its practicality due to its reliance on pre-shared keys for information-theoretic security. Furthermore, these pre-shared keys between each pairs of QKD users in a large network scale poorly, and also make it particularly challenging for a new user joining the network. For this reason, public key infrastructure (PKI) based authentication is more appealing where each user could obtain their digital certificates from a trusted authority, and sign communications using their private key to be verified by anyone in the network.

For quantum-safe alternatives of authentication, PQC-based digital signatures can be used instead \cite{wang2021experimental}. Figure \ref{fig:hybridpk} Step 1, illustrates an authenticated classical channel between two quantum transceivers. Note that reducing the information-theoretic security of the authentication scheme to computational security does not necessarily pose a threat in QKD as long as the attacker can not break the authentication until the key distribution protocol ends \cite{mosca2013quantum}.
Under the assumption that the authentication protocol cannot be broken in real-time, its short-term security suffices to preserve the long-term security guarantee of the keys distributed through QKD.
The duration of this short-term security has been quantified with respect to the temporary trust placed in the intermediary nodes of a trusted repeater network (TRN) setting \cite{solomons2022scalable}. Various demonstrations of integrating PQC-based methods for authentication in QKD have been performed \cite{yang2021all, garms2024experimental, nikolopoulos2024quantum, lauterbach2025post}, and Telsy encryptors are stated to be compatible with PQC signatures \cite{telsy}. While such a hybrid authentication mechanism can, in practice, be applied to other quantum cryptography mechanisms, such as quantum direct communication and quantum oblivious transfer (QOT), an argument analogous to the short-term authentication security for QKD has not been.


Towards a different kind of hybrid authentication approach, there is increasing interest in using physical-layer (PHY) security primitives to provide quantum-resistant solutions \cite{chowdhury2022physical}. One such example is physical unclonable functions (PUF) to replace the pre-shared keys stored in quantum cryptography devices. PUFs are unclonable devices that prevent copying the keys, as opposed to non-volatile memories, and generate secret keys on demand by exploiting inherent structural properties of the PUF material  and the uncontrolled iterations with a probe. Most existing PUF integrations in QKD for authentication, however, also require classical (conventional or PQC) authentication schemes such as digital signatures for the initial communication between the two parties \cite{yang2021all, garms2024experimental, nikolopoulos2024quantum}. This further stresses upon how fundamental authenticated channels are in a communication infrastructure. To further strengthen PUFs' security against classical and quantum machine learning attacks, there are also works with specific constructions of PUFs based on learning with errors (LWE)-hardness \cite{wang2020lattice} and on learning parity with noise (LPN)-hardness assumptions \cite{herder2016trapdoor}. Lastly, there have been proposals for quantum-readout PUFs (QR-PUFs) and quantum PUFs (QPUFs), with the latter being harder to implement \cite{julia2026secure}. To conclude, while PUF-based authentication has mainly been adopted for QKD, it remains interesting for other quantum protocols as well.

\textbf{Combining forces.} Quantum cryptography techniques are currently less mature and practical than PQC for widespread deployment, being more resource-intensive, still in the process of standardization, as well as due to their practical security concerns. Therefore, combining forces through hybrid integration of quantum protocols with PQC-based encryption and KEMs is a practical approach. For example, in a TRN setting for QKD, the key forwarding between trusted nodes could be strengthened using encryption from PQC KEM distributed keys. This was demonstrated in the Berlin OpenQKD testbed \cite{geitz2023hybrid} and then in the Paris backbone network \cite{IDQ}. Moreover, secret keys derived from PQC KEMs can be combined with keys generated from quantum sources to ensure security as long as at least one of the constituent schemes remains secure. Figure \ref{fig:hybridpk} illustrates this combination at Step 2a. In this step, once a new key is obtained from the device hardware, a key management system (KMS) exchanges keys with PQC KEM algorithms to combine them with keys obtained through the quantum protocol. There have been proposals for a hybrid QKD-KEM for integration into TLS using OpenSSL \cite{blanco2025qkd}, as well as for combining PQC, QKD, and classical cryptography into TLS and IPsec \cite{garcia2025enhanced}.

Combining post-quantum and quantum cryptography is, however, non-trivial and remains an active area of research \cite{ricci2024hybrid, hovelmanns2025qkd}, while recommendations from NIST and ETSI already exist \cite{ETSI2025, barker2018recommendation}. In particular, NIST allows a generic key composition scheme under SP 800-56C \cite{barker2018recommendation} in which shared secrets from different key-establishment schemes can be concatenated and then fed into a key derivation function (KDF) that serves as the key combiner. There are also works exploring combining different KEMs. However, note that even though XORing KEM keys was initially accepted \cite{tysowski2018engineering}, it is no longer consistent with NIST recommendations, and was shown insecure against chosen ciphertext attacks \cite{hovelmanns2025qkd}.

Beyond the key distribution problem, other multi-party setting tasks such as secure multi-party computation (MPC) are highly relevant from a network perspective. While hybrid approaches for MPC combining conventional classical and PQC schemes already exist \cite{airin2025hybrid}, constructions integrating PQC and quantum cryptography remain unexplored. Furthermore, given the relevance of OT as a building block for MPC, hybrid OT protocols combining PQC and quantum cryptography should also be explored to assess the practical advantages of such a hybridization. Even though such hybrid OT constructions do not yet exist, they should fit in this category of hybridization. With respect to direct communication protocols, PQC-based encryptions on secret messages could further enhance the security of the transmitted message. Figure \ref{fig:hybridpk} illustrates this combination at Step 2a. This concept was introduced in the secure repeater network (SRN) architecture for QSDC proposed in \cite{long2022evolutionary}, where repeater nodes relay encrypted messages rather than secret keys themselves, as is the case in TRN architectures for QKD. It is worth noting that a key-less QSDC network \cite{qi202115}
does not inherently require the intermediate layers components of key management, derivation, etc., as required by key distribution schemes, and hence, bypasses these layers, as shown in Figure \ref{fig:hybridpk}.

\textbf{Hybrid Infrastructures.} In addition to the aforementioned hybridization through PQC-based authentication and hybrid security strengthening, there are other points of vulnerability in a hybrid network that could benefit from PQC. Since it is impractical (even infeasible in some cases) to secure all network communications using quantum cryptography, a hybrid infrastructure can have QKD protecting the backbone nodes whereas the remaining communications are secured using PQC. These links, depicted in Figure \ref{fig:hybridpk} Step 3, may include the communications between the different KMSs, between KMS and the applications, as well as within the application layer of the network itself. These communications could be protected using encryptions using keys distributed using PQC KEMs as well as authentication based on PQC digital signatures. In \cite{geitz2023hybrid}, a number of hybrid protocols were implemented in a testbed in Berlin, as part of OpenQKD project. Another work discussed a quantum-secure architecture with QKD in the network core and PQC in the edge \cite{384_hoque2024exploring} devices. This design enables the core to keep strong security guarantees, while also enabling end devices to maintain quantum-safe properties at lower costs.



\subsection{Fundamental Building Blocks for Integration}
\label{subsec:integration}

The communication protocols presented previously in Section \ref{subsec:netprotocols} are fundamental building blocks observed in the migration of classical networking protocols to either PQC primitives or practical QKD. However, the abstraction layers used to characterize the integration of classical networking protocols are insufficient to cover more evolved technologies of quantum cryptography \cite{cao2022evolution, li2024survey}. This section discusses fundamental building blocks that enable integrating quantum cryptography protocols in the existing classical optical networks focusing on the first three stages: trusted repeater networks (TRNs), prepare and measure networks (PMN)\footnote{Note that the name prepare and measure in PMN has the same underlying meaning as prepare-and-measure BB84, with the transmitting node having the ability to prepare quantum states and the receiving node that of measuring.} and entanglement distribution networks (EDN) \cite{wehner2018quantum}.


In TRNs, long-distance communication is achieved by placing trust in intermediary nodes, which relay secret information hop by hop between the communicating parties. PMNs constitute the simplest form of quantum networking in which end nodes can prepare and measure quantum states directly over a quantum channel, without requiring entanglement distribution. EDNs extend this capability by enabling the generation and distribution of entanglement between distant nodes, thereby supporting more advanced quantum networking functionalities. These architectures support different classes of protocols: TRNs are sufficient for near-term QKD networks based on trusted nodes; PMNs support prepare-and-measure protocols such as BB84-type QKD, QSDC and some semi-device-independent or other mistrustful cryptography primitives; EDNs enable entanglement-based QKD, networked Bell tests, device-independent protocols, and eventually more advanced quantum-internet applications.

\begin{figure}[h!]
    \centering
\includegraphics[width=0.5\textwidth]{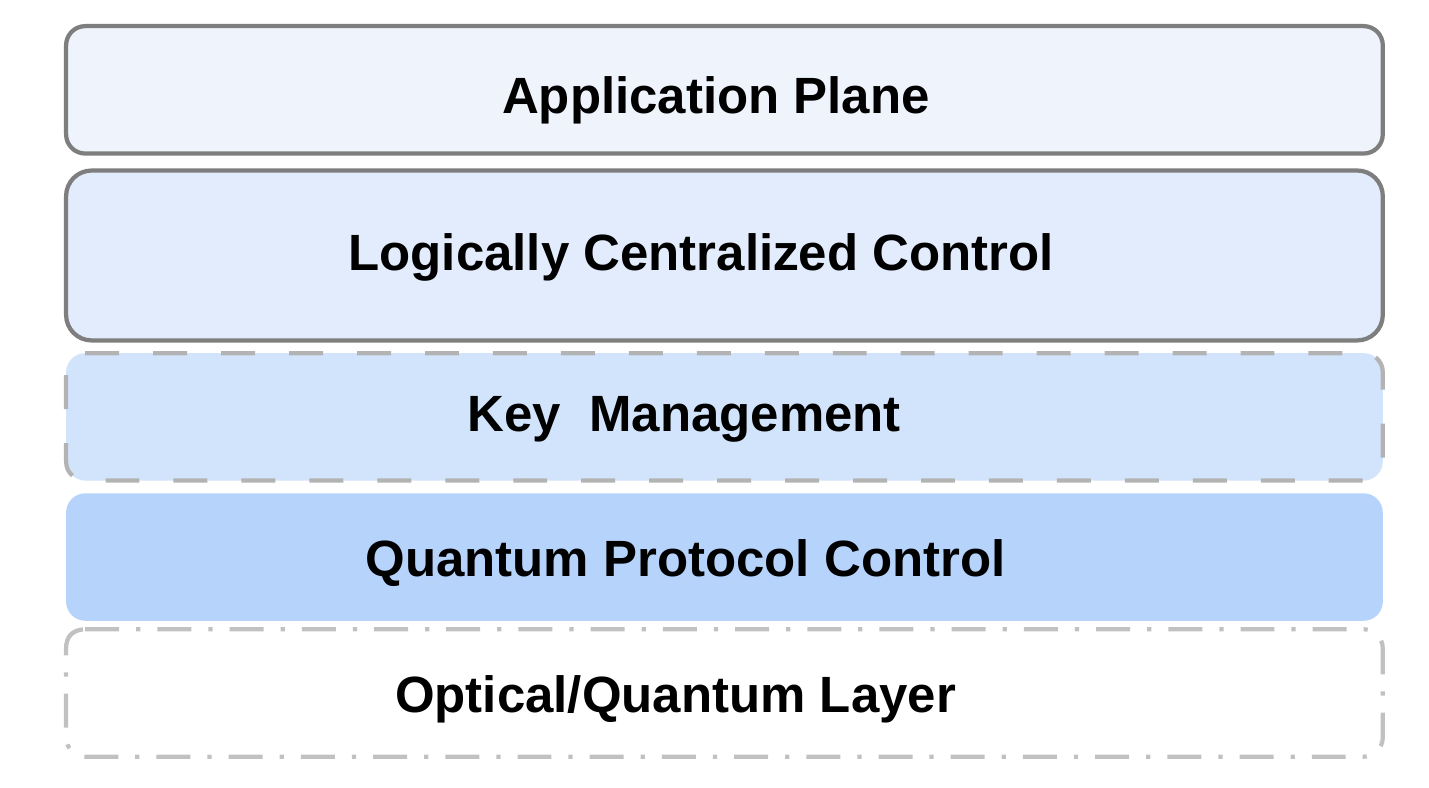}
    \caption{A representation of abstraction layers based quantum communication building blocks and existing literature for SDN/quantum networks management.}
    \label{fig:sdnlayers}
\end{figure}

\begin{figure}
    \centering \includegraphics[width=0.95\linewidth]{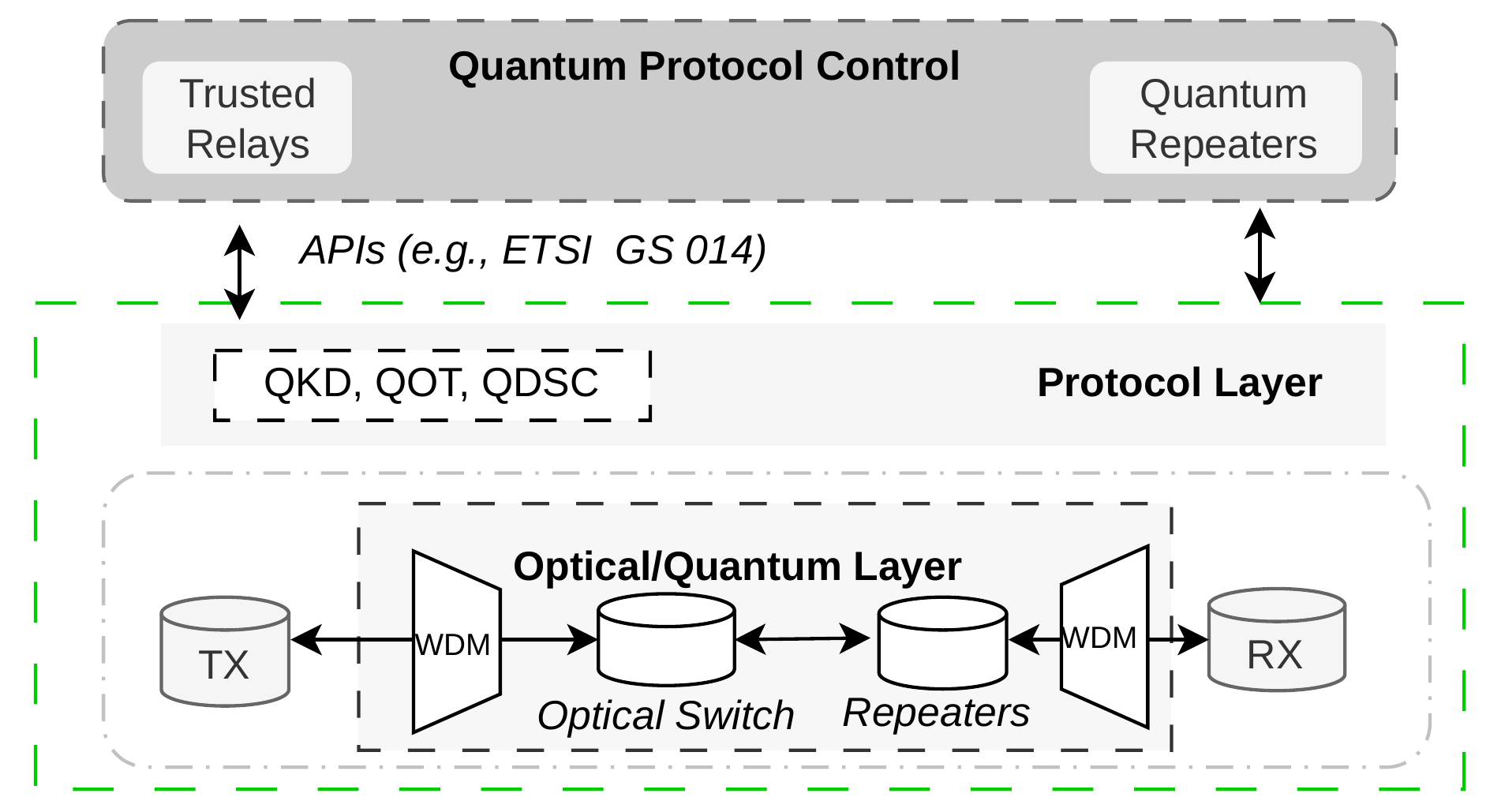}
    \caption{Quantum building blocks: the optical/quantum layer is composed of hardware components for communication; protocol switches enable to switch between different protocols using the same hardware; the protocol control lies in the control of quantum building blocks to build trusted relays or quantum repeaters.  }
    \label{fig:quantum_build}
\end{figure}

Figure \ref{fig:sdnlayers} systematizes a set of abstraction layers distilled from the literature. The figure reflects two worlds: high-level network control and management, and quantum protocol building blocks. In the application plane, standard networking applications can run to support programs that define networking behavior. The application plane interfaces with the controller, which is the network OS, analogous to traditional SDN. The centralized control depends directly on the KMSs and the lower level building blocks. The key management may be directly connected to the quantum layer or to an intermediary layer of trusted relay, which forms a network of trusted devices. Conversely, PMNs and EDNs bypass the need for trusted relays, since they ensure that the underlying hardware can share quantum material directly between the end points. Note that in order to extend the quantum transmission to longer distances in networks beyond TRNs, untrusted repeater nodes, called quantum repeaters, are necessary in the quantum protocol building blocks, as depicted in Figure \ref{fig:quantum_build}. The requirements for the quantum repeaters, for example in terms of storage time, can depend on the exact architecture of the networks. Additionally, the bottom layer represents protocol and hardware building blocks responsible for transmitting quantum states over optical fiber, free-space terrestrial, or satellite links.

Next, we discuss the building blocks present on these layers in a bottom-up manner, i.e., starting from lower-level abstraction layers (optical/quantum layer) up to the highest level of abstraction (application plane). \\


\textit{Quantum Building Blocks:}

\textbf{- Hardware modules.} Most quantum cryptography protocols can be implemented in a modular hardware architecture. A combination of hardware components can serve as fundamental building blocks of the physical layer and can be used to implement different types of quantum protocols. These modules allow the generation and modulation of quantum states at the transmitter, which are subsequently transmitted to the receiver, to be demodulated to decode the information. Typical hardware modules range from pulse carving, intensity modulation, phase modulation, and polarization modulation to interferometric measurements, single-photon detection, and photon-number-resolving detection. Some of these individual modules, such as intensity modulators and phase modulators, are commercial-off-the-shelf devices, which are often operated through graphical user interfaces (GUIs) for ease of configuration. On the other hand, high-speed and flexible system control can be achieved using field-programmable gate arrays (FPGAs), through precise timing synchronization and low-latency electrical control signals. Note that reusing hardware across protocols also requires verifying that the resulting implementation remains compatible with the security proof of each protocol, especially when modulators, detectors, timing electronics, or calibration routines introduce protocol-dependent side channels.

There are various quantum cryptography protocol designs that share a set of common hardware modules for the transmitter and receiver. For example, polarization-based QKD protocols of BB84 and SARG04 can share most of the hardware equipment for their transmitter-receiver modules to allow for a system that dynamically changes between the two protocols. In addition, some QOT protocols \cite{lemus2025performance} share the same hardware as the standard BB84 protocols. A recent experimental demonstration used a set of hardware modules to switch between different protocols for QKD, QDS, QBA, and quantum conference \cite{lu2025fully}. In particular, they implemented a degree-of-freedom (DoF) converter that enables the use of mainstream modulation schemes of polarization, time-bin, and phase. While many modules are commercially sold, their integration and control should be designed for a given set of quantum protocols that they can support. This integration happens in both hardware and software domains.

\textbf{Optical switches.} Going further into the physical layer, additional hardware is required to allow for a flexible communication infrastructure and build larger networks. As in conventional classical optical networks, optical switches play a critical role in quantum cryptography as well. These optical devices extend point-to-point links into a scalable, flexible multi-user network by dynamically reconfiguring the physical optical path.

A simple multi-user network can use optical splitters to allow for a cost-effective solution for passively switching optical paths, as used in passive optical networks (PONs) in a one source to many endpoints $1\times N$ configuration. However, PONs lack flexibility due to passive selection of the end users as well as scalability due to the limited number of users owing to increasing splitter losses with the number of users.

Optical switches allow for cost-effective multi-user networks by enabling dynamic reconfiguration of the $N \times N$ optical transmission paths with lower losses than the passive splitters. While optical switches do not help in extending transmission distances, they solve the issue of a reconfigurable system without the need to trust the intermediate switching node. Small-scale metro networks using optical switching have been demonstrated \cite{tang2018quantum}. While optical switches have already been used in classical optical communication, those same switches cannot directly be used for routing weak quantum signals due to the sensitivity of quantum states to losses and noise. Therefore, specialized optical switches are required to ensure low insertion losses, minimal crosstalk, low polarization-dependencies, fast switching-times, high extinction ratios, and a wide input power dynamic range. 

Different technologies for optical switches exist, such as electro-optic and acoustic-optic switching, that achieve the fastest switching times (nanosecond timescale) at the cost of high insertion losses. Slower (micro-millisecond timescale) opto-mechanical or micro-electro-mechanical (MEMS) switches can provide lower losses. Although opto-mechanical switches were the first to be commercialized, currently there is a large number of vendors selling specific-purpose optical switches based on different switching technologies. Various field trials have been demonstrated over metropolitan networks \cite{ciurana2014quantum, temporao2024rio}.

\textbf{Wavelength division multiplexing/demultiplexing.} 
Quantum signals are ideally transmitted over dark fibers, but the co-existence of quantum signals with other classical communication signals, e.g., for post-processing or synchronization, is seen as a cost-effective deployment solution. Wavelength division multiplexing (WDM) can be used for such solutions and is therefore an important building block for quantum cryptography. WDM allows multiple wavelengths to share the same optical fiber, enabling the coexistence of quantum and classical channels in the same fiber or enabling optical path switching. Recent literature examines the coexistence of quantum and classical channels for key distribution, including DV-QKD and CV-QKD over fiber using WDM  \cite{726_ntanos2020qkd, 393_wang2025facing, 433_milovanvcev2021high, 594_zavitsanos2020qkd, 251_gao2024co}.

WDM can be classified into two different categories: coarse wavelength division multiplexing (CWDM) and dense wavelength division multiplexing (DWDM). CWDM allows for large spacing between different wavelengths, e.g., quantum signal at 1310 nm and the classical at 1510 nm; while dense wavelength division multiplexing (DWDM) has a small channel spacing, mostly used in large data capacity applications. WDMs are already commercialized due to their widespread use by telecommunications companies. However, standard WDMs include optical amplifiers such as an erbium-doped fiber amplifier (EDFA) to enable long-distance propagation, whereas the quantum signal cannot be amplified. Therefore, EDFA bypass schemes are employed in WDMs that are used to transmit both quantum and classical signals over the same fiber, such that only the classical signals are amplified. Moreover, the additional noise due to the relative intensities of the quantum and classical signals is another challenge for WDM-enabled QKD.

The integration of QKD with WDM in optical networks remains an active research area, with numerous field experiments reported. Several field trials have demonstrated QKD with WDM \cite{sasaki2011field}, with some even using DWDM \cite{woodward2021quantum}. Note that the use of WDMs in QKD is not limited to utilizing the same fiber for classical and quantum communication. It extends to the optical path switching capabilities of the system, to have wavelength-selective switches, such as quantum reconfigurable optical add-drop multiplexers (q-ROADM) \cite{wang2019end}. In this way, the system can optimize the path availability to prevent congestion by adding parallel links.

\textbf{Protocol switches.} Another important building block, not much discussed, is that of a protocol switch that allows changing between different quantum protocols. As mentioned above in the hardware module building block, as long as different quantum cryptography protocols utilize the same hardware for the quantum communication stage, there is an advantage in having a reconfigurable system that allows switching between the protocols. Then, depending on the required security levels and communication rates for a given cryptographic application, as well as eavesdropping scenarios, protocol switching could allow supporting different protocols. Recently, an optical payload design for a satellite transmitter was proposed that is capable of switching between QKD and QKPC \cite{mendes2024optical}. For CV-QKD, an experimental demonstration showed the feasibility of protocol switching for CV-QKD \cite{wang2023multi}, while another work demonstrated prototypes for a metropolitan area QKD network \cite{brunner2023cv}. A recent work performed simulations using IBM Qiskit for a protocol switching design for three QKD protocols based on a noise threshold \cite{priyadharshini2025hybrid}. As mentioned above, in \cite{lu2025fully} they demonstrated the experimental implementation for a prepare-and-measure 5-node quantum network with protocol switching capability for a number of quantum protocols, e.g., QKD, QDS, QBA, etc. Another work simulated dynamic protocol switching between classical, post-quantum, and quantum authentication protocols by implementing a real-time threat monitor \cite{zaman2025adaptive}. While this building block is in the research stage only, it will play a crucial role in quantum-secure communication networks, allowing protocol switching based on the end users' application requirements in terms of security levels and performance.

\textbf{Trusted relays.} Inherent physical-layer limitations, such as exponential losses of the weak quantum states, impose fundamental limits on the operational distances of fiber networks. Standard QKD protocols, therefore, allow for practical key generation rates only over a couple of hundred kilometers. Trusted relays, as mentioned above, allow a solution that is scalable to arbitrary distances and numbers of users, provided enough trusted intermediary nodes are placed between the end users. These trusted relay nodes constitute the trusted repeater networks (TRN) \cite{wehner2018quantum}. Each adjacent pair of nodes connected through a quantum channel performs a quantum protocol, establishing pairwise secret keys. Using the secret information established between each pair of adjacent nodes, the source node can then relay its secret information to the destination in a hop-by-hop fashion. Consequently, this solution does not allow for end-to-end transmission of quantum states, and hence, no end-to-end security either, as the intermediary nodes need to be trusted. The relay process occurs over a classical channel by each intermediary node encrypting the received key material using a one-time pad before forwarding it to the next hop. 

Despite the strong security assumption of placing complete trust in the intermediary nodes, TRNs in metropolitan areas have existed for almost two decades \cite{peev2009secoqc, salvail2010security}. However, in order to reduce the risks of employing a trusted relay, some approaches are investigated. The usage of multi-path routing of a key across non-overlapped channels is an interesting possibility. In this solution, the transmitter may split the original key and relay across different paths, while the receiver will merge the different sequences of bytes into a single key \cite{elkouss2013secure}. Other solutions include strengthening the trusted relay nodes. One promising alternative is employing homomorphic encryption to protect the classical processing performed at trusted relay nodes \cite{brazaola2026privacy} or hardware security modules (HSM).

Some notable examples of TRNs with commercial QKD equipment include Cambridge network with Toshiba's commercial DV-QKD systems \cite{dynes2019cambridge}, MadQCI with Huawei's commercial CV-QKD modules as well as IDQuantique's DV-QKD modules \cite{martin2024madqci}, and Hefei network with QuantumCTek DV-QKD modules \cite{chen2021implementation}, among others. Some larger network examples include China network spanning 7,600 km through satellite relay \cite{liao2018satellite}, 4,600 km with integrated fiber and space-based connections \cite{chen2021integrated}, and an inter-European network \cite{ribezzo2023deploying}.

\textbf{Quantum repeaters.} 
To extend transmission distances for quantum-secure communication without placing complete trust in intermediate nodes, one approach is to use QKD protocols that increase transmission distance by design. For example, twin-field QKD (TW-QKD) can overcome the rate-distance limitations of conventional point-to-point QKD \cite{pirandola2017fundamental}. From a large-scale network perspective, this remains insufficient, as state-of-the-art TW-QKD demonstrations are limited to approximately 1000 km \cite{liu2023experimental}. Another option is to employ multiple paths using threshold cryptography techniques. This approach works by making it more difficult for an eavesdropper to attack all the trusted nodes across the multiple paths \cite{salvail2010security, solomons2022scalable}. The ultimate solution would be using quantum repeaters. Since the intermediary nodes in untrusted repeater networks are not trusted, compromising these nodes does not leak any information about the secret keys/messages to be distributed/transmitted between the source and destination nodes. In fact, intermediary nodes never have access to the secret keys or messages, unlike those in TRNs. Therefore, providing the trusted intermediary nodes with the capabilities of quantum repeaters can allow for large scale networks of higher stages than TRNs, providing end-to-end security. 

A key component of quantum repeaters is the creation of long-distance entanglement between the end users through entanglement swapping \cite{wehner2018quantum}. The distribution of entangled photons between end nodes separated by long distances is achieved with the assistance of intermediate nodes, using quantum memories to store parts of the entangled pairs and to perform Bell state measurements. While high-fidelity quantum repeaters are still not available, mainly due to limitations in quantum memories regarding storage time and efficiency, there have been recent demonstrations \cite{liu2026building}. A review on quantum repeaters can be found in \cite{azuma2023quantum}.

The trusted node-free network for 8-users in a  metropolitan area was demonstrated \cite{joshi2020trusted}. More recently, another 5-node network, featuring heterogeneous nodes capable of performing multiple tasks via different quantum protocols \cite{lu2025fully}. Note that small entanglement distribution networks, also capable of supporting device-independent protocols, have also been demonstrated \cite{wengerowsky2018entanglement, fan2022robust}. \\

\textit{Network Control \& Management.}

\textbf{Key Management.} Apart from direct communication protocols such as QSDC, QKPC, etc., most quantum protocols require a key management system (KMS), irrespective of the usage of those keys for encryption, authentication, or privacy-preserving computation tasks. KMS is used to manage the entire life cycle of secret keys, from their generation and storage to demand requests based on different application needs, through renewal and destruction \cite{james2023key}. Therefore, after the two parties finish the quantum protocol, the secret keys can be requested from the quantum hardware to be used by a final application. Note that due to the limited secret key rates, especially for most of the practical applications, quantum hardware could be running continuously in the background. The generated keys are stored in a buffer until they are requested by the application. 

\begin{figure*}[h!]
    \centering    \includegraphics[width=0.9\linewidth]{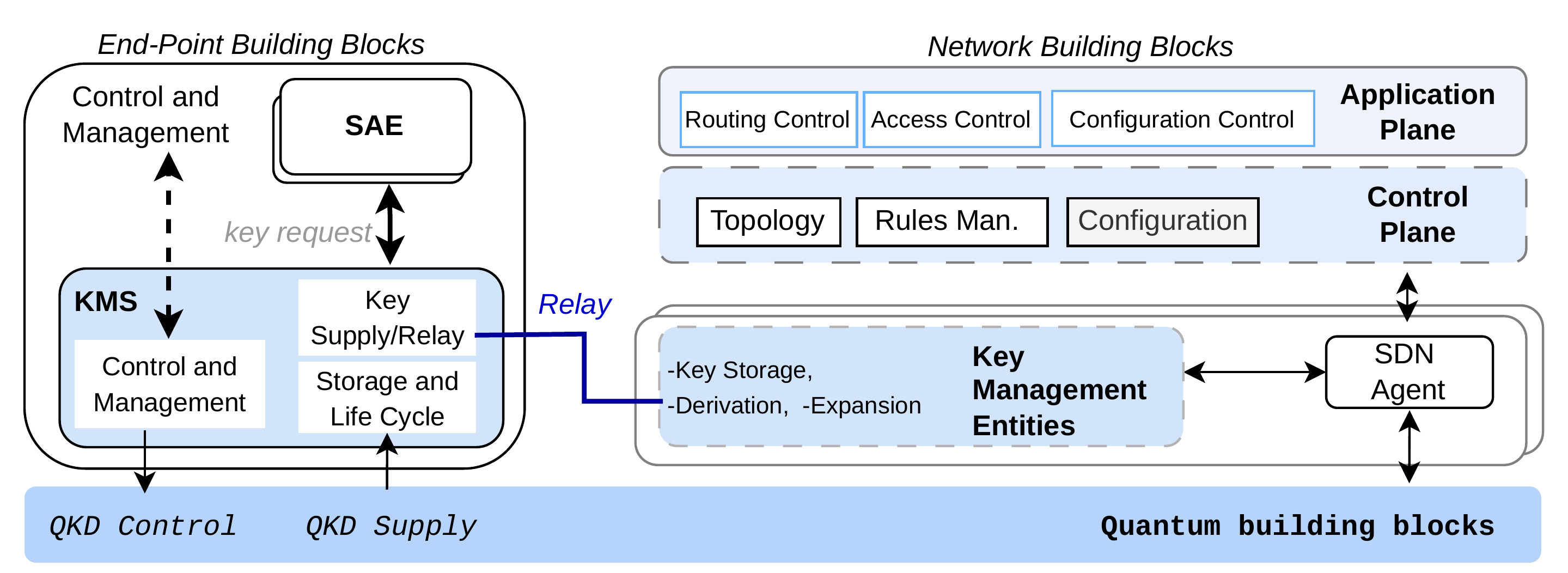}
    \caption{Network control and management: end-point key management inspired on ITU-T standards and components of a SDN over quantum building blocks.  }
    \label{fig:KMS}
\end{figure*}

Currently, several interface specifications for a KMS focusing specifically on QKD are available, including ETSI GS QKD 004 \cite{etsi2020quantum}, ETSI GS QKD 014 \cite{distribution2019protocol}, and Cisco SKIP. A KMS for QKD is shown on the left side of Figure \ref{fig:KMS}, which is a simplified version on ITU-T standards. The keys can be retrieved through this QKD supply interface, along with respective IDs, and stored in the KMS buffers. The keys obtained by the KMS can be relayed to other KMS components hosted in the QKD network, often called key management entities (KMEs) or delivered to a secure application entity (SAE), which uses them for symmetric encryption or authentication. The way a key is relayed (or routed) can be defined by an application over the QKD network controller, which may include quality of service functionalities.

In the case of QOT, the functionality of a KMS will differ from QKD in terms of the storage and distribution of the generated keys. Since the oblivious keys generated through QOT protocols are asymmetric, they must be stored with tags associated with the sender or receiver component. Furthermore, the key requirements differ from those for AES128-192-256 symmetric encryption using QKD keys. It depends on the specific MPC protocol using OTs as sub-protocols, as well as the computation task at hand. Therefore, the KMS must be prepared to handle key demands accordingly. 

As mentioned in Section \ref{sec:concepts}, due to the high OT requirements of most real-world applications, protocols for OT extensions are used to transform a small number of base OTs into a larger number of OTs. This is another feature that a KMS should have, i.e., the ability to perform OT extension protocols \cite{asharov2017more} (as explained in Section \ref{sec:other_PQC}) locally or with another KMS, using the OT keys received from the quantum devices. It is noteworthy that there are efforts in the direction of integrating OT protocols in the QKD interfaces, for example, there are proposals on the extension of ETSI GS QKD 004 functionality to include OT \cite{romero2025enabling}. Lastly, while a KMS for OT will handle oblivious keys, it must also have the standard functionality for QKD. This is because authenticated channels are still required in a multi-party MPC protocol at the application layer. However, since the quantum hardware for most QOT protocols is sufficient to implement QKD, using an initial pre-shared key allows QKD instances to provide keys for future authenticated communication.

\textbf{SDN Controller.} While early QKD deployments primarily focused on establishing secure point-to-point quantum links, the evolution towards large-scale quantum-safe networks requires mechanisms for resource management, service orchestration, interoperability, and dynamic network control. One recurrent concept in the literature of network control and management is software-defined networking (SDN) \cite{SDN, benzekki2016software}, which plays a critical role in the seamless integration of QKD into existing infrastructure, while maintaining flexibility, scalability, programmability, centralized control, and abstraction capabilities. By decoupling the control and data plane logic \cite{mckeown2008openflow} and logically centralizing control, an SDN controller can manage the configurations of the previously discussed devices, such as trusted relays and optical switches. 

Traditionally, SDN is associated with well-defined protocols, such as OpenFlow. In deploying SDN for QKD networks, the SDN control plane lacks standardized protocols, often necessitating rework of ETSI and ITU recommendations. Current efforts employ combinations of Network Configuration Protocol (NetConf) and ETSI 015 to configure low-level devices \cite{paolini2023integrating}. 

The controller interface is often connected to an agent inside the device, which in turn connects to key management entities and the quantum/optical hardware. This allows the controller to perform several tasks, such as managing quantum resources, orchestrating optical links and logical connections, and providing an application interface. 

As QKD networks evolved beyond isolated point-to-point links, SDN assumes a broader orchestration role. Rather than simply controlling network devices, SDN became responsible for coordinating quantum resources distributed across the network. This transition is evident in the Key-as-a-Service framework proposed by Cao et al. \cite{cao2019kaas}, in which key pools and virtual key pools are dynamically assembled and allocated to meet service requirements. Similarly, the software-defined heterogeneous QKD Chaining architecture extends SDN orchestration to multi-protocol environments, enabling the deployment and management of heterogeneous QKD chains composed of different QKD technologies and connectivity patterns \cite{cao2022software}. Network slicing concepts have also been introduced into QKD environments, enabling SDN controllers to allocate quantum resources based on Quality-of-Service requirements and service demands \cite{wang2023network}.


Another research direction focuses on integrating QKD systems into existing optical networking infrastructure. Traditional QKD deployments often rely on dedicated and isolated infrastructures, which limit scalability and increase deployment costs. To address this limitation, several works propose SDN-based frameworks capable of jointly managing classical and quantum network components. Paolini et al. \cite{paolini2023integrating} and Andriolli et al. \cite{andriolli2024software} present SDN architectures that provide unified control over both QKD devices and classical optical network elements while leveraging emerging ETSI and ITU-T standardization efforts. Similar objectives are pursued in the Madrid Quantum Network \cite{martin2019quantum} and the subsequent MadQCI deployment \cite{martin2024madqci}, where SDN is used to support heterogeneous, multi-vendor, and dynamically configurable QKD infrastructures. These studies indicate that SDN can simplify the operational integration of QKD technologies into existing telecommunications ecosystems. 

Secret keys can also be viewed as a network resource that must be managed throughout their lifecycle. This perspective has motivated the development of SDN-assisted key management frameworks. The architecture proposed by Sanz et al. \cite{sanz2025toward} introduces virtual key management systems (vKMSs) and a quantum security controller (QuSeC) to address scalability challenges in trusted relay QKD networks. The framework abstracts multiple underlying key management entities while enabling relay path discovery and policy-based key distribution. In a complementary direction, Tessinari et al. \cite{tessinari2023software} propose securing the SDN control plane itself through QKD-generated keys, effectively creating a feedback loop in which QKD protects the orchestration infrastructure responsible for managing QKD services.


Rather than merely configuring network devices, SDN controllers dynamically modify the operational parameters of QKD systems. Arabul et al. \cite{arabul2021experimental} demonstrate SDN-controlled programmable encryptors capable of changing encryption algorithm and key consumption policies according to application requirements. In parallel, Moreolo et al. \cite{svaluto2025continuous} investigate SDN-enabled CV-QKD systems in flexible optical networks, where quantum channel wavelengths and spectral allocations can be dynamically adjusted to improve coexistence with classical traffic. These approaches illustrate a shift from static QKD deployments to adaptive quantum networking environments that respond to network conditions and service demands.

\textbf{Services and Applications.} The application plane is the highest level of abstraction, where networking applications can be built, abstracting away the low-level configuration details of other layers. The SDN can host applications that perform multiple functionalities, such as traffic engineering, analytics, automation, and policy enforcement. Changing the way keys are routed to preserve QoS requirements is one of the most widely explored applications of quantum cryptography, focusing specifically on QKD \cite{chen2021ada}. The route computation occurs in the application plane, enabling application changes to suit different goals, e.g., bandwidth utilization. Routing is itself an essential task in QKD networks, performed via optical switches in more robust deployments or through a combination of transmitter/receiver hardware on trusted relays.

An application on a centralized SDN controller can dynamically adjust the wavelength to allocate resources and optimize coexistence between classical and quantum channels. The DISCRETION project \cite{bastos2025discretion} demonstrated a real integration between SDN and QKD. The integration between the SDN controller and QKD devices is performed in accordance with the ETSI GS QKD 015 standard. Another application is creating network slices for different tenants, i.e., a virtual, isolated quantum network for each client \cite{wright20215g}. In this scenario, the SDN application can orchestrate and provision resources for each client, accommodating different cryptographic and performance requirements.  Overall, the SDN offers centralized control to manage routing and resource allocation in large-scale QKD networks \cite{398_zavitsanos2023feasibility, 491_gatto2022integration, 818_gatto2021quantum}.

Another example of application is in actively working in the key relay process \cite{james2023key}, which can take the form of hop-by-hop XORing keys before relaying, ensuring transport security. This process is only necessary for the current phase of TRNs. While controlling trusted nodes is an important step toward more incremental deployment of QKD technologies, it requires decryption and re-encryption of keys at every hop. A more robust deployment would rely instead on quantum repeaters. An optical switch like reconfigurable optical add-drop multiplexers (ROADM) allows adding, removing, and altering the destination of a QKD bitstream, maintaining end-to-end security without relying on the security of intermediary nodes. On the other hand, quantum repeaters could allow information to be exchanged over longer distances while maintaining the security guarantees. For these cases, XORing keys is not necessary, but the other applications mentioned, such as traffic engineering, can be envisioned as necessary for scalability. However, the maturity of quantum repeaters, relying on quantum memories inside of QKD deployments, is still in its infancy, while proven practical in real experiments \cite{brunner2023demonstration}. 

\subsection{Summary and Takeaways}

In summary, secure network communication in the quantum era requires solutions spanning multiple protocol layers, with PQC increasingly being adopted across the whole communication stack. This observation suggests that PQC is more mature and deployable than quantum cryptography, making it the primary candidate in current migration efforts that focus on more conservative approaches. On the other hand, while QKD is also being integrated across various layers, the focus of this integration remains predominantly on the lower layers, such as the data link and network layers. Although we also observed QKD at higher abstraction layers (the application layer), the efforts remain comparatively limited. We assume that research on lower layers is more prevalent because current QKD technology seems more natural for point-to-point connections and requires less frequent key updates, which align with the requirements of the existing technology. 

As seen from the identified hybridization techniques, rather than viewed as competing technologies, PQC and quantum cryptography are being regarded as complementary approaches that address different aspects of post-quantum security. The hybrid architectures across multiple protocol layers highlight a migration strategy that leverages the immediate deployability of PQC while incorporating the information-theoretic security guarantees of QKD wherever appropriate. Such integration provides a balanced approach that accommodates current technological constraints while also preparing for security against future quantum-capable adversaries.

\section{Application Domains}
\label{sec:domains}

In this section, we organize the literature identified through our query into a classification of quantum-resistant applications. We present applications that use quantum-safe primitives and classify them by application domain and cryptographic discipline (or scenario). While the broader field of applied cryptography is preparing to migrate to PQC or quantum cryptography solutions, our focus is on applications that are already undergoing such migration in the literature we reviewed. With respect to quantum protocols, the references presented in the various application domains are theoretical unless otherwise specified. This is due to the limited maturity of most quantum cryptography protocols, except QKD.

\begin{figure}[t]
    \centering \includegraphics[width=.95\linewidth]{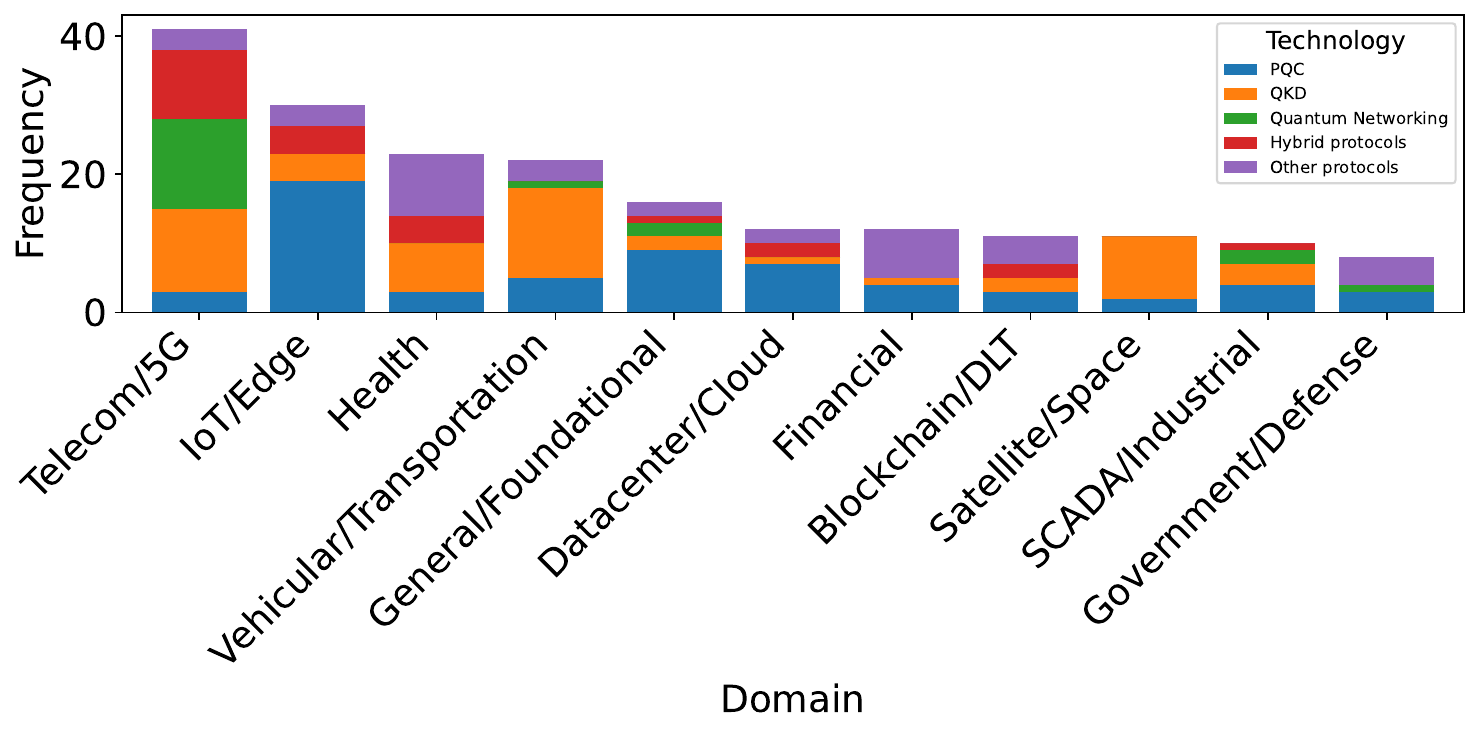}  
    \caption{Application domains identified among the 147 articles selected after Phase~2, grouped by technology type (PQC, QKD and its networking, hybrid and other protocols).}
    \label{fig:domains}
\end{figure}

The distribution of application domains was analyzed among the 147 articles retained after the second phase. Each abstract was coded for the presence of one or more of ten major application domains recurrent in the literature: telecommunications, connected devices and edge computing systems, governmental and defense environments, data center and cloud computing infrastructures, distributed systems and blockchains, industrial control systems and other critical infrastructures, space and satellite communication, vehicular systems and vehicle-to-infrastructure communications, financial systems, and digital health environments. For each domain, the integration strategy was recorded as quantum-based, PQC-based, or a combination of both. The resulting distribution (Figure~\ref{fig:domains}) demonstrates that telecommunications and mobile networks, along with connected devices and edge computing systems, are the most frequently addressed domains, followed by governmental and defense environments and data center and cloud infrastructures. Quantum-based approaches are most prevalent overall, whereas PQC-based and combined strategies are more common in connected devices, edge computing, and governmental contexts. These findings suggest that no single approach currently dominates the development of communication systems resistant to quantum-enabled attacks.

Next, we discuss the application domains mentioned above, clarify how migration takes form, and classify the cryptographic mechanisms employed in each domain.

\subsection{Enabling Infrastructure}
\label{subsec:infra}

We first classify the literature by the application of quantum-safe cryptography to enable application infrastructure.

\begin{figure}[t!]
    \centering    \includegraphics[width=\linewidth]{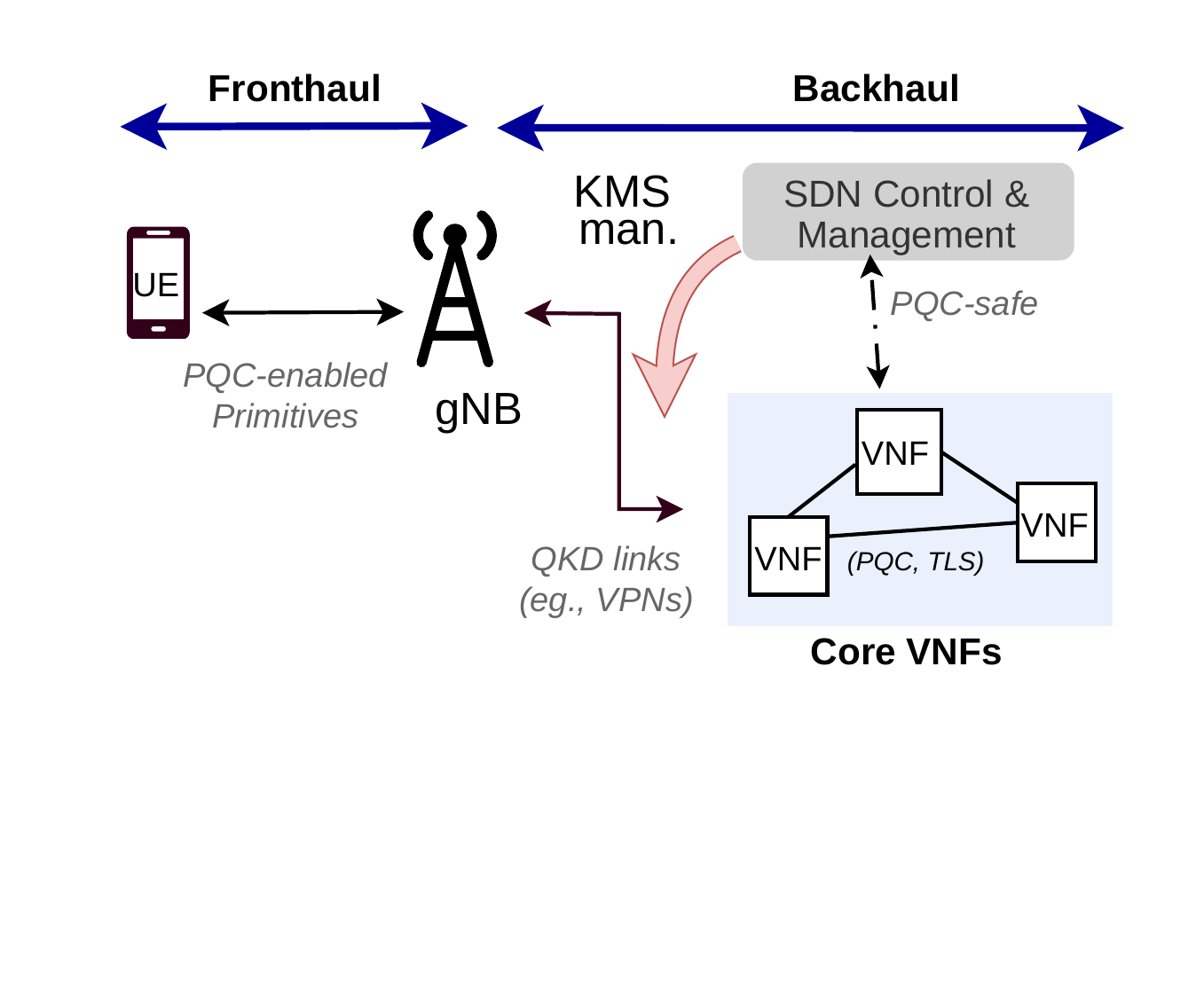}
    \vspace{-2.5cm}
    \caption{Quantum-safe 5G/6G architecture combining PQC/QKD. The user equipment (UE) exchanges data with the next-generation base station (gNB), which interfaces with 5G core virtual network functions via PQC-protected control- and user-plane channels. A dedicated QKD fronthaul/backhaul link generates symmetric keys that are delivered to a KMS coordinated by an SDN controller.} 
    \label{fig:qs_5g} 
\end{figure}

\textbf{Telecom and 5G.} Telecommunications and mobile network infrastructure are a foundational domain due to their role in supporting global connectivity, latency-sensitive services, and large user populations. The 5G core, for example, provides a flexible service-based architecture for telecommunication. The architecture includes services like network slicing, management, forwarding, and routing. Within this domain, these functions can employ post-quantum KEMs (e.g., in TLS) to protect communication between core functions, ensuring confidentiality across both control and data planes \cite{148_electronics13214258}. Beyond efforts to provide security guarantees for the core, we also observed efforts in the Fronthaul, which connects the core to the radio units. There have been various works in the 5G/6G Fronthaul integrating CV and DV-QKD \cite{726_ntanos2020qkd, 393_wang2025facing, 433_milovanvcev2021high, 594_zavitsanos2020qkd, 251_gao2024co}. Moreover, as illustrated in Figure~\ref{fig:qs_5g}, mobile network architectures are evolving to incorporate quantum-secure 5G designs that integrate QKD and PQC at the system level \cite{384_hoque2024exploring, 830_kumar2025quantum}.

Beyond the fronthaul and network core mentioned earlier, which serve as horizontal layers, efforts in the management and orchestration layers are important. Management and orchestration work vertically, spanning both the data and control planes. For example, the management layer can be used to provide efficient key management techniques, focusing on enhancing key generation throughput, enabling inter-domain key exchange, and supporting multiple simultaneous users. To this end, routing strategies for QKD that leverage approaches such as multipath key routing across trusted nodes are investigated. Other approaches are also considered to enable multi-user and multiple-quantum protocols in network architectures using efficient protocol designs, multiplexing techniques, and software-defined control \cite{914_alnaser2025secure, 88_ghourab2025qkdZero-trust, 335_wang2020q-ROADM, cao2022software, 419_giorgetti2025generalized, paolini2023integrating, 489_shokrivahed2024integration, zhou2020measurement, qi202115, brazaola2026privacy}. Specific to QKD, various experiments and feasibility analyses have been conducted to integrate CV and DV-QKD into FSO and optical fiber networks and their management, with a focus on challenges in real-world implementations \cite{brunner2023cv, 818_gatto2021quantum, 491_gatto2022integration, 398_zavitsanos2023feasibility, 593_oliveira2024integration, martin2024madqci, lopez2025enhanced, mendez2026switching}. There have also been efforts to incorporate PQC techniques into various hybridizations, as discussed in Section \ref{subsec:hybrid} \cite{384_hoque2024exploring, garcia2025enhanced, 659_doring2022post_book, lauterbach2025post, blanco2025qkd, 376_garms2024experimental, 540_dowling2020many_book, 900_mendez2024sdn}. There have also been efforts in bringing privacy-preserving features in the relayed QKD network architectures \cite{brazaola2025privacy}.

In addition, PQC is considered for migration in the whole telecom stack \cite{liu2025post}. It is important to note that PQC is more commonly discussed for integration into networking protocols, rather than for changing the underlying physical infrastructure. Fundamentally, the general network protocols and services in the telecom domain can include TLS, MACSec, and VPNs. More specifically, many frameworks are observed, such as WireGuard-QKD, Delta-chat with hybrid PQC and QKD, and others \cite{453_buruaga2025hybrid, 158_prevost2025etsi, 459_bushuev2021immunity, 452_ramu2025hybrid, 522_lyssenko2023leveraging, 945_chen2024security, 847_doberl2023quantum, 730_akshaya2025qsms, 83_network5020020, 668_jasim2025post}. 
More specific examples investigate PQC for protecting TLS communication in the 5G core \cite{scalise2024applied} and the use of hybrid PQC and classical cryptography for user equipment and home network authentication \cite{vuppala2023post}. However, simply integrating standardized NIST can lead to unacceptable overhead. Emerging efforts aim to optimize PQC primitives based on specific details of the 5G architecture. For example, in \cite{darzi2026lightweight}, the authors optimize PQ for base-station broadcast authentication, exploring its structure and operational features.

\begin{ieeeinsight}{Takeaway Message:} The telecom core is observed as the most prominent application domain for quantum cryptography. In this application, strong security guarantees can be achieved with MACs without impacting the flexibility of end-users' connections. 
\end{ieeeinsight}

\textbf{IoT and Edge.} The Internet of Things (IoT) and Edge domain includes applications in which communication and computation are performed on resource-constrained devices with limited compute, memory, and energy capacity. This domain is particularly critical because infrequent update cycles increase the risk of cryptographic obsolescence. Comprehensive research surveys address this domain \cite{kumari2022post, wang2025review, almutairi2025resilience}, focusing on quantum-safe techniques that integrate PQC and, in some cases, QKD to provide key establishment and data protection. 

At the heart of many IoT security frameworks, system-level schemes combine PQC, QKD, blockchain, and Intrusion detection for orchestration, key management, and anomaly detection \cite{146_elkhodr2025ai, 389_biswas2024exploring, 357_biswas2024enhancing, 202_shradha2024beyond, 660_sharma2025post, 848_cibik2025quantum, 831_mujlid2025quantum}. Further, research on embedded systems examines the constrained nature of these devices and the substantial effort required to optimize CPU cycles and memory footprints \cite{513_goes2025key, 674_burstinghaus2020post, 392_stelzer2025extended, 69_ma2024lightweight}. These embedded implementations provide essential cryptographic primitives for applications, including embedded TLS stacks, such as OP-TEE, OpenTitan, and mbed TLS. The development of additional hardware, such as a root of trust based on PQC, is also an active area of investigation. 
Another relevant application of PQC in software security is the adoption of quantum-resistant algorithms within TPM environments \cite{paul2021tpm}. These can be used to create and manage the keys for device authentication, for example. 

Unmanned aerial vehicles (UAVs), commonly known as drones, enhance IoT by providing aerial capabilities. Within the UAV sub-domain, current research focuses on integrating PQC and QKD to secure communication links between aerial vehicles and their ground control stations \cite{epure2025post}. A review on PQC-based security for UAVs can be found in \cite{khan2024future}. These security techniques enable a range of applications, including agricultural monitoring systems \cite{675_minton2025post, 179_bakyt2025application} and surveillance pipelines, which protect video and image data transmitted between sensor devices and processing elements \cite{118_nguyen2023video}.

Nevertheless, another line of research in the mobile and consumer electronics sector evaluates PQC and quantum technologies of QKD and QRNG-assisted applications for smartphones and consumer devices \cite{655_kb2025post, 1042_alibrahim2025unveiling, 618_seyhan2024password}.

\begin{ieeeinsight}{Takeaway Message:} The use of quantum cryptography in edge/IoT differs from what is observed in telecom. Given the high costs of quantum cryptography hardware, this observation is reasonable, once edge/IoT currently includes more accessible mechanisms. Keeping only PQC at the edge is observed more frequently, while the performance bottlenecks of PQC are the subject of research.  
\end{ieeeinsight}

\begin{figure}[t!]
\centering
\includegraphics[width=0.5\textwidth]{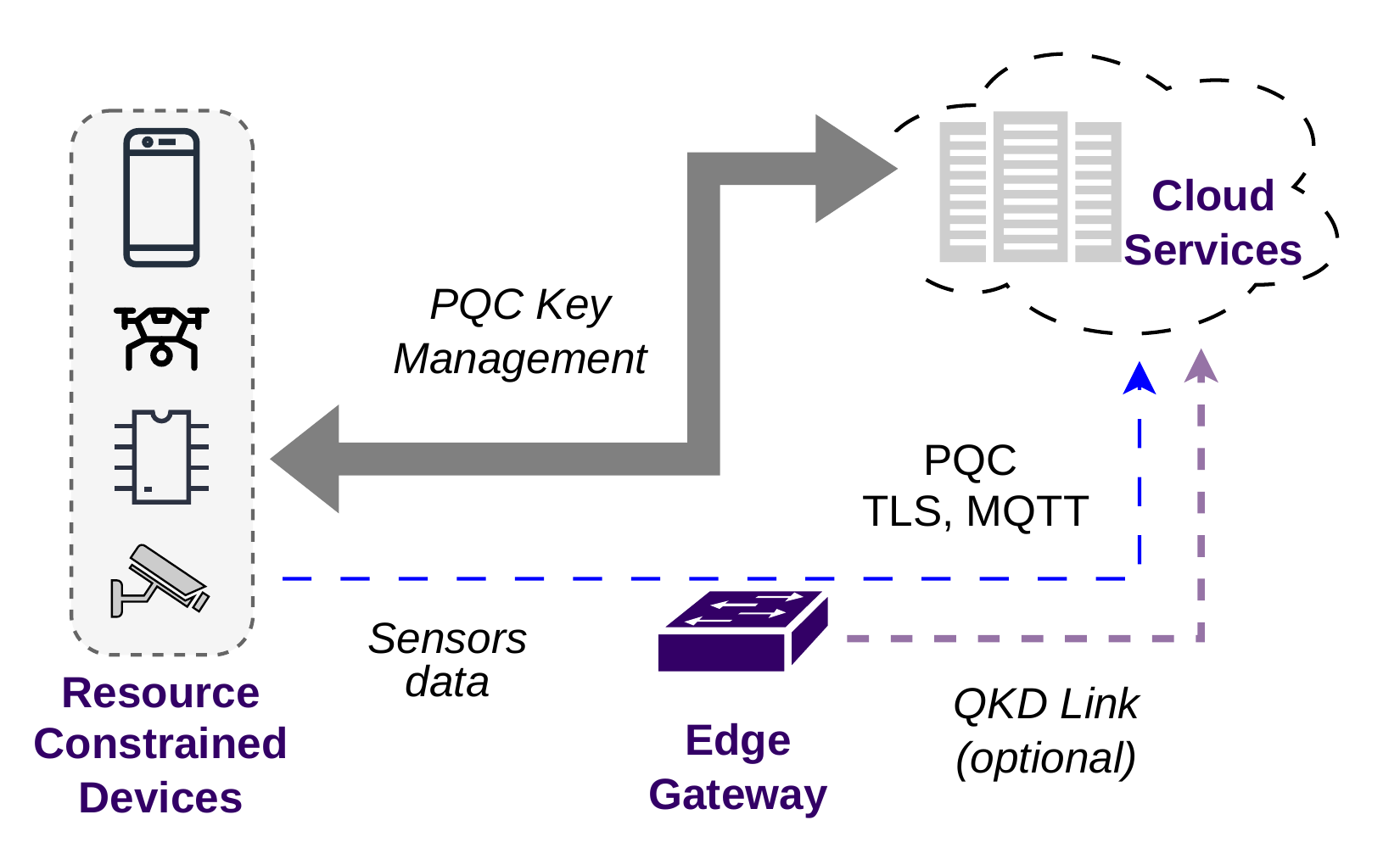}

\caption{IoT and Edge security architecture integrating PQC and optional QKD. Constrained IoT devices employ PQC key-encapsulation mechanisms and digital signatures to protect sensor data delivered to an Edge Gateway, which forwards requests to cloud services using PQC-enabled TLS or PQC-aware MQTT. A dedicated PQC key-management channel provides classical post-quantum key establishment between the gateway and the cloud, while an optional QKD link (bottom dashed path) provides quantum-secure symmetric keys.}
\label{fig:qs_iot}
\end{figure}

\textbf{Datacenter and Cloud.} Datacenter and cloud computing environments are central to modern digital services, concentrating vast amounts of data and computational resources, which makes long-term confidentiality and integrity essential concerns. The domain of datacenters and cloud computing encompasses compute, storage, and transport across cloud networks, providing end-to-end protection for cloud workflows, including secure data storage and transmission from a privacy-preserving point of view \cite{1017_dhinakaran2024towards, 322_swetha2025elevating}. The application of QKD for such end-to-end deployments provides dedicated links between client devices and cloud services. Providing authentication by using a public-key infrastructure (PKI) is also a subject of research in this area \cite{96_tsili2025scalable}. One example is the framework OAuth 2.0, which was adapted to PQC, but with a latency penalty for the communication handshake 
\cite{666_schardong2022post}. Several efforts in this area focus on optimizing or accelerating PQC algorithms and frameworks for deployment. Examples of these include optimizing lattice-based algorithms on data processing units (or SmartNICs) \cite{599_bernstein2022opensslntru} for communication between the cloud and the edge. Other optimizations include using more robust instruction set architectures (ISAs) in modern processors to improve the handshake throughput of key encapsulation mechanisms. 

Further applications include data protection across the entire file system, a database, or a hard disk \cite{redhat_pqc}. However, hard disk or cloud encryption is typically done with symmetric cryptography because it is more efficient. In this case, asymmetric cryptography, the focus of PQC, is used to protect the symmetric key and authenticate users. On the other hand, we have found efforts to employ PQC and QKD to provide cryptographic guarantees for data in transit. In addition, managing data storage also requires preserving security properties. Preserving security during data migration, a common task for cloud services, is important. Similarly, ensuring the authenticity and integrity of stored data is fundamental. This can take the form, for example, of using digital signatures to authenticate the file system \cite{722_anand2024qcrypt}. 

\begin{ieeeinsight}{Takeaway Message:}
Because quantum-safe techniques are expensive in terms of performance, datacenters often have to redesign their infrastructure to consider ways to avoid the latency and throughput penalties. This includes the adoption of SmartNICs, Data Processing Units (DPUs), and specialized accelerators to handle the cryptographic overhead. 
\end{ieeeinsight}

\textbf{Distributed Systems and Blockchains. } Decentralized trust systems encounter significant challenges during the transition to quantum-safe cryptography. In such systems, a distributed network of nodes manages trust functions, including identification and transaction verification, instead of relying on a single centralized authority.

\begin{figure}[t!]
    \centering
    \includegraphics[width=0.45\textwidth]{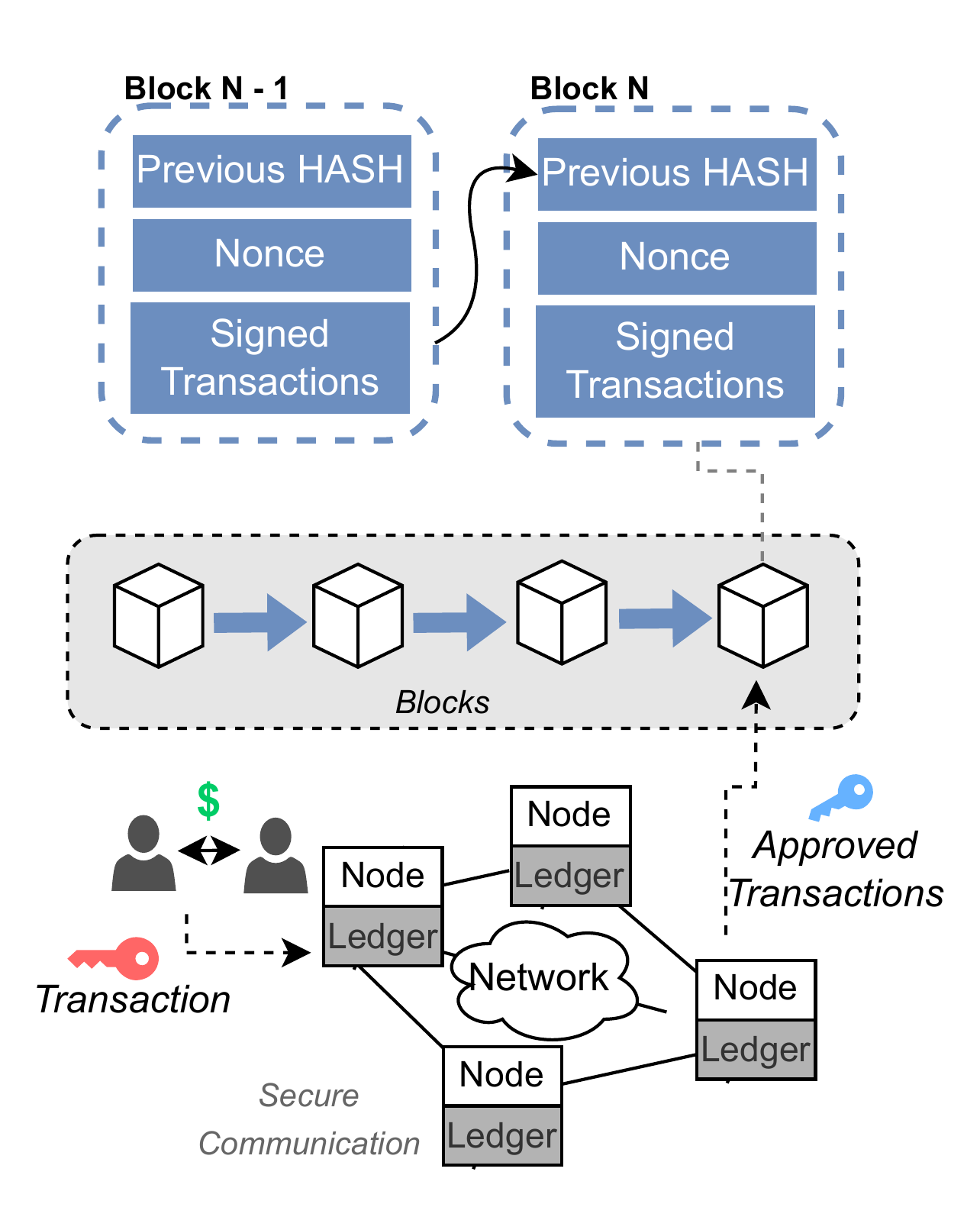}
    \caption{PQC on blockchain systems: Transactions are signed using quantum-safe digital signatures and go through a broadcast using the internet infrastructure. The security of the blockchain relies on the security of the transaction rather than on the encrypted channel. The signed transactions are verified by the blockchain nodes and stored in the blocks.}
    \label{fig:blockchain}
\end{figure}

Blockchains are a prime example of these decentralized systems. Figure \ref{fig:blockchain} illustrates the application of PQC within blockchain environments. Blockchain and distributed ledger technologies (DLT) utilize decentralized data structures that replicate transactions across multiple nodes through consensus mechanisms, thereby ensuring data integrity and traceability.

Blockchains rely on digital signatures to sign and verify transactions between users, and conventional blockchain networks are becoming vulnerable in a post-quantum world \cite{allende2023quantum}. Although blockchains rely on secure communication for the transmission and verification of transactions, transactions are also protected at the application level, requiring the two parties to verify each other through application-specific functionality. Decentralized authentication allows nodes to prove their identity without a centralized trusted service \cite{dutto2022toward}. To reduce vulnerabilities associated with digital signatures, hybrid solutions that integrate quantum-safe primitives such as QKD, PQC, and QRNG are being adopted for authentication, key management, and confidentiality \cite{reddy2025quantum}. Within the Consensus and Security Mechanisms domain, a QKD-assisted Byzantine fault-tolerant (BFT) framework employs hash-based signatures and optimizes the practical Byzantine fault tolerance (PBFT) messaging protocol. In contrast, Hyperledger Fabric supports hybrid X.509 credentials via a modified tool that embeds PQC to reduce the performance impact \cite{905_anand2025secure, 850_abbas2024quantum}.

The use of blockchains for identity is an alternative that also relies on digital signatures for secure identification. Within the identity and self-sovereign identity (SSI) context, existing efforts implement pure and hybrid PQC signatures for decentralized identifiers, verifiable credentials, and verifiable presentations. These efforts can be applied to enhance O-RAN identity management using PQC-based authentication \cite{1030_solavagione2025transition, 647_zeydan2024post}. Regarding Blockchain architecture and PQC for Internet of Things (IoT), Falcon-512, Dilithium, and XMSS are evaluated on embedded systems, employing public-key recovery to reduce transaction size by 17 percent and to assess smart contract latency \cite{1027_marchsreiter2025towards}.

There are also proposals for quantum-resistant blockchain solutions using QKD, QDS, and QRNG \cite{sharma2023securing, weng2023beating, IDQ2018}, as well as upcoming startups such as Quantum Blockchains Inc. in Poland \cite{QBinc}. In exchange/decentralized systems, current efforts outline tentative architectures and simple proofs of concept, aiming to integrate different technologies, such as Blockchains, into a 6G scenario \cite{zeydan2024enhanced}. QKD-enabled fair exchange and the QBEEP model enable secure cross-ledger asset transfer with PQC and homomorphic encryption \cite{39_prajapat2024blockchain, 744_joshi2023quantum}. Recently, there was a proposal for a QSDC-based blockchain for identity verification, message encryption, and consensus, where an optical network-based architecture using only present-day technology was also proposed with simulation results for different network topologies \cite{sun2025quantum}. However, the quantum network is still simulated, and the signing procedure depends on classical hash-based signatures. 

\begin{ieeeinsight}{Takeaway Message:} Different from PQC, the operation of QKD in Blockchain still occurs in the communication domain, thus not contributing specifically to the blockchain system itself. QKD is not responsible for signing transactions; instead, it provides secure communication primitives for nodes. In other words, although QKD can provide security guarantees during transport, it cannot authenticate transactions, so another mechanism must be used. It should be noted that there have been proposals to use QDS and pre-distributed quantum keys in quantum Byzantine agreement (QBA) frameworks, as well as a proof-of-principle experimental implementation in a star-topology network of parties. However, the solution is far from scalable while providing improved security and being fully decentralized.
\end{ieeeinsight} 

\textbf{Knowledge and Software Distribution.} In addition to blockchain, systems for the distribution of learning and for software distribution are of fundamental interest. Federated learning enables the decentralized training of artificial intelligence models without exposing training data. PQC can be integrated into federated learning at both the communication and application levels \cite{li2024enhancing}. This integration may involve secure aggregation mechanisms at the application level that utilize PQC \cite{zhang2025efficient}, multi-party computation, or differential privacy. 

Secure distribution of software across geographically distributed systems employs authentication. This is a fundamental requirement in this context, particularly for modern computer systems that rely on firmware and operating system updates. During the update process, it is essential to verify that new code originates from a trusted entity \cite{banegas2022quantum}. Early efforts to authenticate and ensure source code integrity, such as those undertaken by Red Hat, have incorporated PQC into their products \cite{redhat_pqc}. However, significant challenges persist, especially since many widely used programs are open source and lack valid certificates.

Research also addresses ways to ensure that the software distributed and loaded during boot is trustworthy \cite{wagner2022or}. To this end, the public key of the secure software is stored in the computer hardware, and the private key is kept with the source code. The verification is performed during boot, allowing the signature of the code to be verified, ensuring the code was generated by a trusted maintainer. \\

\begin{figure}[h!]
    \centering
    \includegraphics[width=0.45\textwidth]{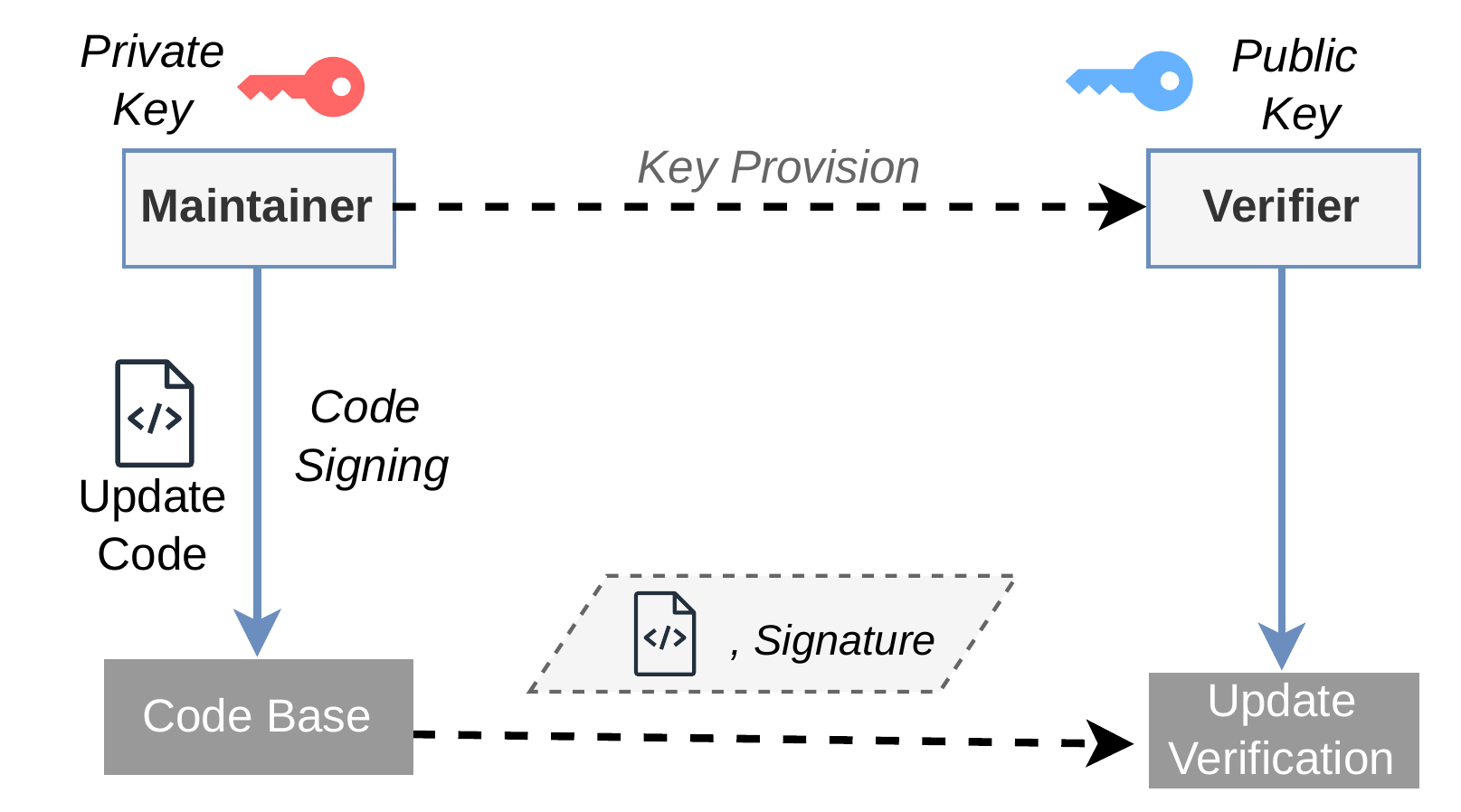}
    \caption{Secure code update example: a maintainer keeps a private key secret and uses it to sign the source code to be updated. The updated source code is sent to the verifier along with a signature, which the verifier can verify with the public key.}
    \label{fig:codeupdate}
\end{figure}

Figure \ref{fig:codeupdate} presents an example of the usage of public cryptography for code signing. In this example, a maintainer signs updates, while a verifier can later verify whether the code was created by an authentic maintainer and whether the update is intact. A similar process can be used to verify IoT devices. In that case, the device is the keeper of private keys, and verifiers are the ones (e.g., brokers) checking the integrity of the device. While code update is a viable candidate for PQC migration, it is not directly applicable for QKD, which does not provide a public key mechanism for providing authentication and integrity checks. 

\begin{ieeeinsight}{Takeaway Message:} Quantum communication, however, was not identified in our review as a current target for migration for authenticating hardware or software. The lack of quantum cryptography in these application domains occurs because authentication is often made by using digital signatures, and the quantum counterpart of digital signatures is not yet in an advanced state.
\end{ieeeinsight}

\textbf{Satellite.} Space communications and satellite systems are characterized by long operational lifetimes, limited physical access, and strong dependence on secure command and telemetry channels. Satellite-based quantum communication represents a critical enabler for global quantum-safe networks, integrating QKD links across space–ground segments to extend secure key exchange beyond fiber limitations. Satellite-based QKD schemes typically employ the sender on the satellite and the receiver on the ground, where prepare-and-measure QKD protocols using a weak coherent pulse (WCPs) sources are more convenient. Studies have been conducted on the design and evaluation of optical ground stations (OGS) for low-earth-orbit (LEO) downlink configurations, defining performance trade-offs between free-space and fiber-coupled architectures and optimizing secure key throughput and user scalability \cite{369_hausler2023evaluation}. 

Notable experimental results with the Chinese-launched satellite Micius have demonstrated satellite-to-ground QKD, entanglement distribution, ground-to-satellite quantum teleportation, and a satellite-relayed intercontinental quantum network over record distances \cite{liao2017satellite, yin2017satellite, ren2017ground, liao2018satellite}, among others. The experiment of an integrated space-to-ground quantum communication network that combined fiber networks with satellite-to-ground links demonstrated the feasibility of large-scale hybrid quantum networks \cite{chen2021integrated}. Moreover, a recent experiment on microsatellite-based real-time QKD using a 23-kilogram payload satellite and a portable OGS has taken a further step towards a global-scale quantum network \cite{li2025microsatellite}. Moreover, European satellite missions of EAGLE-1, QUBE, QuNET, and QoGloQuaN with different satellite and OGS architectures for different QKD protocols and quantum communication networks are also in progress \cite{hausler2025satellite}.

The integration of quantum sensors with key protocols enhances energy monitoring and secure telemetry in space environments through entanglement-based protection of data integrity \cite{837_rana2025quantum}. In application \& development of QKD-based communication, experimental verifications of satellite QKD feasibility, integration of QKD with terrestrial networks, and ongoing standardization for quantum secure communication collectively indicate the maturation of space–ground hybrid architectures toward operational quantum key infrastructures \cite{176_lai2023application}. 

It is also worth noting that there are preliminary studies on PQC implementations for satellite communication networks \cite{xu2024preliminary}, as well as on combining hybrid approaches with asymmetric and symmetric-key techniques \cite{232_bhatt2024comparative}. This includes the usage of PQC algorithms for system update and secure boot.

\begin{ieeeinsight}{Takeaway Message:} 
Remarkable progress in satellite-based quantum communication has been observed recently, ranging from feasibility studies and theoretical modeling to in-orbit demonstrations across various missions. Several milestones were achieved in terms of the quantum protocols, optical link configurations, Earth orbits, satellite dimensions, link budgets, cost-effective solutions, daytime operations, etc. 
\end{ieeeinsight}

\subsection{Critical Application Sectors}
\label{subsec:sectors}

In this section, we present the literature related to applications in critical sectors of society. We made the distinction between the applications in the infrastructure because many applications in this section make use of applications discussed earlier.

\textbf{Health.} Digital health systems and connected medical environments involve highly sensitive personal data and life-critical services, operating under regulatory frameworks that demand sustained confidentiality and availability. The Health domain covers quantum-safe approaches for connected healthcare systems. This domain overlaps with others, addressing, for example, secure communication and storage, as well as clinical needs such as clinical imaging, diagnostics, and telesurgical environments. 

One central sub-domain is the Internet of Medical Things (IoMT), which extends IoT with medical capabilities. Quantum-safe mechanisms for IoMT facilitate secure data exchange among medical devices and gateways by leveraging QKD and PQC to protect biomedical data \cite{832_rana2025quantum, 906_ambika2023secure, 187_sarkar2025artificial, 826_dhinakaran2024quantum, 931_mohamed2024securing}. Another essential component of IoMT is the wireless body area network (WBAN), comprising multiple intelligent sensors distributed around the human body. A privacy-preserving quantum authentication scheme based on QKD was proposed for securely sharing data in WBANs \cite{prajapat2024privacy}. Similarly, real-time systems such as telesurgery systems can leverage PQC and QKD for protecting surgical control \cite{733_bhattacharya2023quant}. Despite the inclusion of architectural designs in these proposals, evaluation on real hardware remains a significant challenge, as most existing work is limited to simulation environments.

Further applications relate to the management of electronic health records (EHRs), which help preserve integrity and privacy. Examples of techniques employed include blockchain-based ledgers with consensus mechanisms for access control and hybrid cloud authentication to ensure data integrity and availability \cite{911_karthikeyan2023secure, 355_venkatesh2024enhancing, 661_roosan2025post, 205_agarwal2024blockchain, 446_yallamelli2025hybrid, 926_bala2025securing}. There are also efforts to share medical images through quantum cryptography, by either using multiple QKD protocols over multiple channels for QKD-secured transmission or a novel quantum image encryption algorithm \cite{556_mallick2025multi, 925_bharamappa2023securing}.

A few software/hardware implementations of quantum cryptography-based PPC solutions for medical data analysis exist. In \cite{815_santos2021quantum, santos2022private}, a MPC protocol based on QOT was implemented in a software testbed for private computation of evolutionary distances between genomic sequences, along with a performance comparison with classical-only solutions. The QOT and QKD protocols were not implemented and only simulated. In \cite{matos2025quantum}, another use case of anomaly detection from a genomic data pool, in addition to phylogenetic trees, was demonstrated in a testbed with commercial CV-QKD equipment, while in \cite{romero2025enabling} an integration of oblivious keys in a QKD SDN system was proposed. It should be noted that the security of QOT using a CV-QKD protocol has not been proven, unlike that of the DV-QKD-based protocol \cite{lemus2025performance}. Recent theoretical works propose methods for privacy-preserving dementia classification and identity verification using quantum-assisted federated learning, homomorphic encryption, and zero-knowledge proofs \cite{838_tanbhir2025quantum, 834_kale2025quantum, garcia2024experimental}. Another recent work proposed a quantum protocol and multiple optical implementation designs for privacy-preserving classification of patients' sensitive data \cite{sulimany2025quantum}.

\begin{ieeeinsight}{Takeaway Message:}
PQC and quantum cryptography-based solutions are critical in the health sector, given the highly sensitive nature of patients' health records. While PQC is easier to implement, particularly in IoMT applications, QKD-based security is used wherever the required infrastructure is available, e.g., a direct fiber connection between healthcare institutions. Moreover, PPC solutions are particularly relevant in the health sector, due to their potential to leverage sensitive data to generate valuable health-related insights without revealing private data.
\end{ieeeinsight}

\textbf{Industrial.} Industrial systems integrate legacy technologies with real-time operational constraints, making security failures capable of directly impacting physical processes. Applications in the industrial sector include supervisory control and data acquisition environments, industrial automation systems, machine-to-machine (M2M) communication, embedded and mechatronic devices, supply chain and enterprise resource planning (ERP) platforms \cite{581_lin2025nsea}, and energy networks. In these contexts, quantum-safe primitives must address the challenges posed by legacy devices, extended operational life cycles, real-time requirements, and the need for resilience.

To investigate the challenges arising from constrained environments, the literature presents benchmarks to assess the viability of PQC for deploying telemetry (e.g., TLS and MQTT) in industrial protocols \cite{191_rampazzo2023assessment}. Additionally, performance analysis of PQC in machine-to-machine (M2M) protocols has been conducted, including key establishment and digital signatures for standard protocols such as the Open Platform Communications Unified Architecture (OPC UA) \cite{1025_paul2022towards}. However, changing the protocol source code is not always possible. The presence of legacy devices and applications complicates the adoption of new cryptography standards. In the lack of control over the device or software code, a proxy layer offers a viable alternative to migrating industrial protocols (such as Modbus) to PQC \cite{716_fabiano2025proxy} while preserving legacy compatibility.

Beyond publish/subscribe and machine-to-machine interactions, the literature also explores QKD deployment across many domains. One example is found in the supply chain and ERP traceability, where the application of quantum-safe blockchains is enhanced by QKD \cite{724_zhang2024qhb} and architectures are evolving to include KEMs. Furthermore, critical applications are increasingly adopting hybrid models. Proposals in mechatronics \cite{764_maurya2024quantum} combine QKD and PQC to enable key derivation, while smart grids \cite{709_tang2020programmable, 345_sree2025enhanced, 329_tang2022enabling} and modular nuclear reactors leverage QKD to provide a robust defense for infrastructure control and telemetry data collection \cite{553_pantopoulou2022monitoring}.

\begin{ieeeinsight}{Takeaway Message:}
Industrial systems hardware often have a long life cycle and are hard to change. Consequently, migrating to quantum-safe approaches often requires hybrid approaches that support legacy systems, keeping them working and leading to the adoption of proxy layers. 
\end{ieeeinsight}

\textbf{Vehicular Networks.} Modern transportation systems increasingly rely on digital communication between vehicles, roadside infrastructure, cloud services, and internal electronic components. These communications support critical functions such as collision avoidance, navigation, remote diagnostics, and energy management in electric vehicles. Because vehicles are highly mobile, operate for many years, and often use embedded devices with limited computational capacity, they are especially vulnerable to long-term cryptographic threats, including those posed by future quantum computers. Vehicular communication environments include connections among vehicles, cloud services, and other vehicles, as well as internal vehicle networks and external systems such as electric charging stations, aviation links, and railway control systems. These environments require strong mechanisms for authentication, key distribution, and data protection. 

QKD and PQC have been proposed to address the needs of vehicular communication environments \cite{shim2021survey}. For example, trust frameworks combine QKD with distributed ledger technologies such as blockchain to ensure both secure key exchange and integrity verification, while lightweight authentication mechanisms allow secure communication across both local and wide-area vehicular networks \cite{731_awan2025qstmf, 881_Phil2025Revolutionizing, 868_cheng2024research}. Within the vehicle itself, internal communication buses connect sensors, control units, and actuators that manage braking, steering, and engine functions. PQC authentication schemes based on physical unclonable functions (PUFs) have been proposed to securely identify electronic control units and protect internal message exchange \cite{718_cultice2022puf}. In electric vehicle charging systems, post-quantum extensions to existing charging communication standards introduce cryptographic agility and hardware-based protection to secure vehicle credentials and payment processes \cite{828_kern2023quantumcharge}. Another work proposed integrating QKD into a vehicle-to-infrastructure (V2I) network for secure communication via software-defined networking \cite{stavdas2024quantum}, whereas \cite{sutradhar2024svqcp} proposed a new secure vehicular quantum communication protocol that does not incorporate V2V and V2I authentication. 

Towards the practical implementation of free-space QKD systems in a V2X setting, various feasibility analyses associated to real-world conditions and proof-of-concept demonstrations were performed, including the drone-based quantum-secure communications through QKD and entanglement-distribution \cite{liu2021optical, conrad2021drone, conrad2023drone, tian2024experimental} and the integration of free-space QKD systems in a V2I application to secure communications within the AirQKD project \cite{fowler2023practical}. Recently, the Kwiat group in Canada achieved multiple milestones, including detailed feasibility analysis and real-world experiments on a US highway between two moving vehicles using FSO QKD systems \cite{conrad2024vehicle, chen2026v2v}, as well as various drone- and vehicle-based demonstrations with SKR in the range of 1.6-20 kbps \cite{conrad2025drone}. More recently, another work on quantum communication for mobile systems demonstrated a compact and robust polarization-entangled photon source weighing 58 g and measuring 36.5 cubic centimeters \cite{wang2026compact}. In \cite{mishra2023development}, the authors proposed a new RLWE-based PQC key-agreement protocol tailored for autonomous vehicles. Beyond road vehicles, similar challenges arise in aviation and railway systems. In aviation, post-quantum authentication and key agreement protocols protect communication between aircraft and ground stations, especially during handovers between coverage zones \cite{669_khan2024post}. In railway environments, including high-speed trains, QKD over free-space optical links has been proposed for secure communication between high-speed mobile systems and base stations, with a quantitative performance analysis under varying visibility conditions \cite{1046_al2024using}.

On the PPC front, a lattice-based privacy-preserving authentication scheme was proposed for vehicular communication in a vehicular ad hoc network (VANET), along with a performance analysis through a theoretical study \cite{dharminder2020lcppa}. Another work combined PQC and ZKP to protect data privacy during verification \cite{495_luo2025intelligent}. A QOT-based MPC protocol for secure lane changes in a vehicular network was demonstrated via simulation \cite{rahmani2023quantum}.

\begin{ieeeinsight}{Takeaway Message:}
Considering advances in modern vehicular networks, particularly autonomous vehicles, PQC and quantum cryptography-based solutions are crucial to achieving quantum-resistant security in critical vehicular infrastructures. Recent experimental efforts, both in proof-of-concept and real-world implementations, demonstrate the growing interest towards the practical realization of quantum-secure technologies in vehicular infrastructures. 
\end{ieeeinsight}

\textbf{Government and Defense.} Governmental and defense environments appear frequently in the literature because of the criticality, longevity, and sensitivity of the information they process, often under strict regulatory and national security requirements. The government and defense domain comprises mission-critical networks and identity services operating under state and defense constraints, demanding quantum-safe keying, authentication, and policy enforcement. Quantum cryptography and communications-based security solutions in military applications are an emerging area of interest due to the long-term security guarantees \cite{krelina2021quantum}. Additionally, quantum sensing, imaging, and radars are also of growing interest despite their theoretical and experimental challenges \cite{karsa2024quantum}. We observe that military networks employ hierarchical multi-plane networks that integrate with a QKD plane under ETSI key standards over a distributed KMS. Real-world field deployments for Portuguese defense entities through the projects of DISCRETION and PTQCI established quantum-secure connections between public bodies and academia within the Lisbon metropolitan network \cite{913_brito2025secure, bastos2025discretion}. The work involved key feed IPsec-based cipher machines that enforce red-black segregation, and extend to SDR tactical radios; a CV-QKD link and TeraFlow SDN extensions in the field. Recent theoretical works useful for future military networks include QKD framework for military and strategic operations, exploring other quantum technologies than QKD, and simulation modeling as a first-order planning reference for a nationwide (Austria as a reference) QKD backbone network \cite{bas2025quantum,raubitzek2026towards, beevi2026quantum}.

The verification of digital IDs and electronic documents is also an important concern in the government domain, and exploring quantum-secure approaches for these tasks is highly relevant \cite{brendel2023post}. Recent efforts investigate how these systems can transition to post-quantum security \cite{A0037_pradel2020post}, ensuring not only authentication of the owner or confidentiality, but also verification of the document's integrity (i.e., that it was not changed or corrupted). The literature presents solutions that include transitions for machine-readable travel documents (MRTDs) and electronic machine-readable travel documents (eMRTDs) for the electronic variant. Examples  \cite{A0037_pradel2020post} present solutions to integrate PQC into both electronic devices and the certificate authority infrastructure. 
A quantum protocol for anonymous voting was proposed using two kinds of entangled states, with a feature of self-tallying \cite{wang2016self}. A generalization to execute tasks of anonymous broadcast and anonymous ranking was also proposed. Another quantum protocol for anonymous voting using primitive quantum protocols was proposed along with simulated experiments \cite{fang2023cloud}. Another practical quantum anonymous voting protocol was proposed based on quantum homomorphic encryption \cite{chen2023practical}. Recently, a complete communication stack implementation of a QOT-based use case for PPC related to defense, where privacy-preserving fingerprint matching against Interpol and the United Nations no-fly lists was demonstrated \cite{ramos2025secure}.

In autonomous systems, a controlled key-distribution and command-authentication scheme for autonomous platforms uses CRYSTALS-Dilithium reinforced by device PUF fingerprints, with latency and bit-error measurements on prototypes \cite{652_partridge2023post}. Together, these works advance deployable quantum-safe capabilities across defense networking, border control, and autonomous systems.

\begin{ieeeinsight}{Takeaway Message:}
Defense applications would typically require higher TRL solutions with environment/scenario and timeline-specific risks of realization. While quantum technologies have the potential to offer significant advances, their exaggerated expectations must be managed, e.g., in quantum radars. A couple of real-world implementations for small military networks were demonstrated, with theoretical efforts to extend them to nationwide networks. For other government applications related to identity verification and anonymous voting, most work remains at the proof-of-concept experimental level, if not purely conceptual.
\end{ieeeinsight}

\textbf{Financial.} Financial systems are included as a distinct domain due to their central role in economic stability, their reliance on long-term data confidentiality, and their exposure to explicit regulatory timelines for adopting security mechanisms resistant to quantum-enabled attacks, which are expected to precede those of most other sectors.

Many recent works have reviewed different PQC-based solutions and their readiness level in the financial sector \cite{kumar2025post, janvciute2025cybersecurity, muzenda2026post}. A recent work proposes a PQC-secured payment platform architecture for cross-border payment systems \cite{mansur2025quantum}. In the banking sector, different PQC-based protocols were implemented in a real-world environment for a comparison \cite{bettale2022post}. Regarding institutional interests, a bank in France also did a trial with PQC secured data exchange \cite{bank}, as well as a joint experiment with the Deutsche Bundesbank and the Bank for International Settlements’ Innovation Hub Eurosystem Centre with NIST selected algorithms \cite{bank_joint}. On the privacy-preserving frontier, the Bank of Canada used a PQC-based zero-knowledge proof of knowledge for securing users' PII credentials in digital transactions \cite{kazmi2022privacy}. The Central Bank of Brazil conducted a feasibility study on applying PQC methods to the Brazilian instant payment system (PIX), including tests and benchmarking \cite{neto2022post}.

Given the relevance of digital platforms to businesses, quantum cryptography-based solutions for online banking, e-commerce, and digital transactions are attractive for the superior security guarantees they offer compared to PQC solutions. Even though these solutions face implementation challenges, they remain interesting from both a research and an implementation point of view \cite{tiberi2023quantum}. A QDS scheme based on MDI-QKD was recently demonstrated for a two-user quantum e-commerce scenario with an untrusted third party \cite{cao2024experimental}. Another recent experimental implementation demonstrated the use of optical quantum memories to implement Weisner's unforgeable quantum money protocol for the scenario of electronic payment \cite{mamann2025quantum}. Another approach to unforgeable quantum money for digital payments was proposed, which did not require quantum storage or trusted third parties \cite{schiansky2023demonstration}.

A quantum protocol for electronic negotiation to facilitate business through electronic trading was proposed \cite{liu2019quantum}. A quantum sealed-bid auction protocol was proposed that does not require a trusted third party to protect the privacy of all non-winning bidders \cite{shi2021quantum}. A recent experimental study demonstrated a QOT-based secure MPC implementation for the private set intersection problem to identify suspicious bank accounts without disclosing any other data \cite{zhang2026experimental}.

\begin{ieeeinsight}{Takeaway Message:}
Despite the computational advantage of quantum computing in the financial sector in terms of portfolio optimization, fraud detection, risk analysis, etc., the severity of its cybersecurity implications cannot be neglected. Both PQC and quantum cryptography-based solutions have been explored in theory and in practice to strengthen the security of communications in e-commerce and digital transactions. Moreover, various PPC-based use cases have found their application in this sector.
\end{ieeeinsight}

\subsection{Summary and Taxonomy}

\providecommand{\maturitycite}[2]{%
  \ifcase#1\cite{#2}%
  \or\textcolor{MaturityConceptual}{\cite{#2}}%
  \or\textcolor{MaturityPoC}{\cite{#2}}%
  \or\textcolor{MaturityReal}{\cite{#2}}%
  \else\cite{#2}%
  \fi%
}

\begin{table*}[ht!]
\centering
\scriptsize
\setlength{\tabcolsep}{2pt}
\renewcommand{\arraystretch}{1.08}
\caption{Deployment domains and technology categories identified in the reviewed literature.}
\label{tab:domains_consolidated}
\begin{tabularx}{\textwidth}{
p{0.20\textwidth}
>{\raggedright\arraybackslash}X
>{\raggedright\arraybackslash}X
>{\raggedright\arraybackslash}X
>{\raggedright\arraybackslash}X
>{\raggedright\arraybackslash}X
}
\toprule
\textbf{Subdomain} & \textbf{PQC} & \textbf{QKD} & \makecell{\textbf{Hybrid protocols}} & \textbf{Quantum Networking} & \textbf{Other Protocols} \\
\midrule
\rowcolor[HTML]{E3F2FD} \multicolumn{6}{l}{\textbf{Telecom/5G}} \\
SDN/QKDN Orchestration &  &  & \maturitycite{3}{453_buruaga2025hybrid,900_mendez2024sdn} & \makecell[l]{\maturitycite{1}{cao2022software},\maturitycite{2}{335_wang2020q-ROADM},\\\maturitycite{3}{419_giorgetti2025generalized,paolini2023integrating}} &  \\
5G/6G & \maturitycite{2}{148_electronics13214258} & \makecell[l]{\maturitycite{1}{726_ntanos2020qkd,393_wang2025facing,594_zavitsanos2020qkd},\\\maturitycite{1}{251_gao2024co},\maturitycite{2}{433_milovanvcev2021high}} &  &  &  \\
Network Protocols and Services & \maturitycite{2}{668_jasim2025post} & \makecell[l]{\maturitycite{1}{452_ramu2025hybrid,945_chen2024security},\\\maturitycite{2}{158_prevost2025etsi,459_bushuev2021immunity,522_lyssenko2023leveraging}} & \makecell[l]{\maturitycite{1}{540_dowling2020many_book},\\\maturitycite{2}{489_shokrivahed2024integration,blanco2025qkd,garcia2025enhanced},\\\maturitycite{2}{lauterbach2025post}} &  &  \\
Optical Transport &  & \maturitycite{1}{brunner2023cv} &  & \maturitycite{3}{398_zavitsanos2023feasibility,491_gatto2022integration,818_gatto2021quantum} & \maturitycite{1}{zhou2020measurement},\maturitycite{2}{brazaola2026privacy},\maturitycite{3}{qi202115} \\
Mobile Network Architecture & \maturitycite{1}{647_zeydan2024post} & \maturitycite{1}{zeydan2024enhanced} & \maturitycite{1}{384_hoque2024exploring} & \maturitycite{1}{830_kumar2025quantum} &  \\
Network Management &  &  & \maturitycite{3}{659_doring2022post_book} & \makecell[l]{\maturitycite{1}{88_ghourab2025qkdZero-trust,mendez2026switching},\maturitycite{2}{593_oliveira2024integration},\\\maturitycite{3}{lopez2025enhanced,martin2024madqci}} &  \\
\addlinespace[2pt]
\rowcolor[HTML]{E3F2FD} \multicolumn{6}{l}{\textbf{IoT/Edge}} \\
Embedded Systems & \makecell[l]{\maturitycite{2}{1027_marchsreiter2025towards,392_stelzer2025extended,69_ma2024lightweight,tasopoulos2023energy},\\ \maturitycite{2}{607_dong2025optimizing,61_eckel2023generic,645_andrade2024post,kim2025optimized},\maturitycite{3}{674_burstinghaus2020post} \maturitycite{2}{paul2021tpm}} &  &  &  &  \\
IoT Frameworks & \makecell[l]{\maturitycite{1}{146_elkhodr2025ai},\\\maturitycite{2}{608_ismail2025optimizing,83_network5020020,paul2021tpm}} & \maturitycite{1}{202_shradha2024beyond,357_biswas2024enhancing} & \maturitycite{1}{389_biswas2024exploring,831_mujlid2025quantum} &  &  \\
Video/Image Security & \maturitycite{2}{118_nguyen2023video} &  &  &  &  \\
Consumer Devices & \maturitycite{1}{660_sharma2025post},\maturitycite{2}{513_goes2025key,618_seyhan2024password} &  & \maturitycite{2}{655_kb2025post} &  & \maturitycite{3}{1042_alibrahim2025unveiling} \\
UAV/UAS & \maturitycite{2}{epure2025post} & \maturitycite{1}{179_bakyt2025application,675_minton2025post} &  &  &  \\
Edge Gateways &  &  & \maturitycite{3}{848_cibik2025quantum} &  &  \\
\addlinespace[2pt]
\rowcolor[HTML]{E3F2FD} \multicolumn{6}{l}{\textbf{Datacenter/Cloud}} \\
P2P/File Sharing & \maturitycite{2}{506_hou2020iota} &  &  &  &  \\
Multi-Cloud Orchestration & \maturitycite{1}{648_zeydan2022post} &  &  &  &  \\
PKI and Authentication & \maturitycite{1}{96_tsili2025scalable} &  &  &  &  \\
Network Protocols and Services & \maturitycite{2}{599_bernstein2022opensslntru} & \maturitycite{2}{730_akshaya2025qsms} & \maturitycite{3}{847_doberl2023quantum} &  &  \\
Cloud Infrastructure & \maturitycite{2}{481_cano2024integrating} &  &  &  & \maturitycite{1}{1017_dhinakaran2024towards,322_swetha2025elevating} \\
Data at Rest & \maturitycite{2}{722_anand2024qcrypt} &  & \maturitycite{1}{914_alnaser2025secure} &  &  \\
\addlinespace[2pt]
\rowcolor[HTML]{E3F2FD} \multicolumn{6}{l}{\textbf{Blockchain/DLT}} \\
Decentralized Exchange &  & \maturitycite{1}{39_prajapat2024blockchain} & \maturitycite{1}{744_joshi2023quantum} &  & \maturitycite{1}{IDQ2018},\maturitycite{2}{weng2023beating} \\

Blockchain Architecture & \maturitycite{2}{850_abbas2024quantum} &  & \maturitycite{2}{reddy2025quantum} &  & \maturitycite{1}{dutto2022toward,sun2025quantum} \\
Consensus &  & \maturitycite{1}{905_anand2025secure} &  &  &  \\
Self-Sovereign Identity & \maturitycite{2}{1030_solavagione2025transition} &  &  &  &  \\
\addlinespace[2pt]
\rowcolor[HTML]{E3F2FD} \multicolumn{6}{l}{\textbf{Knowledge/Software Distribution}} \\
Federated Learning  & \maturitycite{2}{li2024enhancing, zhang2025efficient} &  &  &  & \\
Software Update &  \maturitycite{2}{banegas2022quantum, wagner2022or} &  &  &  & \\
\addlinespace[2pt]
\rowcolor[HTML]{E3F2FD} \multicolumn{6}{l}{\textbf{Health}} \\
Diagnostics &  &  &  &  & \makecell[l]{\maturitycite{1}{815_santos2021quantum,838_tanbhir2025quantum,romero2025enabling},\\\maturitycite{1}{santos2022private,sulimany2025quantum},\maturitycite{3}{matos2025quantum}} \\
Telesurgery &  & \maturitycite{1}{733_bhattacharya2023quant} &  &  &  \\
IoMT & \maturitycite{1}{911_karthikeyan2023secure},\maturitycite{2}{187_sarkar2025artificial,931_mohamed2024securing} & \maturitycite{1}{826_dhinakaran2024quantum,832_rana2025quantum,906_ambika2023secure} &  &  & \maturitycite{1}{prajapat2024privacy} \\
Medical Imaging &  & \maturitycite{1}{556_mallick2025multi,925_bharamappa2023securing} &  &  &  \\
EHR &  & \maturitycite{1}{926_bala2025securing} & \makecell[l]{\maturitycite{1}{205_agarwal2024blockchain,355_venkatesh2024enhancing,446_yallamelli2025hybrid}\\\maturitycite{1}{661_roosan2025post}} &  &  \\
Identity Verification &  &  &  &  & \maturitycite{1}{834_kale2025quantum},\maturitycite{3}{garcia2024experimental} \\
\addlinespace[2pt]
\rowcolor[HTML]{E3F2FD} \multicolumn{6}{l}{\textbf{SCADA/Industrial}} \\
Microgrid &  &  &  & \maturitycite{2}{329_tang2022enabling,709_tang2020programmable} &  \\
Modbus/OPC UA & \maturitycite{2}{1025_paul2022towards,716_fabiano2025proxy} &  &  &  &  \\
Industrial M2M & \maturitycite{2}{191_rampazzo2023assessment} &  &  &  &  \\
Critical Infrastructure &  & \maturitycite{1}{553_pantopoulou2022monitoring} & \maturitycite{1}{764_maurya2024quantum} &  &  \\
Smart Grid &  & \maturitycite{1}{345_sree2025enhanced} &  &  &  \\
Supply Chain/ERP & \maturitycite{2}{581_lin2025nsea} & \maturitycite{1}{724_zhang2024qhb} &  &  &  \\
\addlinespace[2pt]
\rowcolor[HTML]{E3F2FD} \multicolumn{6}{l}{\textbf{Vehicular/Transportation}} \\
In-Vehicle/CAN Bus & \maturitycite{1}{718_cultice2022puf} &  &  &  &  \\
V2X/IoV & \maturitycite{1}{mishra2023development} & \makecell[l]{\maturitycite{1}{sutradhar2024svqcp,731_awan2025qstmf,868_cheng2024research,881_Phil2025Revolutionizing},\\\maturitycite{1}{chen2026v2v},\maturitycite{2}{conrad2024vehicle,fowler2023practical}} &  & \maturitycite{1}{stavdas2024quantum} & \maturitycite{1}{495_luo2025intelligent,dharminder2020lcppa,rahmani2023quantum} \\
EV Charging & \maturitycite{2}{828_kern2023quantumcharge} &  &  &  &  \\
Aviation and Railway & \maturitycite{2}{669_khan2024post} & \maturitycite{1}{1046_al2024using} &  &  &  \\
UAVs &  & \makecell[l]{\maturitycite{1}{conrad2021drone,conrad2023drone},\\\maturitycite{2}{tian2024experimental,wang2026compact},\maturitycite{3}{conrad2025drone}} &  &  &  \\
\addlinespace[2pt]
\rowcolor[HTML]{E3F2FD} \multicolumn{6}{l}{\textbf{Satellite/Space}} \\
Satellite-based communication & \maturitycite{1}{xu2024preliminary} & \makecell[l]{\maturitycite{1}{369_hausler2023evaluation},\maturitycite{3}{chen2021integrated,liao2018satellite}\\\maturitycite{3}{li2025microsatellite,ren2017ground,liao2017satellite,yin2017satellite}} &  &  &  \\
Sensing and Imaging & \maturitycite{3}{232_bhatt2024comparative} & \maturitycite{1}{837_rana2025quantum} &  &  &  \\
\addlinespace[2pt]
\rowcolor[HTML]{E3F2FD} \multicolumn{6}{l}{\textbf{Government/Defense}} \\
Identity and E-Documents & \maturitycite{1}{brendel2023post},\maturitycite{2}{A0037_pradel2020post} &  &  &  & \maturitycite{2}{ramos2025secure} \\
Autonomous Systems & \maturitycite{2}{652_partridge2023post} &  &  &  &  \\
Military Networks &  &  &  & \maturitycite{1}{bas2025quantum,raubitzek2026towards,beevi2026quantum},\maturitycite{3}{913_brito2025secure,bastos2025discretion} &  \\
Anonymous Voting/Broadcast &  &  &  &  & \maturitycite{1}{chen2023practical,fang2023cloud,wang2016self} \\
\addlinespace[2pt]
\rowcolor[HTML]{E3F2FD} \multicolumn{6}{l}{\textbf{Financial}} \\
E-commerce/Digital transactions & \makecell[l]{\maturitycite{1}{mansur2025quantum},\maturitycite{2}{bank},\\\maturitycite{2}{neto2022post},\maturitycite{3}{bettale2022post}} & \maturitycite{1}{bank_joint} &  &  & \makecell[l]{\maturitycite{1}{kazmi2022privacy,liu2019quantum,shi2021quantum}\\\maturitycite{2}{cao2024experimental,mamann2025quantum},\maturitycite{3}{schiansky2023demonstration}} \\
Fraud Detection &  &  &  &  & \maturitycite{2}{zhang2026experimental} \\
\bottomrule
\end{tabularx}
\vspace{2pt}
\footnotesize{Reference colors indicating the maturity levels:
\textcolor{MaturityConceptual}{M1 = Conceptual};
\textcolor{MaturityPoC}{M2 = Proof-of-concept};
\textcolor{MaturityReal}{M3 = Real-world implementation}.}
\end{table*}

Table \ref{tab:domains_consolidated} categorizes the literature discussed in this section and the corresponding migrated applications by the domains (or disciplines) in which quantum-resistant primitives are employed. We classify these domains as enabling technologies and infrastructures, and the sectors that use these infrastructures to provide end applications. In addition to covering the main enabling technologies of PQC, QKD, and its networking, and hybrid protocols, we also identified the usage of other protocols of QOT, QRNG, QDS, PPC, etc. It is important to note that some domains (e.g., IoT and vehicular systems) span both infrastructure and application layers, reflecting their dual role as both communication substrates and end-user systems. For the literature listed in the table, we also classify their maturity level in three categories, in the increasing order of technological advancement as: 1) Conceptual, 2) Proof-of-Concept implementation, and 3) Real-world implementation. The lowest level of conceptual work (depicted in blue) involves modeling and simulation, while the intermediate level of proof-of-concept work (depicted in orange) involves experimental implementations in a controlled laboratory environment. The most advanced level of real-world implementations (depicted in green) includes experiments, deployments, and prototyping.

\section{Research Directions}
\label{sec:challenges}

This section categorizes the challenges and corresponding research directions identified for both the migration to post-quantum cryptography and quantum cryptography. We focus on three axes: efficiency, deployment, and security challenges. 

\definecolor{Mercury}{rgb}{0.901,0.901,0.901}
\begin{table*}
\centering
\caption{Summary of challenges and mitigation strategies in PQC .}
\label{tab:systematic_challenges_pqc}
\begin{tblr}{
  width = \linewidth,
  colspec = {Q[90]Q[387]Q[463]},
  row{1} = {Mercury,font=\bfseries},
  cell{2}{1} = {r=3}{},
  cell{5}{1} = {r=3}{},
  cell{8}{1} = {r=3}{},
  hlines,
  vlines,
}
Category        & Specific Challenge                                                                                                                                                    & Mitigation / Research Direction                                                                                                                                                                                                                                                                                     \\ \hline
\textit{Efficiency}      
                & {Computational overhead and latency overheads,~\\particularly in resource-constrained and IoT devices.}          
                
                & {\labelitemi\hspace{\dimexpr\labelsep+0.5\tabcolsep}Protocol stack customization (integrating transport and crypto layers),~\\
                \labelitemi\hspace{\dimexpr\labelsep+0.5\tabcolsep}Replacing digital signatures with KEMs for authentication.}\\
                
                & {Large key and ciphertext sizes causing memory overload,~\\reduced throughput, and packet fragmentation.}
                
                & {\labelitemi\hspace{\dimexpr\labelsep+0.5\tabcolsep}New algorithmic strategies (algebraic simplifications, metric-based reductions),~\\
                \labelitemi\hspace{\dimexpr\labelsep+0.5\tabcolsep}Accelerated polynomial multiplication.}\\
                
                & {High energy consumption due to algorithm complexity,~\\posing a challenge for battery-powered devices such as smart cards.}
                
                & {\labelitemi\hspace{\dimexpr\labelsep+0.5\tabcolsep}Hardware accelerators (GPUs, SmartNICs, dedicated hardware),\\
                \labelitemi\hspace{\dimexpr\labelsep+0.5\tabcolsep}Optimized implementations for specific processors (e.g., ARM, RISC-V).}\\
                
\textit{Deployment}      
                & {Software/vendor incompatibility across versions and the high risk of cryptographic downgrade attacks to weaker legacy schemes.}
                
                & {\labelitemi\hspace{\dimexpr\labelsep+0.5\tabcolsep}Backward-compatible protocol engineering,\\
                \labelitemi\hspace{\dimexpr\labelsep+0.5\tabcolsep}Cross-vendor interoperability verification,\\
                \labelitemi\hspace{\dimexpr\labelsep+0.5\tabcolsep}Secure handshake designs.}\\
                
                & {Tedious and error-prone manual selection of algorithms\\due to complex resource trade-offs (e.g., memory vs. energy)\\~requiring low-level expertise.}
                
                & {\labelitemi\hspace{\dimexpr\labelsep+0.5\tabcolsep}Software abstractions,\\
                \labelitemi\hspace{\dimexpr\labelsep+0.5\tabcolsep}Modular code design (separating primitives from libraries/protocols),\\
                \labelitemi\hspace{\dimexpr\labelsep+0.5\tabcolsep}Automated migration APIs guided by PQC compliance policies.}\\
                
                & {Transitioning~ production systems to new standards without downtime, primitive lifecycle management amidst inconsistent standards, and workforce training gaps.}
                
                & {\labelitemi\hspace{\dimexpr\labelsep+0.5\tabcolsep}Continuous monitoring,\\
                \labelitemi\hspace{\dimexpr\labelsep+0.5\tabcolsep}Automated verification mechanisms for cryptographic compliance,\\
                \labelitemi\hspace{\dimexpr\labelsep+0.5\tabcolsep}Standardized lifecycle management frameworks.}\\
                
\textit{Vulnerabilities}
                & {Lack of formal theoretical proofs of hardness (e.g., NP-Hard), making PQC algorithms susceptible to future classical or quantum algorithmic breakthroughs.}
                
                & {\labelitemi\hspace{\dimexpr\labelsep+0.5\tabcolsep}Classical-hybrid (PQC + public-key) cryptography\\
                \hspace*                {0.42\leftmargin}\labelitemii\hspace{\dimexpr\labelsep+0.5\tabcolsep}~combining PQC with classical key-exchange (e.g., Kyber+ECDH),\\
                \hspace*{0.42\leftmargin}\labelitemii\hspace{\dimexpr\labelsep+0.5\tabcolsep}~hybrid certificates with dual signatures (e.g., Dilithium+ECDSA).}\\
                
                & {Potential compromise of PQC mechanisms by novel quantum algorithms \textit{and} the availability of a CRQC rendering classical-hybrid constructions vulnerable.}
                
                & {\labelitemi\hspace{\dimexpr\labelsep+0.5\tabcolsep}Deploying QKD alongside PQC as the ultimate backup for long-term information-theoretic security,\\
                \labelitemi\hspace{\dimexpr\labelsep+0.5\tabcolsep}Utilizing secure authentication schemes; pre-shared or hybrid.}\\
                
                & {Side-channel attacks (SCA) exploiting physical parameter~\\leakages (e.g., power consumption, cache access patterns).}
                
                & {\labelitemi\hspace{\dimexpr\labelsep+0.5\tabcolsep}Noise insertion techniques (masking to hide correlated data)~\\
                \labelitemi\hspace{\dimexpr\labelsep+0.5\tabcolsep}Introducing redundant operations to achieve runtime independence.}
\end{tblr}
\end{table*}

\subsection{PQC Challenges and Opportunities}

\subsubsection{Efficiency}
\label{subsec:pqc_eff}

PQC algorithms introduce computational overhead that can turn routine
operations, such as encryption and signing, into bottlenecks, particularly
during connection establishment. This challenge is most acute in
resource-constrained devices common to embedded systems and IoT networks,
where increased latency and high processing power consumption can degrade
overall application performance beyond just cryptographic operations.

Protocol stack customization offers one path to mitigating this overhead.
For instance, QUIC consolidates transport-layer and cryptographic handshakes
into a single exchange, avoiding the full round-trip cost of a traditional
TLS handshake over TCP. At the algorithm level, replacing digital signatures
with KEMs for authentication can further reduce computational
load~\cite{schwabe2021more}.

Memory and bandwidth consumption are additional concerns in PQC deployment.
Classical schemes such as RSA and ECC typically rely on compact key sizes; RSA commonly uses 2048-bit keys, whereas PQC schemes such as ML-KEM
may require between 1,000 and 6,000 bytes depending on the security level
and parameter set. These larger key and ciphertext sizes place greater
pressure on memory and reduce transmission efficiency, as they increase
packet fragmentation and lower throughput. The impact is particularly
pronounced in protocols that operate over UDP, where packets are already
more susceptible to fragmentation and loss, and where bandwidth constraints
are tighter.

The computational and memory demands of PQC algorithms also translate into
higher energy consumption~\cite{roma2021energy, tasopoulos2023energy},
posing a critical challenge for battery-powered and energy-constrained
devices such as smart cards and IoT sensors. Hardware accelerators offer a
promising mitigation path: offloading cryptographic operations to
energy-efficient, purpose-built hardware can substantially reduce power
consumption compared to software-only implementations. However, designing
accelerators that are both efficient and adaptable to future algorithmic
updates, given the evolving nature of PQC standardization, remain an open
challenge.

The choice of accelerator depends heavily on the application context. In
communication systems, SmartNICs equipped with dedicated cryptographic
modules can perform PQC operations at line rate~\cite{cano2024line,
lawo2024falcon, singh2025performance}. In 5G/6G networks, dedicated hardware can accelerate key
distribution and management at scale. GPU-based acceleration represents
another viable approach, particularly for parallel workloads~\cite{gupta2020pqc}. The full hardware implementation in reconfigurable platforms such as FPGAs \cite{xing2021compact} is also an interesting research direction and subject of constant investigation. 

Algorithmic optimization represents a complementary strategy to hardware
acceleration to improve PQC efficiency. Techniques such as algebraic
simplifications and metric-based reductions can decrease key sizes without
compromising security~\cite{loidreau2017new}, while energy-efficient algorithm designs must be rigorously verified to ensure both correctness and resistance to side-channel attacks. Crucially, these optimizations are often hardware-aware and tailored to the constraints and instruction sets
of the target platform. For example, extending the ISA of RISC-V processors and the underlying processor microarchitecture to accelerate PQC algorithms is explored in \cite{alkim2020isa}.

Several lines of work illustrate this direction. Kim et al.~\cite{kim2025optimized} optimize ML-KEM for constrained MQTT environments, demonstrating that
protocol-specific tuning can yield meaningful efficiency gains. Abdulrahman
et al.~\cite{abdulrahman2022faster} target ARM processors, exploiting
architecture-specific instructions to accelerate core ML-KEM operations.
At a finer granularity, Zeng et al.~\cite{zeng2024implementation} focus on
accelerating polynomial multiplication, a dominant cost in lattice-based
schemes within ML-KEM, while Huang et al.~\cite{huang2024yet} address
operation-level optimizations tailored to IoT devices. Together, these
hardware and algorithmic strategies form a multi-layered approach to making
PQC deployment practical across diverse and constrained environments.

\subsubsection{Deployment}
\label{subsec:pqc_dep}

Migrating to PQC algorithms involves non-trivial software adaptation that
presents challenges on multiple fronts. A key concern is backward
compatibility: many systems continue to run older software versions, creating
interoperability issues across protocol versions and vendor implementations.
This heterogeneity exposes a critical attack surface, and adversaries can
exploit version mismatches to force downgrade attacks, coercing endpoints
into using deprecated or weaker cryptographic schemes~\cite{ott2019identifying,
alnahawi2023state}.

Algorithm selection is another non-trivial aspect of PQC migration. Existing
PQC schemes inherently trade one resource for another, e.g., a scheme may be
memory-efficient but computationally expensive or fast in execution but
demanding in key storage. Selecting the right algorithm, therefore, requires
co-analysis of the target application, the underlying hardware platform, the
available cryptographic APIs, and the security and performance requirements
of the deployment context. 

Manual migration to PQC is inherently tedious and error-prone, requiring low-level cryptographic expertise that is rarely available at scale. Two complementary strategies can ease this burden: abstraction and automation.
Well-designed abstractions simplify the deployment by decoupling cryptographic
concerns from application logic. In particular, this means separating algorithm
primitives from library implementations, protocol bindings, and
application-level frameworks, exposing only stable, reusable interfaces that
can be integrated without deep cryptographic knowledge. This separation of
concerns reduces the risk of misuse and makes future algorithmic updates more manageable.

Automation complements abstraction by reducing the manual effort required
to migrate existing code-bases. Cryptographic agility APIs, for instance,
can abstract over code dependencies and runtime environments, enabling
systems to transition to new PQC schemes without modifying application
code directly. Instead, migration policy, including algorithm selection
and compliance with PQC standards, is expressed declaratively, shifting
the burden from developers to the underlying infrastructure. 

Beyond abstraction and automation, sustainable PQC adoption requires attention to lifecycle management, human factors, and operational continuity. Cryptographic primitives must be actively maintained throughout their lifecycle, from initial deployment to deprecation, ensuring that keys, certificates, and algorithm configurations remain valid and auditable over time.
Human factors present an equally significant barrier. Insufficient training among practitioners involved in PQC adoption can slow migration and increase the risk of misconfiguration. Inconsistent or ambiguous standards compound this challenge, requiring significant expertise to interpret and reconcile conflicting guidance across vendors, protocols, and deployment scenarios.
Operational continuity is an additional concern for systems already in production. Transitioning to new cryptographic standards must be achievable without system downtime, which requires careful planning of rollout strategies such as hybrid schemes or phased migrations. Equally important is the provision of monitoring and verification mechanisms that continuously validate whether migration has been implemented correctly and in compliance with organizational PQC policies.

Interoperability and backward compatibility are critical enablers of
successful PQC migration. Interoperability requires that implementations
from different vendors correctly implement the same cryptographic standards
so that heterogeneous systems can communicate without friction, e.g., a
browser client negotiating a PQC handshake with a third-party server. Correct and consistent implementation across
parties is therefore essential, as a single non-compliant endpoint can break the entire exchange.

Backward compatibility introduces an additional layer of complexity,
particularly in hardware-constrained environments. Unlike software, hardware deployed in the field is often under third-party control and cannot be updated or patched by the original developers. This limits the degree to which backward compatibility can be enforced and places the migration burden asymmetrically on the software stack. Nevertheless, where backward compatibility is achievable (through hybrid schemes, negotiation mechanisms, or versioned protocol extensions), it provides a practical bridge that allows systems to transition incrementally to PQC standards without requiring simultaneous updates across all endpoints. 

Thus, these deployment challenges demonstrate the need for coordinated efforts among standardization bodies, tool developers, and system operators to ensure a smooth and secure transition to PQC.


\subsubsection{Vulnerabilities}
\label{subsec:pqc_vuln}

Although current PQC algorithms are widely believed to be quantum-resistant, neither theoretical nor practical guarantees exist that they will remain so indefinitely. Existing schemes are conjectured to be computationally hard, but hardness proofs are absent. Specifically, there is no guarantee that the underlying problems are NP-hard. This leaves open the possibility that future quantum, or even classical, algorithms considered secure today could be vulnerable. Moreover, side-channel attacks pose an equally serious threat: rather than solving the underlying cryptographic problem directly, an adversary can recover secret key material by exploiting information leaked through physical or implementation-level observables, such as timing, power consumption, or electromagnetic emissions.

While conventional cryptography can already no longer be safely used, migrating entirely to PQC algorithms is also considered risky due to the possibility of unforeseen cryptanalytic advances against these relatively new algorithms. One conservative strategy is to combine conventional public-key cryptography with PQC algorithms. We refer to such a construction as classical-hybrid to distinguish it from the hybrid constructions we have mentioned so far that combine quantum cryptography with PQC. Using classical-hybrid construction for key exchange would, for example, be pairing elliptic-curve Diffie-Hellman (ECDH) key exchange with a PQC scheme, such as ML-KEM, to ensure that security is preserved even if one of the two components is compromised. If a future cryptographically relevant quantum computer (CRQC) compromises ECDH, the security is preserved by ML-KEM. Conversely, if a classical cryptanalytic breakthrough compromises ML-KEM before the advent of a CRQC, ECDH continues to provide security. The same can be adopted for signatures by using hybrid certificates with dual public keys and dual, one each from the PQC and public-key scheme. For instance, IETF is already drafting standards for such a hybrid authentication within existing Internet protocols \cite{compositeMLDSA}.
Such constructions provide a practical transitional safeguard, maintaining
security guarantees across both the classical and post-quantum threats
landscapes during the migration period.

On the other hand, the discovery of novel quantum
algorithms capable of breaking current PQC assumptions as well as the availability of a CRQC would simultaneously
undermine both PQC and hybrid PQC-classical constructions, since both
components would be vulnerable. Such attack threats need proactive and layered countermeasures. Therefore, scenarios requiring long-term security, information-theoretically secure alternatives of quantum cryptography would represent the only solution,
as their security rests on physical principles rather than computational
hardness assumptions. For example, a hybrid key-exchange scheme combining QKD with PQC would be the only solution to boost PQC security against future quantum attacks. However, authentication in QKD
protocols cannot rely solely on PQC either and must instead depend on hybrid signatures or
pre-shared keys and information-theoretically secure authentication schemes. This introduces practical
bootstrapping challenges around secure key distribution and scalability,
which remain open problems in the broader QKD deployment landscape.

Side-channel attacks pose a significant challenge for PQC implementations. Unlike invasive and semi-invasive approaches, non-invasive side-channel attacks infer sensitive information by observing physical parameters, e.g., power consumption, timing, and electromagnetic emissions, during system operation \cite{chowdhury2022physical}. Concrete examples include attacks that exploit metadata generated by cryptographic protocols, such as cache access patterns and energy consumption traces. Countermeasures tend to be attack-specific and therefore incomplete: power analysis attacks, for instance, can be mitigated by injecting dummy noise to obscure correlated data, while timing-based leakage can be reduced by introducing redundant operations that render runtime independent of secret information. However, applying one countermeasure rarely neutralizes the full spectrum of side-channel threats, leaving implementations potentially vulnerable to alternative attack vectors.

\subsection{Quantum Cryptography Challenges and Opportunities}

\definecolor{Mercury}{rgb}{0.901,0.901,0.901}
\begin{table*}
\centering
\caption{Summary of challenges and mitigation strategies in  quantum cryptography.}
\label{tab:systematic_challenges}
\begin{tblr}{
  width = \linewidth,
  colspec = {Q[68]Q[370]Q[450]},
  row{1} = {Mercury,font=\bfseries},
  cell{2}{1} = {r=4}{},
  cell{6}{1} = {r=5}{},
  cell{11}{1} = {r=4}{},
  hlines,
  vlines,
}
Category        & Specific Challenge                                                                                                                                             & Mitigation / Research Direction                                                                                                                                                                                                                                                                                                                                                       \\ 
\textit{Efficiency}      
                & Low secret key rate (SKR) and transmission throughput.
                
                & Physical layer upgrades: high system operating rates, high-efficiency detectors (SNSPDs), higher-dimensional protocols, multiplexing techniques and minimal channel loss regime operation.\\
                
                & Limitations in software post-processing, storage/retrieval overhead in commercial devices, and security degradation when key expansion using key derivation functions (KDF).
                
                &{\labelitemi\hspace{\dimexpr\labelsep+0.5\tabcolsep}Real-time hardware-accelerated post-processing,\\
                \labelitemi\hspace{\dimexpr\labelsep+0.5\tabcolsep}Optimized APIs and storage standards,\\
                \labelitemi\hspace{\dimexpr\labelsep+0.5\tabcolsep}Key derivation functions,\\
                \labelitemi\hspace{\dimexpr\labelsep+0.5\tabcolsep}Restricted eavesdropping protocols.}\\
                
                & Even worse scaling of communication rates in most direct communication protocols due to design.
                
                & Transitioning to higher communication rate quantum keyless private communication (QKPC), with higher noise-tolerance allowing daylight operations; exploring efficient capacity achieving forward error correction (FEC).\\
                
                & Engineering and routing bottlenecks in large-scale wavelength division multiplexing (WDM) networks, wavelength planning constraints, and key-pool scheduling.
                
                & {\labelitemi\hspace{\dimexpr\labelsep+0.5\tabcolsep}Optimal placement of Trusted Repeater Nodes,\\
                \labelitemi\hspace{\dimexpr\labelsep+0.5\tabcolsep}Multi-path QKD with different key reconstruction approaches,\\
                \labelitemi\hspace{\dimexpr\labelsep+0.5\tabcolsep}Quantum key pools,\\
                \labelitemi\hspace{\dimexpr\labelsep+0.5\tabcolsep}Dynamic integer-linear programming (ILP)-based optimizations.}\\

\!\textit{Deployment}
                & High capital and operational expenditure (CapEx/OpEx) of quantum links, the intermediary nodes as well as the equipment at each node (e.g., single photon detectors).
                
                & Exploring simpler protocol designs, e.g., simplified-BB84 QKD, on-off-keying (OOK)-modulation QKPC (for line-of-sight satellite links) and photonic-integrated circuits (PICs)-based transceivers for cost-effective setups; cost-optimized network architectures.\\
                
                & Advanced hardware dependencies in protocols (e.g., quantum memory in QSDC/QDS) or impractical trusted third-party verification architectures.
                
                & {\labelitemi\hspace{\dimexpr\labelsep+0.5\tabcolsep}Developing quantum memory-free schemes (often with other assumptions),~\\
                \labelitemi\hspace{\dimexpr\labelsep+0.5\tabcolsep}Alternatives such as optical delay lines-based storage, until further advances in quantum memories.                
                }\\

                & Lack of direct compatibility and multi-layer interoperability between commercial QKD devices from different vendors.
                
                & Implementing intermediary KMEs to translate, manage, and distribute secret keys while adapting to application demands.\\
                
                & Physical transmission distance limitations and complex orchestration of global heterogeneous quantum networks.
                
                & {\labelitemi\hspace{\dimexpr\labelsep+0.5\tabcolsep}Combining ultra-low-loss fibers, satellite free-space links, placement of repeater nodes,\\
                \labelitemi\hspace{\dimexpr\labelsep+0.5\tabcolsep}Optimizing routing, SKR demands, wavelength planning, etc.}\\
                
                & Lack of commercial standards despite early milestones, and a lack of mechanisms for quantum cryptographic agility across distinct security models.
                
                & {\labelitemi\hspace{\dimexpr\labelsep+0.5\tabcolsep}Advancing standardization efforts via global bodies\! (ETSI, ITU),\\
                \labelitemi\hspace{\dimexpr\labelsep+0.5\tabcolsep}Designing flexible network architectures that allow on-the-fly switching between quantum protocols.}\\

\!\!\!\!\textit{Vulnerabilities} 
                & Side-channel attacks (SCA) exploiting flaws in optical as well as electronic components of non-ideal devices (optical state preparation, electromagnetic radiation or power analysis on single-photon detectors and FPGAs, etc.).
                
                & {\labelitemi\hspace{\dimexpr\labelsep+0.5\tabcolsep}Hardware/software patching in commercial devices for known attacks,\\
                \labelitemi\hspace{\dimexpr\labelsep+0.5\tabcolsep}Regular upgrades through hack-and-pack approach,\\                %
                \labelitemi\hspace{\dimexpr\labelsep+0.5\tabcolsep}Randomizing operational parameters by the user,\\
                \labelitemi\hspace{\dimexpr\labelsep+0.5\tabcolsep}Exploring protocols that are inherently secure and consider devices as a ``black box'', namely DI and MDI-QKD.}\\
                
                & Different security features depending on the application, such as stand-alone vs composable security, information-theoretic vs computational, etc.
                
                & {\labelitemi\hspace{\dimexpr\labelsep+0.5\tabcolsep}Composable security under assumptions, and stand-alone security alternatives otherwise,\\
                \labelitemi\hspace{\dimexpr\labelsep+0.5\tabcolsep}In case of impossibility results for unconditional secure quantum protocols, exploring quantum resources in computationally secure ones.}\\

                & Software-layer vulnerabilities where distributed keys are consumed, including DDoS susceptibility in KMS, OS/application retrieval flaws, and memory exposure.
                
                & {\labelitemi\hspace{\dimexpr\labelsep+0.5\tabcolsep}Implementation of robust memory protection mechanisms,\\
                \labelitemi\hspace{\dimexpr\labelsep+0.5\tabcolsep}Secure cryptographic library designs,\\
                \labelitemi\hspace{\dimexpr\labelsep+0.5\tabcolsep}Classical network security defenses tailored for KMS.}\\
                
                & Practical security bottlenecks in establishing and securing the classical authenticated channel required by quantum protocols.
                
                & Relying on short-term secure classical links, pre-shared key infrastructures, physical-layer security primitives (e.g., physical unclonable functions) and resilient PQC-based authentication primitives.
\end{tblr}
\end{table*}


\subsubsection{Efficiency} 
\label{subsec:quant_eff}

One of the primary efficiency challenges facing quantum communication protocols is the low transmission rate imposed by high channel losses over long distances. Quantum signals, typically single photons or weak coherent pulses carrying a very low average photon number, are inherently susceptible to loss and noise. In optical fibers, attenuation increases exponentially with distance, severely constraining achievable communication ranges; state-of-the-art ultra-low-loss fibers achieve an attenuation of approximately 0.16 dB/km at the telecom wavelength of 1550 nm. Free-space links offer certain advantages over fiber-based transmission, but their performance is heavily influenced by environmental factors such as atmospheric turbulence, weather conditions, and background illumination, all of which introduce additional losses and signal fluctuations. Beyond channel losses, intrinsic experimental losses, most notably limited detector efficiency, further degrade overall system performance. Single-photon detectors introduce additional constraints through dark-count noise and detector dead time, both of which place practical upper bounds on achievable transmission rates. 

In the context of QKD implementations, these channel and detector limitations translate directly into low secret key generation throughput. While QKD-generated keys are commonly used to seed efficient symmetric encryption schemes such as AES, many applications may demand higher key generation rates. This requirement becomes particularly important when perfect secrecy is desired: one-time-pad encryption mandates keys of equal length to the plaintext and strictly prohibits key reuse, meaning that achieving perfect secrecy at data rates of gigabits per second or higher would require secret key rates (SKR) far beyond what current state-of-the-art QKD systems can deliver. Closing this gap remains one of the central open challenges in the practical deployment of quantum-secure communications. 

Most approaches to increase QKD throughput operate in the physical layer \cite{kodukhov2023boosting, li2023high}. Representative strategies include raising the system clock rate, deploying state-of-the-art single-photon detectors, using ultra-low-loss optical fibers, frequency multiplexing, multi-mode and multi-core fiber architectures, higher-dimensional QKD encoding, and protocols with greater tolerance to QBER. Among these, single-photon detector performance, characterized by detection efficiency, dead time, jitter, after-pulsing probability and dark count rate, is among the principal implementation bottlenecks. Superconducting nanowire single-photon detectors (SNSPDs) excel in all three metrics \cite{marsili2013detecting}, and the current benchmark of 115.8 Mbps SKR over 10 km has been achieved using multi-pixel SNSPDs, which enable even higher photon detection rates \cite{li2023high}. Complementary gains have been demonstrated through wavelength and spatial multiplexing techniques, as well as hybrid architectures that combine both \cite{dynes2016quantum, xavier2020quantum, joshi2020trusted}. However, most of these approaches come at the cost of increased resource overhead. From a fundamental point of view, repeaterless QKD over a direct pure-loss optical channel is fundamentally constrained by the PLOB bound in terms of the maximum achievable SKR \cite{pirandola2017fundamental}, though careful protocol design can push practical implementations closer to this theoretical limit.

Beyond physical-layer improvements, advances in classical post-processing can also meaningfully enhance QKD system performance. Faster hardware-accelerated post-processing enables real-time operation and reduces latency, which is critical for applications with stringent SKR requirements. In the software domain, several factors impose additional constraints on secret key throughput. The mechanisms by which keys are stored and retrieved from QKD devices can introduce non-trivial overhead, motivating the need for well-designed APIs, protocols, and standardized interfaces. The use of key derivation functions (KDFs) seeded with QKD-generated key material offers another avenue to increase effective SKR \cite{chen2024q}, though this comes at the cost of reduced cryptographic security relative to purely QKD-generated keys, a trade-off that must be carefully evaluated against application requirements. That is, while KDF-based key expansion can increase the amount of application keying material derived from QKD output, it does not increase the information-theoretic secure SKR of the QKD system; instead, it changes the resulting security guarantee from purely information-theoretic to one that additionally relies on computational assumptions. Finally, QKD protocols operating under relaxed eavesdropping models have been proposed \cite{pan2020secret, ghalaii2022realistic} as a means to achieve higher key generation rates, making them suitable for use cases that prioritize throughput over the strongest security guarantee. 

In quantum protocols for direct communication, transmission rates correspond to the rate at which information is transferred between users since no encryption step is required. A direct comparison with QKD is therefore not entirely fair; nonetheless, the direct communication protocols of QSDC offer no clear advantage over QKD in this respect. In fact, the two-way block-based transmission scheme used in QSDC protocols incurs higher channel losses than some QKD protocols \cite{pan2023free, pan2024evolution}, so most existing QSDC schemes scale with communication rate no better than QKD does. Under its more restrictive physical-layer adversarial model, QKPC can achieve substantially higher communication rates and greater background-noise tolerance than QKD in relevant free-space regimes \cite{vazquez2021quantum}. This tolerance also enables QKPC to operate in daylight at higher rates than QKD \cite{mendes2025quantum,neto2026daylight}, whereas QKD usually requires top-of-the-line single-photon detectors and filtering techniques, as noted above. Finally, practical direct communication protocols generally require forward error correction (FEC), where efficient error-correcting codes and their hardware-accelerated implementations can further improve message transmission rates and reduce latency.

Multi-path QKD is an avenue worth exploring, in which multiple paths between end users can enhance SKRs; the challenges here involve managing trust assumptions about trusted repeater nodes and designing flexible key reconstruction schemes accordingly \cite{wang2023segment}. WDM networks, in turn, require careful wavelength planning within a tight spectral band using DWDM techniques to fully utilize the available wavelength range in the optical fiber connections. However, this can create bottlenecks in large networks depending on demand and traffic constraints. A second set of challenges concerns efficiently managing the buffers in which keys are stored and finding optimal scheduling to route them. For example, the concept of a quantum key pool (QKP) has been proposed for optimal timely scheduling of generated secret keys \cite{cao2017key, cao2018time}. Better solutions, in both the physical and logical domains, remain relevant research directions — for instance, finding optimal ways to multiplex classical and quantum signals over the same fiber \cite{xavier2020quantum}, which can be formalized as integer linear programming (ILP) problems to dynamically adapt fiber multiplexing.

\subsubsection{Deployment}
\label{subsec:quant_dep}

The implementation of quantum protocols inherently requires specialized hardware and complex electronics with precise control, irrespective of technological maturity. Even QKD, currently the most mature quantum technology, continues to face challenges in the configuration and deployment of practical systems, relying on dedicated optical transmitters - typically attenuated coherent-state laser sources in practical prepare-and-measure systems, or entangled photon sources in entanglement-based architectures - and sensitive photodetectors such as single-photon avalanche diodes (SPADs). Many QKD field trials have further used state-of-the-art SNSPDs to extend achievable communication distances; while commercially available SPADs are already considerably more expensive than conventional telecommunication photodetectors, SNSPDs entail an even greater cost increase, along with cryogenic temperature requirements (on the order of a few Kelvin). Consequently, the large-scale deployment of QKD networks remains a challenge from both economic and operational perspectives. These constraints motivate the exploration of quantum protocols with simpler setups. QKPC is one such example, combining standard modulation techniques such as on-off keying (OOK), widely used in classical optical communication systems, with a minimal number of SPADs at the receiver end, resulting in a relatively cost-effective and simple solution. QKPC was originally proposed specifically for a satellite-to-ground scenario \cite{vazquez2021quantum}. However, this design choice aligns well with practical requirements, as compact and resource-efficient transmitters are particularly advantageous for satellite platforms \cite{mendes2024optical}. QKPC's restriction to line-of-sight (LoS) free-space links should not necessarily be regarded as a limitation, since satellite links are expected to play an important role in extending quantum-secure communication to global scales.

Certain quantum protocols may also require advanced quantum hardware. A first example is QSDC, where block transmission of quantum states requires storing photons in quantum memories between the eavesdropping-detection and information-encoding steps \cite{pan2023free, pan2024evolution}. Some demonstrations have used optical delay lines for short storage times, while other memory-free schemes have been proposed; the latter, however, require pre-shared keys between the two users \cite{sun2018design}, posing a separate practical challenge between two arbitrary users in a network. Early QDS proposals relied on long-term quantum memories and quantum-state comparison operations. Memory-free QDS protocols were subsequently developed, but practical challenges remain, including multi-user distribution, trust assumptions, transferability, and in some schemes the requirement for trusted third parties or pre-designated verifiers \cite{du2025chip, cao2024experimental}. With respect to commercial equipment for deployment in quantum communication networks, interoperability among commercial QKD systems presents an important practical limitation. Devices developed by different vendors or based on distinct standards are often not directly compatible, further complicating large-scale deployment, with interoperability challenges arising at multiple layers of the system architecture. A standardized key management entity (KME) layer can abstract vendor-specific QKD systems from applications by exposing interoperable key-management and key-delivery interfaces, while handling buffering, allocation and distribution of generated keys. 

In addition, existing technology imposes limits on transmission distance, which is critical from a global network perspective. Common solutions for achieving longer distances include ultra-low-loss optical fibers, free-space satellite links, and intermediary trusted or untrusted repeaters. Finding the best combination of techniques for a given deployment, considering distance and throughput, is itself a research challenge, as is the methodology for doing so across different scenarios. Additional important aspects include strategizing routing, wavelength allocation, deployment and operational expenditures, resource allocation, link availability, and repeater node placement, among others \cite{zhao2018resource, zhang2023routing}.
While cryptographic agility has mostly been addressed in the context of PQC, where hybrid PQC–quantum protocols are appealing, it is worth exploring for quantum cryptography as well \cite{galambos2025hardware}. Building on some of the building blocks discussed in the previous section, network architectures should also incorporate methods to enable crypto agility among different quantum protocols. This is even more relevant for quantum protocols that differ in their security assumptions and corresponding communication rates, making their use scenario-specific. 

Lastly, with respect to future widespread deployment of quantum technologies, an important challenge is that of standardization. This is being tackled through standardization efforts by various entities, such as the European Telecommunications Standards Institute (ETSI) in Europe and the International Telecommunication Union (ITU) globally. Despite substantial progress \cite{saez2024current} in ETSI, ITU-T and other standardization bodies, broad cross-vendor interoperability, mature certification ecosystems, harmonization across standards bodies, and widely adopted deployment profiles remain incomplete.

\subsubsection{Vulnerabilities} 
\label{subsec:quant_vuln}

Many quantum cryptographic protocols admit information-theoretic proofs under explicit assumptions about the devices, channels, laboratories, and the adversary. Practical vulnerabilities arise when an implementation deviates from the model used in the security proof \cite{xu2020secure}. Side-channel attacks (SCA) exploiting physical vulnerabilities are a big threat for all quantum protocols, similar to classical and post-quantum ones. These include the timing, power-monitoring and electromagnetic emission-based attacks, as mentioned previously.

QKD systems, consisting of optics and electronics components, have had longstanding research on studying side-channel attacks by finding vulnerabilities in these components. There can further be attacks in the post-processing layers in quantum protocols, in hardware or software. Non-ideal optical devices for state preparation such as weak coherent pulses, as well as measurement such as SPADs can result attacks in the optical domain. Whereas attacks based on electromagnetic emissions can be used on SPADs, while other electronic components such as FPGAs can be analyzed for both power consumption and electromagnetic emissions \cite{durak2021attack, baliuka2023deep, pantoja2024electromagnetic}. Many of theses threats' countermeasures are being investigated in the literature as they are identified. In fact, some of these optical side-channel attacks were demonstrated against commercial QKD equipment \cite{zhao2008quantum, lydersen2010hacking}. Although many vendors have mitigated a large class of known attacks associated to some QKD components in their newer generations of equipment by modifying the hardware and/or software \cite{lim2015random}, those patches may not work for future unknown attacks, requiring regular upgrades.

The ultimate fix is to use protocols whose security make no assumptions on the working of different devices, i.e., consider them as ``black boxes". Device-independent QKD (DI-QKD) removes the need to characterize the internal operation of the quantum devices, at the price of additional operational assumptions and demanding experimental requirements. While DI-QKD provides security without making any assumptions on the devices, its implementation is extremely challenging \cite{zhang2022device}: a recent proof-of-principle experiment demonstrated positive finite-size DI-QKD key generation over 10 km, with positive asymptotic rates predicted beyond 100 km \cite{liu2026building}. There are other types of DI protocols, which are not full DI meaning that we do not drop all assumptions about the devices, but we stick to minimal assumptions which are physically reasonable. For example, in semi-device independent QKD (SDI-QKD), usually a minimal assumption is made about the source, for example that the dimension of the states produced by it is upper bounded. Measurement Device-Independent QKD (MDI-QKD) is substantially more practical than DI-QKD and eliminates detector side channel attacks while retaining assumptions on the devices used for state-preparation. This improved security generally comes with a secret key rate penalty relative to optimized conventional prepare-and-measure QKD. Regarding other quantum protocols, there have been proposals for MDI versions of QSDC \cite{zhou2020measurement} as well as QDS \cite{roberts2017experimental, zhang2021twin}.

While cryptographic primitives such as QOT, are very useful for serving as a fundamental building block for advanced cryptographic tasks such as MPC, the security of the resulting ``composed'' protocols should also be ensured. That is, when a primitive is employed as a subroutine within a larger protocol, the security of the overall construction becomes critical. Composable security guarantees that the security properties of individual components are preserved when they are integrated into more complex protocols, typically under security assumptions such as common random reference string (CRS) \cite{canetti2000universally}. However, proving composable security is not always feasible. In such cases, only standalone security can be established, demonstrating that the protocol satisfies its intended security properties when executed in isolation, without guaranteeing that these properties are preserved under arbitrary composition. Analogously, for the cases where impossibility results are shown for information-theoretic security of some quantum protocols, such as QOT and bit-commitment, computational hardness assumptions of public-key cryptography can be replaced by weaker ones \cite{grilo2021oblivious, lemus2025performance}.

On the software layers, where the keys distributed through QKD are consumed, we identified only a small portion of research, leaving space for DDoS to the KMSs, for example. The way the keys are retrieved by the key management systems, OS or application themselves, can also be susceptible to threats. Memory should be protected, and the libraries and applications should also be prepared to avoid attacks. This is already common for existing classical cryptography \cite{citation-nsdi}, and may be no different for QKD. 

Lastly, as mentioned in the previous section \ref{subsec:hybrid}, security of the classical authenticated channel used in most quantum  protocols is another challenge from a practical point of view. While only short-term security of such classical links suffices, from a network perspective, establishing such a link and securing it still poses as a big challenge. 



\section{Conclusion}

The potential emergence of fault-tolerant quantum computers may break widely used cryptographic algorithms. However, numerous challenges and limitations remain in adopting both quantum-safe cryptography mechanisms. In this work, we examine common building blocks adopted in the migration to quantum-safe mechanisms.  We also systematically analyzed applications that have adopted quantum-resistant schemes in their cryptographic protocols and classified them by the application domain, cryptographic scheme, and maturity level. We also discussed challenges and future research directions across different aspects, including efficiency, deployment, and security.

\textbf{Acknowledgements.} This work was partially funded by the project ``QuIIN Integração CV-QKD com Redes Clássicas'' supported by QuIIN - Inovação Industrial Quântica, Centro de Competência EMBRAPII CIMATEC em Tecnologias Quânticas, with financial resources from the PPI IoT/Manufatura 4.0 under MCTI call for proposals No. 053/2023, established with EMBRAPII. The authors used ChatGPT and Gemini
to assist with small editing. All content remain the original work of the authors. This work is funded by national funds through FCT – Fundação para a Ciência e a Tecnologia, I.P., and, when eligible, co-funded by EU funds under project/support UID/50008/2025 – Instituto de Telecomunicações, with DOI identifier https://doi.org/10.54499/UID/50008/2025, and by the
European Union through the Iberian Quantum Communication Infrastructure project, IberianQCI (CEF-DIG2024-EUROQCI-WORKS, Grant No. 101249593).

\bibliographystyle{ieeetr}
\bibliography{IEEEexample}

\section*{List of Abbreviations}
\begin{description}[leftmargin=5em, style=nextline]
\item[AES] Advanced Encryption Standard \item[AI] Artificial Intelligence \item[API] Application Programming Interface \item[CAN] Controller Area Network \item[CV-QKD] Continuous-Variable Quantum Key Distribution \item[DLT] Distributed Ledger Technology \item[DNS] Domain Name System \item[DOI] Digital Object Identifier \item[DV-QKD] Discrete-Variable Quantum Key Distribution \item[ECC] Elliptic Curve Cryptography \item[ETSI] European Telecommunications Standards Institute \item[EV] Electric Vehicle \item[FIPS] Federal Information Processing Standards \item[HE] Homomorphic Encryption \item[HQC] Hamming Quasi-Cyclic \item[IETF] Internet Engineering Task Force \item[ILP] Integer Linear Programming \item[IoT] Internet of Things \item[IoV] Internet of Vehicles \item[IPsec] Internet Protocol Security \item[ISO] International Organization for Standardization \item[ITU] International Telecommunication Union \item[KEM] Key Encapsulation Mechanism \item[KME] Key Management Entity \item[KMS] Key Management System \item[L-OTS] Lamport One-Time Signature \item[LDACS] L-band Digital Aeronautical Communication System \item[LWE] Learning With Errors \item[MAC] Message Authentication Code \item[ML-DSA] Module-Lattice-Based Digital Signature Algorithm \item[ML-KEM] Module-Lattice-Based Key Encapsulation Mechanism \item[MPC] Multi-party Computation \item[MQTT] Message Queuing Telemetry Transport \item[NIST] National Institute of Standards and Technology \item[NTT] Number Theoretic Transform \item[OCS] Optical Circuit Switching \item[OT] Oblivious Transfer \item[PKI] Public Key Infrastructure \item[PPC] Privacy-Preserving Computation \item[PQC] Post-Quantum Cryptography \item[PUF] Physical Unclonable Function \item[QKD] Quantum Key Distribution \item[QOT] Quantum Oblivious Transfer \item[QoS] Quality of Service \item[QOT] Quantum Optical Networks \item[QRNG] Quantum Random Number Generator \item[RSA] Rivest-Shamir-Adleman \item[SCADA] Supervisory Control and Data Acquisition \item[SDN] Software-Defined Networking \item[SH-DSA] Stateless Hash-Based Digital Signature Algorithm \item[SSH] Secure Shell \item[SVP] Shortest Vector Problem \item[TEE] Trusted Execution Environment \item[TLS] Transport Layer Security \item[TNN] Trusted-Node Network \item[TPM] Trusted Platform Module \item[V2X] Vehicle-to-Everything \item[VNF] Virtualized Network Function \item[VPN] Virtual Private Network \item[WDM] Wavelength Division Multiplexing \item[ZKP] Zero-Knowledge Proof 
\end{description}
\end{document}